\documentclass[twoside,12pt]{article}
\usepackage[english]{babel} 
\usepackage{amsmath,amsfonts,amsthm,bm} 
\usepackage{epsfig}
\usepackage{cite}
\newcommand{\be}{\begin{equation}}
\newcommand{\ee}{\end{equation}}
\newcommand{\bea}{\begin{eqnarray}}
\newcommand{\eea}{\end{eqnarray}}

\newcommand{\epgreact}{e p \to e p \gamma}

\newcommand{\vcsreact}{\gamma^* p \to  \gamma p}
\newcommand{\vcsreactnucleon}{\gamma^* N \to N \gamma}
\newcommand{\eppionreact}{e p \to e p \pi^0}
\newcommand{\qcm}{q_\mathrm{cm}}
\newcommand{\qpr}{q_\mathrm{cm}^{\prime}}
\newcommand{\qprpowminusone}{q_\mathrm{cm}^{\prime \ -1}}
\newcommand{\qprpowzero}{q_\mathrm{cm}^{\prime \, 0}}
\newcommand{\qprpowone}{q_\mathrm{cm}^{\prime }}
\newcommand{\qprpowtwo}{q_\mathrm{cm}^{\prime \, 2}}
\newcommand{\thcm}{\theta_{\mathrm{cm}}}
\newcommand{\phicm}{\varphi}
\newcommand{\cthcm}{\cos \theta_{\mathrm{cm}}}
\newcommand{\qsqtild}{\tilde{Q} ^2}
\newcommand{\qtild}{\tilde{Q}}
\newcommand{\qzerotild}{ \tilde{q}_{0 \mathrm{cm}}}
\newcommand{\pllptte}{P_{LL} -P_{TT} / \epsilon}
\newcommand{\ptt}{P_{TT}}
\newcommand{\pll}{P_{LL}}
\newcommand{\plt}{P_{LT}}
\newcommand{\pltperpprim}{P_{LT}^{' \perp}}
\newcommand{\pltz}{P_{LT}^{z}}
\newcommand{\pltzprim}{P_{LT}^{' z}}
\newcommand{\pltperp}{P_{LT}^{\perp }}
\newcommand{\pttperp}{P_{TT}^{\perp}}
\newcommand{\pttperpprim}{P_{TT}^{' \perp}}
\newcommand{\pzcm}{{\cal P}_z^\mathrm{cm}}
\newcommand{\pxcm}{{\cal P}_x^\mathrm{cm}}
\newcommand{\pycm}{{\cal P}_y^\mathrm{cm}}
\newcommand{\picm}{{\cal P}_i^\mathrm{cm}}

\newcommand{\la}{\Lambda_{\alpha}}
\newcommand{\lb}{\Lambda_{\beta}}

\newcommand{\aeq}{\alpha_{\mathrm{E1}}(Q^2)}
\newcommand{\bmq}{\beta_{\mathrm{M1}}(Q^2)}
\newcommand{\ale}{\alpha_{\mathrm{E1}}}
\newcommand{\bem}{\beta_{\mathrm{M1}}}
\newcommand{\ohigher}{{\cal O}(q_\mathrm{cm}^{\prime \, 2})}
\newcommand{\ohigherdr}{{\cal O}(q_\mathrm{cm}^{\prime \, 2})_{\mathrm{DR}}}
\newcommand{\vll}{V_{LL}}
\newcommand{\vlt}{V_{LT}}
\newcommand{\sigbhb}{d \sigma_{\mathrm{BH+Born}}}

\newcommand{\sigexp}{d \sigma_{\mathrm{exp}}} 
\newcommand{\sigdr}{d \sigma_{\mathrm{DR}}}
\newcommand{\siglex}{d \sigma_{\mathrm{LEX}}} 
\newcommand{\sigpol}{d \sigma } 
\newcommand{\deltam}{\delta {\cal{M}}}
\newcommand{\aqed}{\alpha_{QED}}
\newcommand{\mnucleon}{M_N}
\newcommand{\rsqaz}{\langle r^2 _{\alpha_{E1}} \rangle }
\newcommand{\rsqbz}{\langle r^2 _{\beta_{M1}} \rangle }
\begin{document}

\title{ \vspace{1cm} Virtual Compton Scattering \\
and Nucleon Generalized Polarizabilities \\
}
\author{ H.\ Fonvieille,$^{1}$ B.\ Pasquini,$^{2,3}$ N.\ Sparveris$^{4}$ \\
\\
$^1$Universit\'e Clermont Auvergne, CNRS/IN2P3, LPC, Clermont-Ferrand, France \\
$^2$Dipartimento di Fisica Nucleare e Teorica, Universit\'a degli Studi di Pavia  \\
$^3$ Istituto Nazionale di Fisica Nucleare, Sezione di Pavia, Italy \\
$^4$ Temple University, Philadelphia, PA 19122, USA }
\maketitle

\begin{abstract}
This review gives an update on virtual Compton scattering (VCS) off the nucleon, $\vcsreactnucleon$, in the low-energy regime. We recall the theoretical formalism related to the generalized polarizabilities (GPs) and model predictions for these observables. We present the GP extraction methods that are used in the experiments: the approach based on the low-energy theorem for VCS and the formalism of Dispersion Relations. We then review the experimental results, with a focus on the progress brought by recent experimental data on proton GPs, and we conclude by some perspectives in the field of VCS at low energy.
\end{abstract}

\tableofcontents

\ \\

\section{Introduction}

Virtual Compton Scattering (VCS) on the nucleon became a well-identified field of hadron physics  in the 1990's. After first conceptual attempts motivated by the Pegasys and Elfe{~\cite{Arvieux:1995kw} projects,
the field built itself on two different energy regimes: the near threshold  with the concept of generalized polarizabilities (GPs), and  the  high-energy, high-$Q^2$ regime of deeply virtual Compton scattering. 
The whole field has seen a continuous and fruitful development with a wealth of new observables to explore the nucleon structure. This review focuses on VCS at low energy and the generalized polarizabilities (GPs) of the nucleon,  a topic that has seen substantial progress in recent years.   
In light of new experimental data, a consistent picture of the proton scalar GPs starts to emerge and it is an appropriate time to review our knowledge of these observables.

This article  will be usefully complemented by previous reviews, addressing the subject in more or less depth. Through the two main references connected to the present article, i.e., Refs.~\cite{Guichon:1998xv,Drechsel:2002ar}, 
and a non-exhaustive list of other reviews~\cite{dHose:2000vgm,Fonvieille:2004rb,HydeWright:2004gh,dHose:2006bos,Downie:2011mm,Hagelstein:2015egb,Pasquini:2018wbl}, 
the reader can seize the temporal evolution of the field.
The present article combines a summary of the theoretical framework and the experimental status. The new aspects concern the focus on the use of the Dispersion Relation (DR)  model \cite{Drechsel:2002ar} in VCS experiments, and an update of experimental results.

\section{Notations and Kinematics}
\label{sec-notations-and-kinematics}

In all of the following, we consider a proton target, although the same considerations are applicable to the neutron target. The particle four-momentum vectors are denoted as: $k^\mu$ and $k'^\mu$ for the incoming and scattered electrons, $q^\mu$ and $q'^\mu$ for the virtual photon and final real photon, $p^\mu$ and $p'^\mu$  for the initial and final protons. 
The modulus of the three-momenta is denoted as $q=\vert\mathbf q\vert$ , etc. Variables are indexed ``lab'' (or not indexed) in the laboratory frame, where the initial proton is at rest. They are indexed ``cm'' in the center-of-mass frame (c.m.) of the (initial proton + virtual photon), i.e. the c.m. of the Compton process $\vcsreact$.

The kinematics of the  ($\epgreact$) reaction are defined by five independent variables
\footnote{Note that this is up to a rotation of the leptonic plane by the azimuthal angle $\varphi'_{e \, \mathrm{lab}}$ of the scattered electron. If specified, $\varphi'_{e \, \mathrm{lab}}$  is then a sixth independent variable.}. 
We will adopt the most usual set of variables:  $(\qcm, \qpr, \epsilon, \cthcm, \phicm )$ where $\epsilon$ is the virtual photon polarization parameter, i.e.,  $\epsilon = 1 / [ 1+2 {q_{\mathrm{lab}}^2 \over Q^2} \tan^2 (\theta'_{e \, \mathrm{lab}}/2) ]$, and
$\qcm$ and $\qpr$ are the three-momentum modulus of the virtual photon and final photon in the c.m., respectively. $\thcm$ and $\phicm$ are the angles of the Compton process, i.e., the polar and azimuthal angles of the outgoing real photon w.r.t. the virtual photon in the c.m.
\footnote{Other sets of variables are sometimes used, such as $(k_{\mathrm{lab}}, k'_{\mathrm{lab}}, \theta'_{e \, \mathrm{lab}})$ or $(Q^2, \sqrt{s}, \epsilon )$ for the leptonic vertex, and $t$ instead of $\cthcm$.}, 
see Fig.~\ref{fig-kinematics-1}. The triplet ($\qcm, \qpr, \epsilon )$ defines the leptonic vertex $e \to e' \gamma^*$.  The c.m. total energy is  $W=\sqrt{s}$, and $\mnucleon$ is the nucleon mass.

At fixed beam energy, the five-fold differential cross section is  
$d^5 \sigma / ( d k'_{\mathrm{lab}} d \cos \theta'_{e \, \mathrm{lab}} d \varphi'_{e\, \mathrm{lab}} d \cthcm d \phicm)$  and will be denoted $d \sigma$ for simplicity.


Since the GPs are defined from the VCS amplitude  in the limit of $\qpr = 0$, keeping $\qcm$ fixed (cf. Sect.~\ref{sec-the-nonborn-amplitude-and-the-gps}), a number of kinematical variables are also defined in this limit. They are designated with a tilde in the low-energy expansion (LEX) formalism. Among them, we find the photon virtuality $Q^2$, which takes the form: $\qsqtild =  2 \mnucleon \cdot ( \sqrt{\mnucleon ^2 + \qcm ^2} - \mnucleon )$
\footnote{We also find ${\tilde Q} = \sqrt{\qsqtild}$, and $\qzerotild$ defined in the Appendix.}. 
Therefore $\qcm$ and $\qsqtild$ are equivalent variables. Throughout this article we will use the notation ``$Q^2$'' everywhere for simplicity, knowing that when $\qpr \to 0$ a ``$\qsqtild$'' is meant instead
\footnote{This is for instance the case for all experimental values of $Q^2$ in Sect.~\ref{sec-experiments}. }. 
GPs and structure functions will thus depend equivalently on $\qcm$ or $Q^2$.

\begin{figure}[tb]
\begin{center}
\begin{minipage}[t]{11 cm}
\epsfig{file=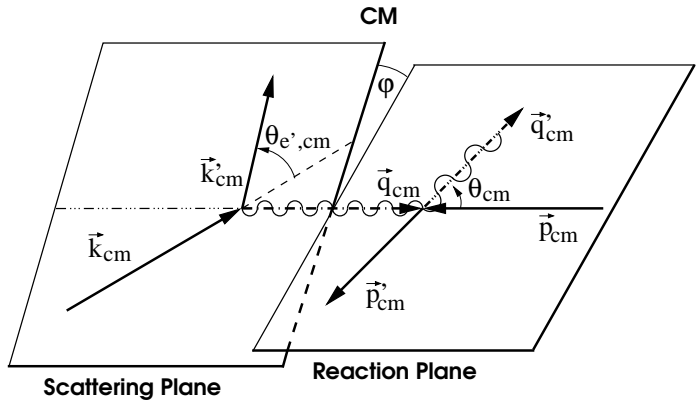,scale=0.5}
\end{minipage}
\begin{minipage}[t]{16.5 cm}
\caption{
Kinematics of the $(\epgreact )$ reaction, showing the scattering (or leptonic) plane, the reaction (or hadronic) plane, and the Compton scattering in the the center-of-mass system. For polarization experiments, axes are defined such that $\hat z _{cm}$ is along $\mathbf{q}_{\mathrm{cm}}$  and $\hat y _{cm}$ is orthogonal to the scattering plane. Figure taken from ref.~\cite{Janssens:2008qe}. 
\label{fig-kinematics-1}}
\end{minipage}
\end{center}
\end{figure}

\section{Theoretical framework}
\label{sec-theoretical-framework}

This section reviews the formalism related to the nucleon GPs. 
We first outline some basic properties of polarizabilities and their generalization to finite $Q^2$ (Sect.~\ref{sec-from-rcs-to-vcs}). Then the main ingredients of the low-energy theorem in VCS are summarized in Sect.~\ref{sec-vcs-at-low-energy-and-the-gp-formalism-the-let}. A synthetic overview of model predictions for GPs is given in Sect.~\ref{sec-theoretical-models-for-nucleon-gps}, and the DR formalism is presented in more details in Sect.~\ref{sec-the-dr-model}. We then move closer to  experimental aspects with Sects.~\ref{sec-the-epg-cross-section-and-the-gp-effect} to ~\ref{sec-qed-radiative-corrections-to-vcs}.

\subsection{From real to virtual Compton scattering}
\label{sec-from-rcs-to-vcs}

The polarizabilities of a composite object are fundamental characteristics of the system, just as its mass or shape. Among all the known properties of the nucleon,  polarizabilities have  the unique status of characterizing the nucleon dynamical response to an external electromagnetic (EM) field, describing how easy the charge and magnetization distributions inside the nucleon are distorted by the EM field.
Real Compton scattering (RCS) experiments, performed since more than 50 years, have accumulated an impressive amount of knowledge on these nucleon polarizabilities, and the field is still at the forefront of hadron physics, see for instance the recent reviews~\cite{Hagelstein:2015egb,Pasquini:2018wbl}.

It is well known that the static dipole electric $(\ale )$ and magnetic $(\bem )$ polarizabilities of the proton are very small quantities: $\ale = 11.2 \pm 0.4$ and $\bem = 2.5 \pm 0.4$ in units of $10^{-4}$ fm$^3$~\cite{Tanabashi:2018oca}, testifying to the strong binding force of QCD.  The smallness of $\bem$ relative to $\ale$ is generally understood as coming from two large contributions, of para- and dia-magnetic nature, which are of opposite sign and cancel to a large extent.

In VCS the incoming real photon is replaced by a virtual, space-like photon of four-momentum transfer squared $Q ^2$, produced by an incoming lepton. The virtual photon momentum $q$ sets the scale of the observation, while the outgoing  real photon momentum $q'$ defines the size of the EM perturbation. 
Polarizabilities are then generalized to $Q^2 \ne 0$ and acquire a meaning analogous to form factors: their Fourier transform will map out the spatial distribution density of the polarization induced by an EM field. The RCS polarizabilities are then seen as the ``net result'' of such spatial distributions, while the role of GPs is to give access to the details of these spatial dependencies.

Quoting the illustrative sentence of Ref.~\cite{Guichon:1998xv}: 
 `` ... {\it VCS at threshold can be interpreted as electron scattering by a target which is in constant electric and magnetic fields.} The physics is exactly the same as if one were performing an elastic electron scattering experiment on a target placed in between the plates of a capacitor or between the poles of a magnet.''

The GP formalism was first introduced in Ref.~\cite{Arenhoevel:1974twc} for the case of nuclei. The idea  was that nuclear excitations could  be studied in a more complete way with an incoming virtual photon, instead of a real photon. To this aim, the concept of polarizabilities as a function of excitation energy and momentum transfer was introduced. The formalism was later applied to the nucleon case in ~\cite{Guichon:1995pu}.

In contrast to elastic form factors, which are sensitive only to the ground state of the nucleon, polarizabilities (and GPs) are sensitive to its whole excitation spectrum, with the excited states contributing virtually.
For instance, in a non-relativistic approach, the electric polarizability is obtained from the quadratic Stark effect calculated at the second-order in perturbation theory as
%
\bea
 \ale \ = \ 2 \sum_{N^* \ne N} \
{ \vert
< N^* \vert D_z \vert N > \vert ^2 
\over 
E_{N^*} - E_N},
\label{alpha-nr}
\eea
%
where  $D_z$ is the electric dipole moment operator and $N^*$ indicates a nucleon resonance.

Furthermore, the polarizabilities (and GPs) are particularly suited to address the widely-used picture of the nucleon as  a quark core surrounded by a pion cloud, since both of these components can be ``seen'' and interpreted to some extent in the Compton observables, using hadron structure models. The physical content of the GPs will be discussed in more detail in Sect.~\ref{sec-theoretical-models-for-nucleon-gps}.

\subsection{VCS at low energy 
and the GP formalism: the LET}
\label{sec-vcs-at-low-energy-and-the-gp-formalism-the-let}

The pioneering work of Ref.~\cite{Guichon:1995pu} opened a new era of investigation of nucleon structure, by establishing the physics case of VCS off the nucleon for the first time and providing a way to access GPs through experiments.
We give here an overview of the formalism, the results of which will be further exploited in the experimental section.

\subsubsection{Amplitudes for the photon electroproduction process}
\label{sec-amplitudes-epgamma}

The VCS process is accessed via the exclusive photon electroproduction reaction. As the virtual photon needs to be produced by a lepton beam, VCS is always accompanied by the so-called Bethe-Heitler (BH) process, where the final photon is emitted by the lepton instead of the nucleon.
Fig.~\ref{fig-theo-graphs-1} shows the different amplitudes contributing to the  $(\epgreact)$ process: the BH graphs or bremsstrahlung from the electron(s), the VCS Born graphs or bremsstrahlung from the proton(s), and typical diagrams for the resonance excitation and non-resonant $\pi N$ contribution in the $s$-channel entering the VCS  non-Born (NB) term.

The BH and VCS Born amplitudes are entirely calculable in QED, with the nucleon electromagnetic form factors $(G_E, G_M)$ as  inputs. The non-Born amplitude  $T^{\mathrm{NB}}$ contains the physics of interest and is parametrized at low energy by the nucleon GPs. The three amplitudes add up coherently to form the total photon electroproduction amplitude: 
%
%
\bea 
T^{\epgreact} \ = \ T^{\mathrm{BH}} \ + \  T^{\mathrm{Born}} \ + \  T^{\mathrm{NB}} \ = \ T^{\mathrm{BH}} \ + \  T^{\mathrm{VCS}} \ . 
\label{eq:splitting}
\eea

The $t$-channel  involving the $\pi^0$ exchange is conventionally included in the non-Born term, i.e., in the GPs.
As has been pointed
out in~\cite{Guichon:1995pu,Scherer:1996ux}, the splitting in Eq.~(\ref{eq:splitting}) is not unique. Contributions
which are regular in the limit $q'\rightarrow 0$ and separately gauge
invariant can be shifted from the Born amplitude to the non-Born
amplitude and vice versa. Therefore, when calculating the GPs, 
one has to specify which Born
terms have been subtracted from the full amplitude, since different
Born terms lead to different numerical values of the
GPs. In our calculation we use the Born amplitude
as defined in~\cite{Guichon:1995pu}.

Finally, one will note that the virtual photon four-momentum differs in the BH process ($q_{\mathrm{BH}}^\mu = p'^\mu - p^\mu$) and in the VCS process ($q_{\mathrm{VCS}}^\mu  = k^\mu - k'^\mu$), since one has $k^\mu + p^\mu = k'^\mu + p'^\mu + q'^\mu$.

\begin{figure}[tb]
\begin{center}
\begin{minipage}[t]{12 cm}
\epsfig{file=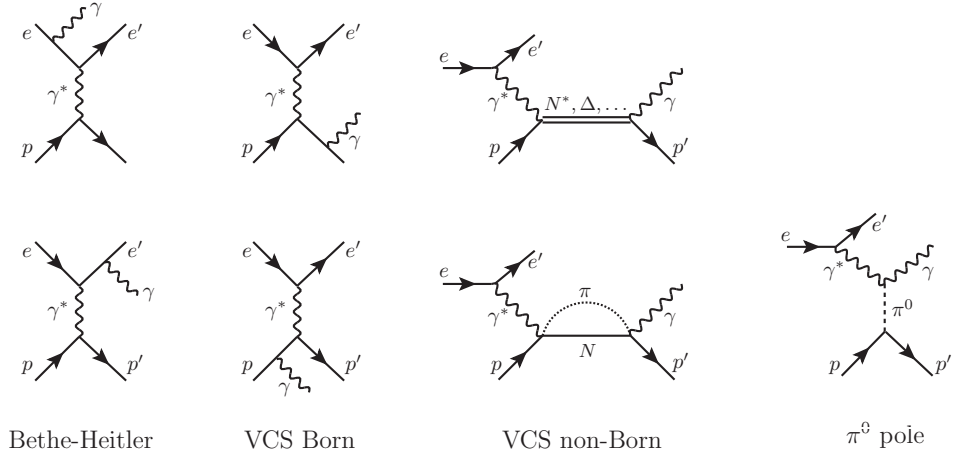,scale=0.8}
\end{minipage}
\begin{minipage}[t]{16.5 cm}
\caption{
Diagrams for the BH term, the VCS Born contribution,  a typical resonance excitation and non-resonant pion-nucleon contribution entering the non-Born term and  the $\pi^0$-pole contribution in the $t$ channel.
\label{fig-theo-graphs-1}}
\end{minipage}
\end{center}
\end{figure}

\subsubsection{The non-Born amplitudes and the GPs}
\label{sec-the-nonborn-amplitude-and-the-gps}

We briefly recall how the nucleon GPs are introduced in the work of Ref.~\cite{Guichon:1995pu} and later works. A multipole expansion of the non-Born amplitude $H_{\mathrm{NB}}$ is performed  in the c.m. frame, yielding the multipoles $H_{\mathrm{NB}}^{ ( \rho ' L' , \rho L ) S } (\qpr, \qcm )$. Here $L \,(L')$ represents the angular momentum of the initial (final) electromagnetic transition in the $(\vcsreact )$ process, whereas $S$ differentiates between the spin-flip ($S=1$) or non spin-flip ($S=0$) transition at the nucleon side. $[ \rho \,(\rho ') = 0,1,2 ]$ characterizes the longitudinal $(L)$, electric $(E)$ or magnetic $(M)$ nature of the initial (final) photon.

GPs are obtained as the limit of these multipoles when $\qpr$ tends to zero, at arbitrary fixed $\qcm$. At this strict threshold, the final photon has zero frequency, its electric and magnetic fields are constant (``static field'') and the GPs represent the generalization at finite $\qcm$ of the polarizability in classical electromagnetism.

For small values of $\qpr$ one may use the dipole approximation ($L'=1$), corresponding to electric and magnetic final-state radiation that is dipolar only. In this case, angular momentum and parity conservation lead to ten different dipole GPs~\cite{Guichon:1995pu}. However, it was shown~\cite{Drechsel:1997xv,Drechsel:1998zm} that nucleon crossing combined with charge conjugation symmetry reduces the number of independent GPs to six. Table~\ref{tab-six-dipole-gps} gives the usually adopted set of the six lowest-order GPs: two scalar, or spin-averaged, or spin-independent GPs ($S=0$) and four spin-dependent, or spin-flip, or vector, or more simply spin GPs ($S=1$). They depend on $\qcm$, or equivalently on $Q^2$ (cf. Sect.~\ref{sec-notations-and-kinematics}). The notation in column 2 of Table~\ref{tab-six-dipole-gps} will be used throughout this article.
The two scalar GPs, electric and magnetic, are thus defined  as:
%
\bea
\begin{array}{lll}
\aeq &  =  & - {e^2 \over 4 \pi} \cdot  \sqrt{3 \over 2} \cdot P^{(L1,L1)0} (Q^2) \\
\bmq &  =  & - {e^2 \over 4 \pi} \cdot  \sqrt{3 \over 8} \cdot P^{(M1,M1)0} (Q^2)
\label{formula-aeq-bem}
\end{array}
\eea
%
with  $e^2 / 4 \pi = \aqed = 1/137$. At $Q^2 = 0$ they coincide with the RCS polarizabilities $\ale$  and $\bem$.

\begin{table}[h]
\begin{center}
\begin{minipage}[t]{16.5 cm}
\caption{
The standard choice for the six independent dipole GPs. Column 1 refers to the original notation, and column 2 to the more standard multipole notation. Column 3 gives the correspondence in the RCS limit, defined by $Q^2 \to 0$ or $\qcm \to 0$. }
\label{tab-six-dipole-gps}
\end{minipage}
\begin{tabular}{|l|l|c|c|}
\hline
\  $P^{(\rho ' L', \rho L ) S } (\qcm)$ \ & \ $P^{(f,i)S} (\qcm) \ $ 
& RCS limit  ($Q^2 \to 0$) \\
\hline
 \, \, $P^{(01,01)0}$ &  \, \, $P^{(L1,L1)0}$ & 
$ - {4 \pi \over e^2} \sqrt{2 \over 3} \ \ale $   \\
 \, \,  $P^{(11,11)0}$ &  \, \, $P^{(M1,M1)0}$ & 
$ - {4 \pi \over e^2} \sqrt{8 \over 3} \ \bem $   \\
  \, \, $P^{(01,01)1}$ &  \, \, $P^{(L1,L1)1}$ & 0      \\
  \, \, $P^{(11,11)1}$ &  \, \, $P^{(M1,M1)1}$ & 0    \\
 \, \,  $P^{(01,12)1}$ &  \, \, $P^{(L1,M2)1}$ & 
$ - {4 \pi \over e^2} {\sqrt{2} \over 3} \ \gamma_{E1M2} $   \\
  \, \, $P^{(11,02)1}$ &  \, \, $P^{(M1,L2)1}$ & 
$ - {4 \pi \over e^2} { 2 \sqrt{2} \over 3 \sqrt{3}} \gamma_{M1E2} $ \\
\hline
\end{tabular}
\end{center} 
\end{table}

\smallskip
\smallskip
\smallskip

An alternative approach to analyzing low-energy VCS process  was proposed in  Ref.~\cite{Lvov:2001zdg} for zero-spin targets, and it was further extended to spin-$1/2$ targets  (and therefore to the spin-dependent GPs) in Ref.~\cite{Gorchtein:2009wz}. 
It is based on a Lorentz covariant description of the Compton amplitudes, which are expanded on a basis written in terms of electromagnetic field strength tensors.  Working with this basis, one finds a different set of GPs with respect to the definition from the  multipole expansion in the  c.m. frame.
In particular, in addition to the dipole electric and magnetic  polarizabilities, one finds a transverse electric polarizability, which describes rotational displacements of charges inside the hadron. As such, this polarizability  does not contribute to the induced charge polarization and  shows up at higher-order in the soft final-momentum limit of the VCS amplitude.

\subsubsection{The Low-Energy Theorem}
\label{sec-the-low-energy-theorem}

The work of Ref.~\cite{Guichon:1995pu} laid the foundations of the low-energy theorem (LET) and the low-energy expansion (LEX), for VCS in the unpolarized case. It provided for the first time a way to access GPs through experiments, via the $(\epgreact )$ reaction. The review article~\cite{Guichon:1998xv} contains in addition the LET for the doubly polarized case
\footnote{This case is exposed in Sect.~\ref{sec-double-spin-asymmetry}, with  related formulas in the Appendix.}, 
and puts in perspective two other regimes of VCS: the hard scattering  and Deeply Virtual Compton Scattering (DVCS).

The LET is directly inspired from the low-energy theorem of Low~\cite{Low:1958sn} and states that, in an expansion in powers of $\qpr$ (keeping $\qcm$ fixed) the first term of the BH and Born amplitudes is of order $\qprpowminusone$, while the first term of the non-Born amplitude is of order $\qprpowone$}:
%
\bea
\begin{array}{lll}
T^{\mathrm{BH}} \ + \  T^{\mathrm{Born}} &  =  & 
{ \displaystyle b_{-1} ( \qcm, \epsilon, \thcm,  \phicm ) 
\over
 \displaystyle \qpr
}
 \ + \ {\cal O} ( \qprpowzero )  \ ,    \\
 T^{\mathrm{NB}} &  =  & 
 b_{1} ( \qcm, \epsilon, \thcm,  \phicm ) \cdot \qpr
 \ + \ {\cal O} ( \qprpowtwo )  \ .   
\label{amplitudes-formula-1}
\end{array}
\eea
%

The LEX formula then yields for the photon electroproduction cross section below the pion production threshold:
%
%
\bea
\begin{array}{lllll}
d\sigma & =  & \sigbhb \ + \  \Phi  \cdot \qpr  \cdot  \Psi_0   \ + \ \ohigher 
& = & \siglex \ + \ \ohigher  \ , \\
 \Psi_0  & =  & \vll  \cdot ( \pllptte ) + \vlt \cdot \plt  \ , 
\label{lexformula-1}
\end{array}
\eea
%
%
where $\sigbhb$ is the BH+Born cross section. As stated in Sect.~\ref{sec-amplitudes-epgamma}, this cross section is entirely calculable in QED and just needs the knowledge of the nucleon elastic form factors $G_E$ and $G_M$.  It contains no polarizability effect, and serves as an important cross section of reference throughout the whole formalism.

The next term of the formula, $( \Phi  \cdot \qpr  \cdot  \Psi_0 )$, is where the GPs first appear in the expansion. $\Psi_0$ is the first-order polarizability term, obtained from the interference between the BH+Born and non-Born amplitudes at the lowest order; it is therefore  of order  $\qprpowzero$, i.e., independent of $\qpr$. 
The term $( \Phi  \cdot \qpr )$ is a phase-space factor (see the Appendix for details), in which an explicit factor $\qpr$   has been factored out in order to emphasize the fact that, when  $\qpr$ tends to zero,  $ ( \Phi  \cdot \qpr  \cdot  \Psi_0)$  tends to zero and the whole cross section tends to $\sigbhb$.

We will denote the cross section $\siglex$  of Eq.~(\ref{lexformula-1})  as the ``LEX cross section'', obtained by neglecting the $\ohigher$ term. The latter represents all the higher-order terms of the expansion and contains GPs of all orders. Below the pion production threshold, $\sigbhb$ is essentially the dominant part of the cross section, $\Psi_0$ is  the leading polarizability term and the higher-order terms $\ohigher$ are expected to be negligible.

The $\Psi_0$ term contains three VCS response functions, or structure functions: $\pll, \plt$, and $\plt$, which are the following combinations of five of the six lowest-order GPs
\footnote{The sixth GP, $P^{(M1,L2)1}$, appears in the polarized structure function $\pltperpprim$; cf. the Appendix.}: 
%
%
%
%
%
\bea
\begin{array}{lll}
\pll (Q^2)  & = &   
-2 \sqrt{6} \mnucleon \cdot G_E^p(Q^2) \cdot P^{(L1,L1)0}(Q^2) \ ,  \\
\ptt (Q^2)  & = & 
  -3 G_M^p(Q^2) \cdot  { \qcm ^2 \over {\tilde q^0} } \cdot
{\big ( } \, 
 P^{(M1,M1)1}(Q^2)  - {\sqrt{2}} {\tilde q^0} \cdot 
P^{(L1,M2)1}(Q^2) 
\, {\big ) } \ , \\
\plt (Q^2)  & = & \sqrt{3 \over 2} \cdot {\mnucleon \cdot  \qcm \over \qtild } \cdot 
G_E^p(Q^2) \cdot   P^{(M1,M1)0}(Q^2)    + 
{\bigg [ } \, 
 {3 \over 2} \, {  \qcm \sqrt{ Q^2} \over \tilde q^0 } \cdot G_M^p(Q^2) \cdot P^{(L1,L1)1}(Q^2)  
\, {\bigg ] } \ .
\label{formula-sfs-combinations-of-gps-1}
\end{array}
\eea
%
%

The general structure of each term, of the type [ nucleon form factor $\times$ GP ], originates from the  (BH+Born)-(NB) interference. The indices $(LL,TT,LT)$ refer to the longitudinal or transverse nature of the virtual photon polarization in the (BH+Born) and (NB) amplitudes. In contrast to RCS where the spin polarizabilities appear at a higher order than the scalar ones, in VCS the spin and scalar GPs appear at the same order in $\qpr$.

Here we will emphasize three features of Eq.~(\ref{formula-sfs-combinations-of-gps-1}): $i)$  $\pll$ is proportional to the electric GP, $ii)$ $\plt$ has a spin-independent part that is proportional to the magnetic GP, plus a spin-dependent part $\plt \, _{spin}$, and $iii)$  $\ptt$ is a combination of two spin GPs.
Using the LEX, experiments have extracted so far the two combinations $\pllptte$ and $\plt$ of  Eq.~(\ref{lexformula-1}), at fixed $\qcm$ and $\epsilon$. In particular, the separation of  $\pll$ and $\ptt$, which requires measurements at different values of $\epsilon$, has not been investigated yet experimentally.

Finally, we remind the kinematical dependence of each term in Eq.~(\ref{lexformula-1}). The $( \Phi  \cdot \qpr )$ factor depends on $\qcm, \qpr$ and $\epsilon$ but not on $\thcm$ and $\phicm$.  The $\vll$ and $\vlt$ coefficients (see the Appendix for their definition) depend on  $\qcm, \epsilon, \thcm$ and $\phicm$ but not on $\qpr$. The structure functions depend only on $\qcm$ or $Q^2$.

\subsection{Theoretical models for nucleon generalized polarizabilities}
\label{sec-theoretical-models-for-nucleon-gps}

Virtual Compton scattering has been investigated in various theoretical frameworks. Studies in different models, based on  complementary assumptions, help to unravel the mechanisms coming into play in the GPs. The very first predictions of GPs have been calculated in the framework of a non relativistic constituent quark model (NRCQM)~\cite{Guichon:1995pu,Liu:1996xd}, which was reviewed in Ref.~\cite{Pasquini:2000ue}. This model has been  extended in Ref.~\cite{Pasquini:1997by} to include relativistic effects by considering  a Lorentz covariant multipole expansion of the VCS amplitude and a light-front relativistic calculation of the nucleon excitations. 
In the constituent quark models, the GPs are expressed as the sum of the product of transition form factors of nucleon resonances over the whole spectrum, weighted by the inverse of the  excitation energy. The actual calculations truncate the sum to few intermediate states corresponding to  the $\Delta(1232)$ and the main resonances in the second resonance region. As discussed in Ref.~\cite{Pasquini:2000ue}, this truncation necessarily leads to a violation of gauge invariance. Furthermore, the Compton tensor   satisfies the constraint due to photon crossing at the real photon point, but it does not respect the nucleon crossing symmetry. As a consequence,  constituent quark models  predict  ten, and not six, independent GPs. Despite these limitations, calculations from the constituent quark model have been helpful for providing a first order-of-magnitude estimate for the nucleon resonance contributions to GPs.

More refined evaluations of the resonance contribution to GPs have been obtained  in the framework of an effective lagrangian model (ELM), based on a fully relativistic effective Lagrangian framework, which contains  baryon resonance contributions as well as $\pi^0$ and $\sigma$ exchanges in the  $t$-channel~\cite{Vanderhaeghen:1996iz,Vanderhaeghen:1997bx}. A similar model using a coupled-channel  unitary approach has been adopted also in Ref.~\cite{Korchin:1998cx}.

All these calculations are complementary to the theoretical approaches emphasizing pionic degrees of freedom and chiral symmetry, such as  the linear sigma model (LSM) and chiral effective field theories. Although the LSM is not a very realistic description of the nucleon, it is built on all the relevant symmetries like Lorentz, gauge and chiral invariance. Thanks to this, the calculation of the GPs within the LSM, performed in the limit  of infinitely large sigma mass~\cite{Metz:1996fn,Metz:1997fr}, pointed out for the first time  the existence of  relations between the VCS multipoles, beyond the usual constraints of parity and angular momentum conservation~\cite{Drechsel:1996ag,Drechsel:1997xv}, as discussed in Sect.~\ref{sec-the-nonborn-amplitude-and-the-gps}.

\begin{figure}[t]
\begin{minipage}[t]{16.5 cm}
\begin{center}
\epsfig{file=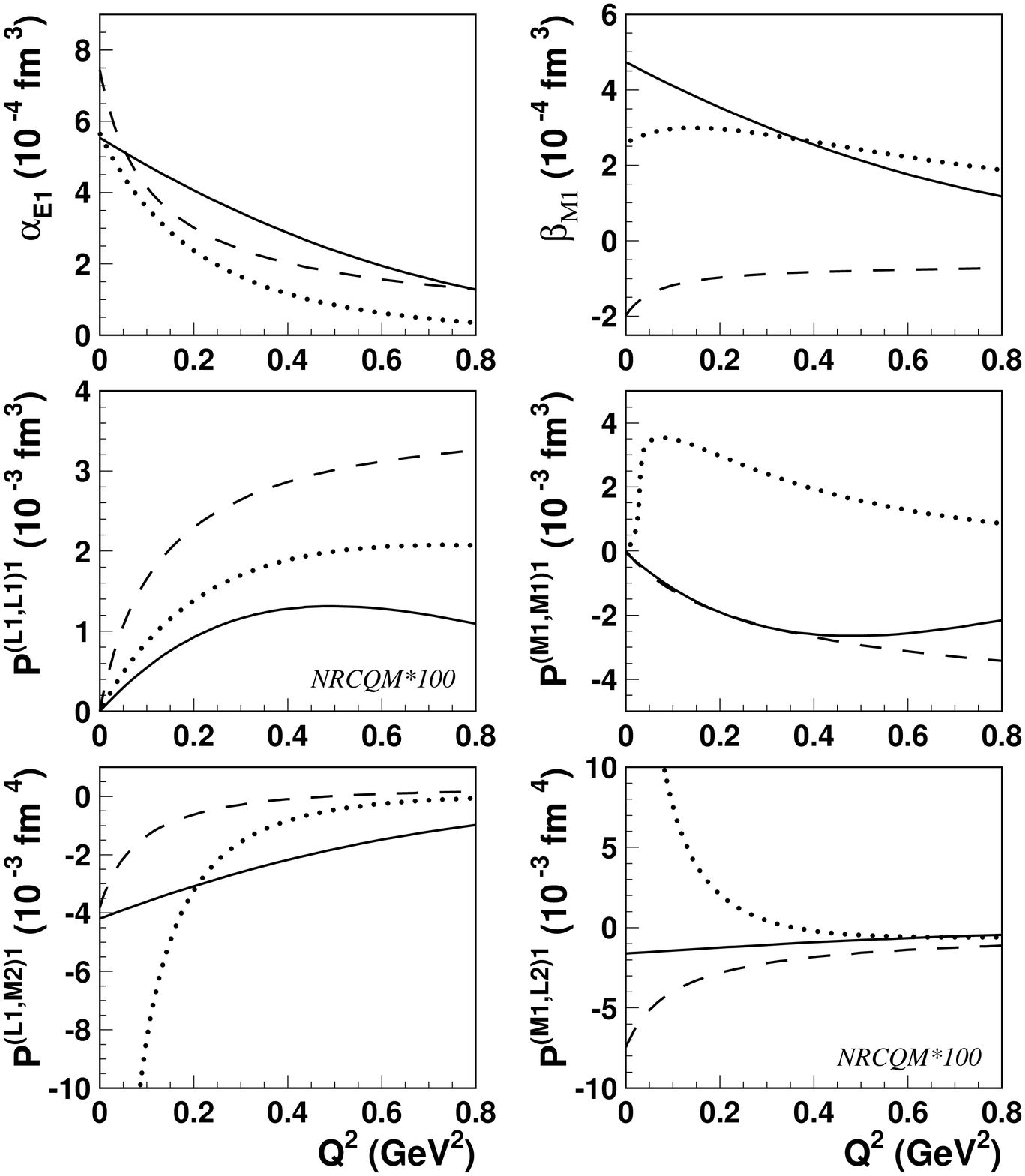,scale=0.45}
\end{center}
\end{minipage}
\begin{center}
\begin{minipage}[t]{16.5 cm}
\caption{
Results for GPs in different model calculations as a function of squared momentum transfer $Q^2$. Full lines: non-relativistic constituent quark model~\cite{Pasquini:2000ue}; dashed lines: linear sigma model~\cite{Metz:1996fn,Metz:1997fr}; dotted lines: effective Lagrangian model~\cite{Vanderhaeghen:1996iz,Vanderhaeghen:1997bx}. The non-relativistic CQM results for $P^{(L1,L1)1}$ and $P^{(M1,L2)1}$ have been multiplied by a factor 100.
\label{fig-theo-gp-1}}
\end{minipage}
\end{center}
\end{figure}

In Fig.~\ref{fig-theo-gp-1} we show the results for the GPs in the NRCQM of Ref.~\cite{Pasquini:2000ue}, in the LSM~\cite{Metz:1996fn,Metz:1997fr}, and in the ELM~\cite{Vanderhaeghen:1996iz,Vanderhaeghen:1997bx}. In the  NRCQM the excited states of the nucleon are given by resonances, and the $Q^2$ behavior of the GPs is  determined by the electromagnetic transition form factors. 
In contrast to this, the LSM describes the excitation spectrum as pion-nucleon scattering states with quite a different $Q^2$ dependence. In the ELM, one sees  both  resonant and non-resonant contributions at work. In the case of the scalar GPs, the LSM predicts a rapid variation at small momentum transfer, and a smaller one at higher momentum. On the contrary, in the NRCQM and in the ELM the scalar GPs show a rapid fall-off in $Q^2$, with a gaussian shape in the case of the NRCQM due to the assumed parametrization for the electromagnetic transition form factors. 
All the calculations underestimate the scalar electric polarizabilities $\ale (0)$ at the real photon point. In the case of the magnetic GP $\bmq$, the pion cloud gives rise to a positive slope at the origin, while the $N\Delta$ transition form factor determines the paramagnetic contribution, which decreases as function of $Q^2$. The LSM  describes only the negative diamagnetic  contribution, whereas the NRCQM takes into account only the positive paramagnetic contribution. The interplay of the two  competing effects can be observed  in the ELM, in particular at $Q^2=0$ where the ELM prediction is in very good agreement with the recent evaluation of $\bem (0)$ from the Particle Data Group (PDG)~\cite{Tanabashi:2018oca}.

While the $\sigma$ exchange strongly influences the numerical values of the scalar polarizabilities,  it does not contribute to the vector polarizabilities. On the other hand, the $\pi^0$-exchange in the $t$-channel (the anomaly diagram, see Fig.~\ref{fig-theo-graphs-1}) is irrelevant in the spin-independent case, but very important for calculations of the vector GPs. Since the anomaly is gauge invariant and regular in the soft photon limit, it could also be considered as part of the Born amplitude instead of the non-Born amplitudes, contrary to what has been done in Ref.~\cite{Guichon:1995pu}. In Fig.~\ref{fig-theo-gp-1}, and in the following, the contribution to the spin GPs  from the anomaly is not shown
\footnote{This contribution is usually not shown because it can be calculated in terms of the $\pi NN$ coupling constant and the $\pi^0\gamma^*\gamma$ form factor, as in Eq.~(240) of Ref.~\cite{Drechsel:2002ar}, and it does not contain information on the nucleon structure.}. 
All the models satisfy the model-independent constraints $P^{(M1,M1)1}(0)=0$ and $P^{(L1,L1)1}(0)=0$ due to photon-crossing symmetry. The NRCQM and LSM  predict the same sign for all the vector GPs, while $P^{(M1,M1)1}$ and $P^{(M1,L2)1}$ have opposite sign in the ELM. Furthermore, the results for $P^{(L1,L1)1}$ and $P^{(M1,L2)1}$ are substantially larger in the LSM and ELM than in the NRCQM, indicating that here the non-resonant background is more important than the nucleon resonances. The variation of the spin GPs at low $Q^2$ is very different in all the three models, except for $P^{M1,M1)1}$ in the LSM and in the NRCQM at variance with the ELM.

Systematic calculations of pion-cloud effects became possible with the development of chiral perturbation theories (ChPTs), an expansion in the external momenta  and the pion mass (``$p$"-expansion). In such theories, one constructs the most general VCS amplitude consistent with electromagnetic gauge invariance, the pattern of chiral symmetry breaking in QCD, and Lorentz covariance, to a given order of the small parameter $p\equiv\{P, m_\pi\}/\Lambda$. Here, $P$ stands for each component of the four-momenta of the photons and of the three-momenta of the nucleons, while $\Lambda$ is the breakdown scale of the theory. 
There exist different variants of ChPT calculations. The pioneering  calculation for VCS  was performed with only nucleons and pions as explicit degrees of freedom, with the effects of the nucleon-resonance encoded in a string of contact operators~\cite{Hemmert:1996gr,Hemmert:1997at}. Furthermore, this work used  the heavy-baryon (HB) expansion for the nucleon propagators, which amounts to making an expansion in $1/M_N$ along with the expansion in $p$. 
The first HBChPT calculations have been performed at ${\cal O}(p^3)$~\cite{Hemmert:1996gr,Hemmert:1997at}, and have  recently been  extended to ${\cal O}(p^4)$  for all the spin GPs and to ${\cal O}(p^5)$ for some of them in Refs.~\cite{Kao:2002cn,Kao:2004us}. Since the excitation energy of the $\Delta(1232)$ is low, it may not be justified to ``freeze" the degrees of freedom of this near-by resonance. The inclusion of the $\Delta(1232)$ as an explicit degree of freedom in the  calculation of the VCS process has  first been addressed in Ref.~\cite{Hemmert:1999pz}, by introducing the excitation energy of the $\Delta(1232)$ as an additional expansion parameter (\lq\lq$\epsilon$ expansion\rq\rq). 
Subsequently, a different counting has been proposed in Ref.~\cite{Pascalutsa:2002pi} (``$\delta$-expansion"), and it was employed for the VCS process in Ref.~\cite{Lensky:2016nui} using a manifestly Lorentz invariant version of baryon chiral perturbation theory (BChPT). The two schemes mainly differ in the counting of  the $\Delta(1232)$ excitation energy $\bar{\Delta}=M_\Delta- \mnucleon$, compared with the pion mass $m_\pi$. In the $\epsilon$-expansion, they enter at the same order  ($\bar \Delta\sim m_\pi$), while in the $\delta$-expansion $m_\pi\ll \bar\Delta$. We refer to the original works for further details. The predictions for the GPs in HBChPT and in BChPT will be discussed in the following section, in comparison with the DR results.

\subsection{The dispersion relation formalism}
\label{sec-the-dr-model}

Historically, DRs  have been considered for the first time for the VCS process in Ref.~\cite{BergLindner}. Recently, the formalism has been reviewed  in Ref.~\cite{Pasquini:2001yy}, using a different set of VCS amplitudes that avoid numerical artefacts due to kinematical singularities. Following the derivation in Refs.~\cite{Drechsel:1997xv,Drechsel:1998zm},  the VCS Compton tensor is parametrized in terms of twelve independent functions $F_i(Q^2,\nu,t)$, $i=1,\dots,12$, which depend on  three kinematical invariants, i.e.,  $Q^2$,  $t$, and the crossing-symmetric variable $\nu=(s-u)/(4M_N)$. 
In terms of these invariants, the limit $q'\rightarrow 0$ at finite three-momentum $q$ of the virtual photon corresponds to $\nu\rightarrow 0$ and $t\rightarrow -Q^2$ at finite $Q^2$. The GPs can then be  expressed in terms of the non-Born contribution to the VCS invariant amplitudes, denoted as $F_i^{\mathrm{NB}}$,  at the point $\nu = 0$, $t=-Q^2$ for finite $Q^2$ (the explicit relationships between the $F_i^{\mathrm{NB}}$ and the GPs can be found in  Ref.~\cite{Pasquini:2000pk}). The $F_i$  functions are free of poles and kinematical zeros, once the irregular nucleon pole terms have been subtracted in a gauge-invariant fashion, and are even functions of $\nu$, i.e. $F_i(Q^2,-\nu,t)=F_i(Q^2,\nu,t)$.~Assuming an appropriate analytic and high-energy behavior, these amplitudes fullfil unsubtracted DRs in the variable $\nu$ at fixed $t$ and $Q^2$:
%
%
\bea
\mathrm{Re}\, F^{\mathrm{NB}}_{i}(Q^2,\nu,t)=F_{i}^{\mathrm{pole}}(Q^2,\nu,t)-F^{\mathrm{Born}}_{i}(Q^2,\nu,t)+ \frac{2}{\pi}\mathcal{P}\int_{\nu_{thr}}^{+\infty}{\rm d}\nu'\frac{\nu'\mathrm{Im}\, F_i(Q^2,\nu',t)}{\nu'^2-\nu^2},
\label{DR-VCS}
\eea
%
%
where  the Born contribution $F_{i}^{\mathrm{Born}}$ is defined as  in~\cite{Guichon:1998xv,Guichon:1995pu}, whereas $F_{i}^{\mathrm{pole}}$ denotes the nucleon pole contribution (i.e., energy factors in the numerators are evaluated at the pole position)
\footnote{The pole and Born contributions differ only for the $F_1$, $F_5$ and $F_{11}$  amplitudes.}. 
Furthermore, $\mathrm{Im} \,F_i$ are the discontinuities across the $s$-channel cuts, starting at the pion production threshold $\nu_{thr}=m_\pi+(m_{\pi}^2+t/2+Q^2/2)/(2  \mnucleon  )$. However, such unsubtracted DRs require that at high energies ($\nu\rightarrow \infty$) the amplitudes $\mathrm{Im} \, F_i$ drop fast enough so that the integral of Eq.~(\ref{DR-VCS}) is convergent and the contribution from the semicircle at infinity can be neglected. The high energy behavior of the amplitudes  in the limit of  $\nu\rightarrow \infty $ at fixed $t$ and $Q^2$ can be deduced from Regge theory~\cite{Pasquini:2001yy}. 
It follows that  the unsubtracted dispersion integral in Eq.~(\ref{DR-VCS}) diverges for the $F_1$ and $F_5$ amplitudes.  In order to obtain useful results for these two amplitudes, we can use finite-energy sum rules, by restricting the unsubtracted integral in a finite range $-\nu_{\mathrm{max}}\le\nu\le\nu_{\mathrm{max}}$ and closing the contour of the integral by a semicircle of finite radius $\nu_{\mathrm{max}}$ in the complex plane, with the result
%
%
\bea
  \mathrm{Re}\, F^{\mathrm{NB}}_{i}(Q^2,\nu,t)=F_{i}^{\mathrm{pole}}(Q^2,\nu,t)-F^{\mathrm{Born}}_{i}(Q^2,\nu,t)+ \frac{2}{\pi}\mathcal{P}\int_{\nu_{thr}}^{\nu_{\mathrm{max}}}{\rm d}\nu'\frac{\nu'\mathrm{Im}\, F_i(Q^2,\nu',t)}{\nu'^2-\nu^2} +F_i^{\mathrm{as}}.
\label{DR-VCS2}
\eea
%
%
In Eq.~(\ref{DR-VCS2}), the \lq\lq asymptotic term\rq\rq $F_i^{\mathrm{as}}$ represents the contribution along the finite semicircle of radius $\nu_{\mathrm{max}}$ in the complex plane,  which is replaced by a finite number of energy-independent poles in the $t$ channel.  At $\nu=0$ and $t=-Q^2$, the difference between the pole and Born contributions is vanishing for all the amplitudes and the GPs  can be evaluated directly  by unsubtracted DRs through the following integrals 
%
%
\bea
 F_i^{\mathrm{NB}}(Q^2,\nu=0,t=-Q^2)=\frac{2}{\pi}\int_{\nu_0}^{\nu_{\mathrm{max}}}d\nu'\frac{\mathrm{Im}_s\, F_i(Q^2,\nu',t=-Q^2)}{\nu'}+F_i^{\mathrm{as}},\label{DR-VCS3}
 \eea
%
%
%
where the asymptotic contribution $F_i^{\mathrm{as}}$ enters only for $i=1, 5$.

The $s$-channel integrals in Eqs.~(\ref{DR-VCS})-(\ref{DR-VCS3}) can be  evaluated by  expressing  the imaginary part of the amplitudes through the unitarity relation, taking into account all the possible intermediate states which can be formed between the initial $\gamma^* N$ and final $\gamma N$ states. As long as we are interested in the energy region up to the $\Delta(1232)$ resonance, we may restrict ourselves to only the dominant contribution from the $\pi N$ intermediate states,  setting the upper limit of integration to $\nu_{\mathrm{max}}=1.5$ GeV. This dispersive $\pi N$ contribution will be denoted with $F_i^{\pi N}$  in the following. 
In the actual calculation, we evaluate $F_i^{\pi N}$ from the pion photo- and electro-production amplitudes of the phenomenological MAID analysis (MAID2007 version)~\cite{Drechsel:1998hk,Drechsel:2007if}, which contains both resonant and non-resonant pion production mechanisms. It turns out that the residual dispersive contributions beyond the value $\nu_{\mathrm{max}} = 1.5$ GeV is relevant mainly for the amplitude $F_2$, while it can be neglected for the other amplitudes.

Once the dispersive  integrals are evaluated, we need  a suitable parametrization of the energy-independent functions for the asymptotic contribution to the $F_1$ and $F_5$ amplitudes, and for the higher-dispersive corrections to $F_2$. The asymptotic contribution to the $F_5$ amplitude is saturated from the $t$-channel  $\pi^0$ exchange (the anomaly diagram, see Fig.~\ref{fig-theo-graphs-1}), that is calculated according to Ref.~\cite{Pasquini:2001yy}. 
The asymptotic contribution to $F_1$ can be described phenomenologically as the $t$-channel exchange of an effective $\sigma$ meson. The $Q^2$-dependence of this term is unknown and can be parametrized in terms of a function directly related to the magnetic dipole GP $\beta_{\mathrm{M1}}(Q^2)$ and fitted to VCS observables. Analogously, the higher-energy dispersive contribution to $F_2$ can effectively be accounted for with an energy-independent function, at fixed $Q^2$ and $t = -Q^2.$ This amounts to introducing an additional fit function, which is directly related to the sum of the electric and magnetic dipole GPs. 
In conclusion, the contributions beyond the dispersive $\pi N$ integrals can be recast in terms of the following two functions
%
%
\bea
\alpha_{E1}(Q^2)-\alpha_{E1}^{\pi N}(Q^2)=\left(\alpha_{E1}^{\mathrm{exp}}-\alpha_{E1}^{\pi N}\right)f_\alpha(Q^2),\quad
\beta_{M1}(Q^2)-\beta_{M1}^{\pi N}(Q^2)=\left(\beta_{M1}^{\mathrm{exp}}-\beta_{M1}^{\pi N}\right)  f_\beta(Q^2), 
\label{GPs-parametrization}
\eea
%
%
where $\alpha_{E1}$ and  $\beta_{M1}$ are the RCS polarizabilities, with superscripts $\mathrm{exp}$ and $\pi N$ indicating, respectively, the experimental value~\cite{Tanabashi:2018oca} and the $\pi N$ contribution evaluated from unsubtracted DRs. In Eq.~(\ref{GPs-parametrization}),  $f_\alpha(Q^2)$ and $f_\beta(Q^2)$ are  fit functions, with the constraints $f_\alpha(0)=f_\beta(0)=1$. Their functional form  is unknown and should be adjusted by a fit to the experimental cross sections.  However, in order to provide predictions for VCS observables, we adopt the following parametrization
%
%
\bea
f_\alpha(Q^2)=\frac{1}{(1+Q^2/\Lambda_\alpha^2)^2},\quad f_\beta(Q^2)=\frac{1}{(1+Q^2/\Lambda_\beta^2)^2},\label{asymp}
\eea
%
%
where the mass scale parameters $\Lambda_\alpha$ and $\Lambda_\beta$ are  free parameters, not necessarily constant with $Q^2$.
%
%
%
%
%

\begin{figure}[tb]
\begin{minipage}[t]{16.5 cm}
\begin{center}
\epsfig{file=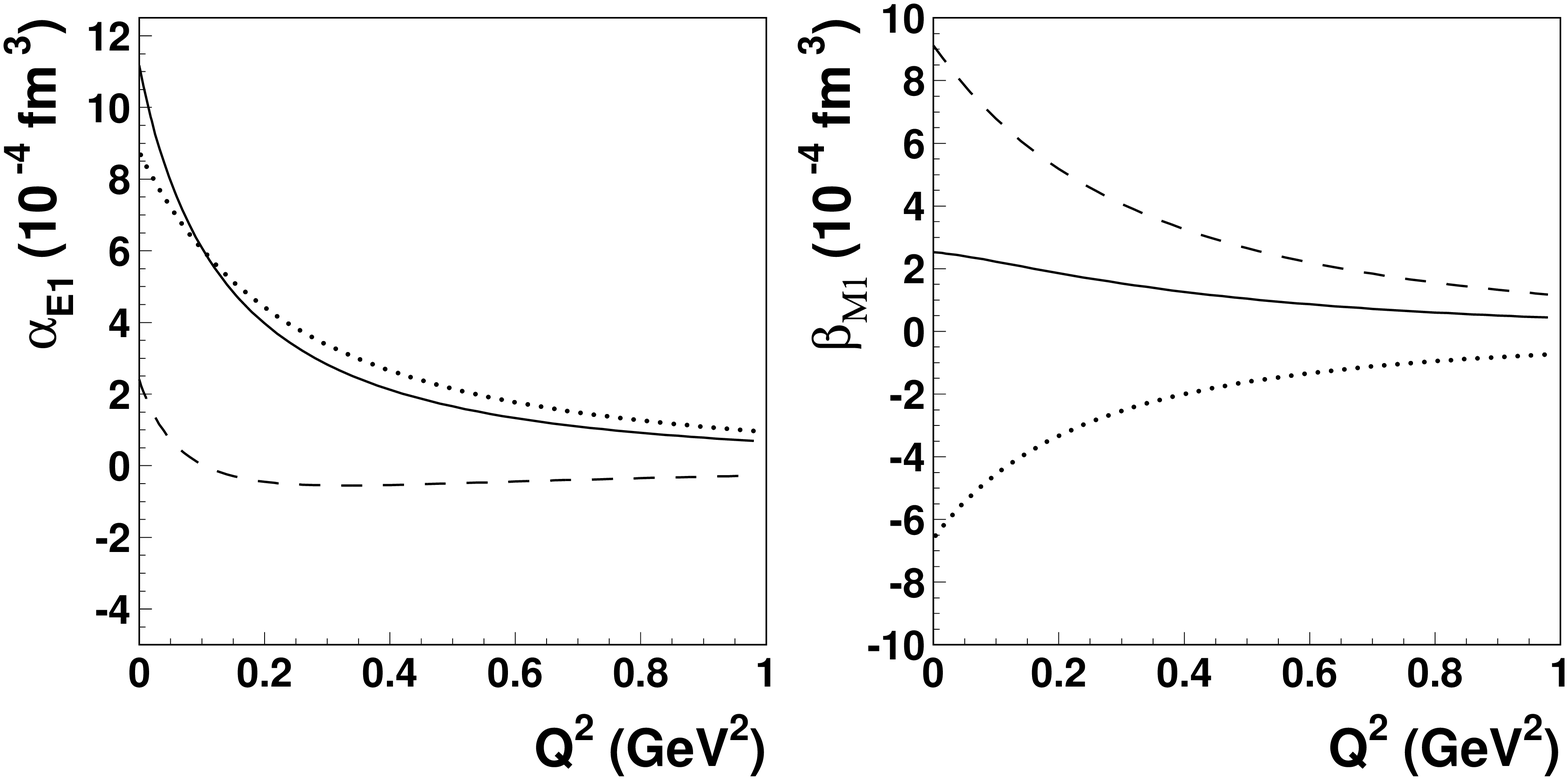,scale=0.35}
\end{center}
\end{minipage}
\begin{center}
\begin{minipage}[t]{16.5 cm}\caption{
DR results~\cite{Pasquini:2001yy,Drechsel:2002ar} for the scalar GPs $\alpha_{\mathrm{E1}}(Q^2)$ (left panel) and $\beta_{\mathrm{M1}}(Q^2)$ (right panel) as function of the four-momentum transfer $Q^2$. The dashed curves show the dispersive $\pi N$ contribution, the dotted curves correspond to the asymptotic contributions from Eq.~(\ref{asymp}) with $\Lambda_\alpha=0.7$ GeV and $\Lambda_\beta=0.7$ GeV, respectively.  The solid curves are the total results.
\label{fig-theo-ab-1}}
\end{minipage}
\end{center}
\end{figure}
%
%
%
%
%
%
In Fig.~\ref{fig-theo-ab-1}, we show the DR predictions for the scalar GPs as function of $Q^2$, along with the separate contributions from the $\pi N$ dispersive term and the asymptotic term. The RCS values at $Q^2=0$ are fixed to the PDG values~\cite{Tanabashi:2018oca}. The electric GP is dominated by a large positive asymptotic contribution. The $\pi N$  dispersive contribution adds up to the asymptotic term in the RCS limit and smoothly decreases at higher $Q^2$. The magnetic GP results from a large dispersive  $\pi N$ (paramagnetic) contribution, dominated by the $\Delta(1232)$ resonance, and a large asymptotic (diamagnetic) contribution with opposite sign, leading to a relatively small net result.

\begin{figure}[t]
\begin{minipage}[t]{16.5 cm}
\begin{center}
\epsfig{file=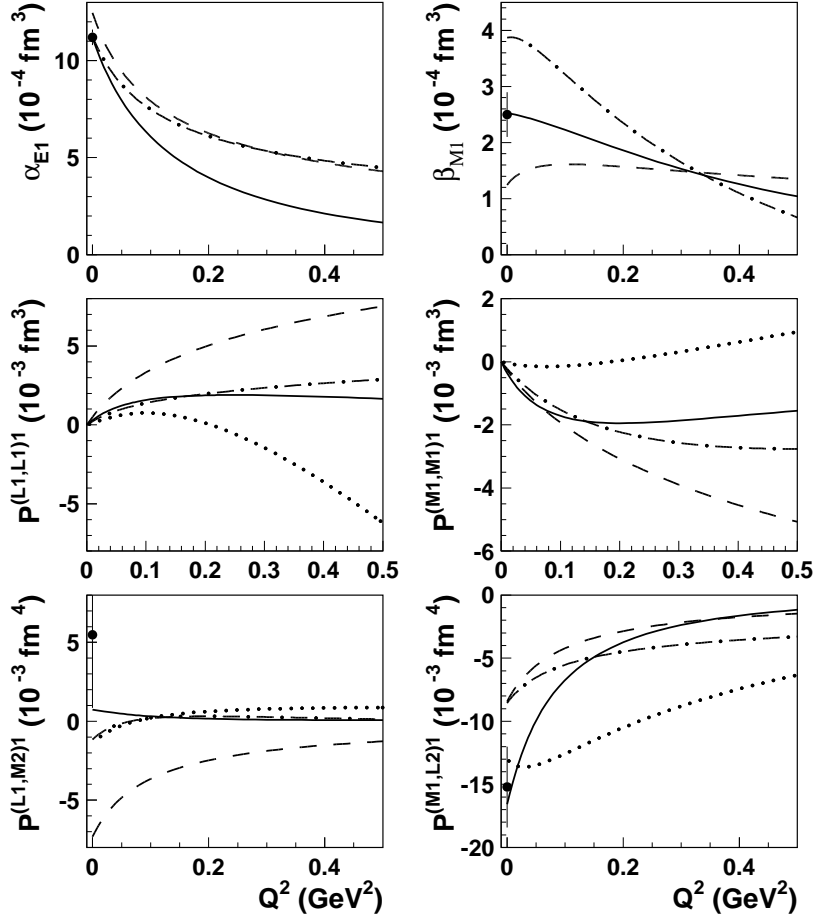,scale=0.45}
\end{center}
\end{minipage}
\begin{center}
\begin{minipage}[t]{16.5 cm}
\caption{
Comparison of HBChPT, BChPT and DR calculation of the GPs.  Solid curves: DRs~\cite{Pasquini:2000pk,Pasquini:2001yy}; Dashed-dotted curves: BChPT ${\cal O}(p^3)+{\cal O}(p^4/\bar\Delta)$~\cite{Lensky:2016nui}; Dashed curves: HBChPT ${\cal O}(p^3)$; Dotted curves: HBChPT at ${\cal O}(p^4)$~\cite{Kao:2002cn,Kao:2004us}. The experimental points at $Q^2=0$ for the scalar polarizabilities are from PDG~\cite{Tanabashi:2018oca} and for the spin GPs are from the MAMI measurements~\cite{Paudyal:2019mee}.
\label{fig-theo-gp-2}}
\end{minipage}
\end{center}
\end{figure}

In Fig.~\ref{fig-theo-gp-2} we compare the DR predictions for the  GPs with the results from  covariant BChPT at ${\cal O}(p^3)+{\cal O}(p^4/\bar\Delta)$ and the calculation within HBChPT at ${\cal O}(p^3)$ for the scalar GPs and within HBChPT both at  ${\cal O}(p^3)$ and at ${\cal O}(p^4)$ for the spin-dependent GPs. 
For the electric polarizability, the calculations within HBChPT and BChPT are very similar. The HBChPT result in the RCS limit is also in good agreement with the DR calculation, fixed to the PDG value~\cite{Tanabashi:2018oca}. However,  both the calculation within HBChPT and BChPT deviate from the DRs at increasing $Q^2$, with a softer fall-off in $Q^2$. 
For the magnetic polarizability, the  calculation in HBChPT does not fully account for the positive paramagnetic component coming from $\Delta(1232)$ degrees of freedom, at variance with the covariant BChPT. However, we notice that the BChPT results are quite different from the DR results in the whole $Q^2$ range. At $Q^2=0$ the DR value is fixed to the current PDG evaluation~\cite{Tanabashi:2018oca}, while the BChPT predicts a substantially larger value~\cite{Lensky:2009uv,Lensky:2015awa}, i.e. $\beta_{\mathrm{M1}}=3.7\times 10^{-4}$ fm$^3$.

In the spin-dependent sector, by comparing the HBChPT results at the lowest and at the next order, we notice large corrections at next order. The corrections to the leading order bring the HBChPT results toward the DR and BChPT estimates only in the case of  the $P^{(L1,M2)1}$ GP. For the GP $P^{(M1,L2)1}$, the sizeable large correction at the real photon point brings the HBChPT results close to the dispersive results, but the dependence on $Q^2$ remains quite different. Similarly, we notice sizable differences between DRs and HBChPT for the two GPs that vanish at $Q^2=0$, $P^{(L1,L1)1}$ and $P^{(M1,M1)1}$. 
Instead, the results for these GPs from BChPT are more similar to the DR calculation, especially in the low $Q^2$ region. Model-independent constraints on the slope of the $P^{(L1,L1)1}$ and $P^{(M1,M1)1}$ GPs at $Q^2=0$ can be obtained through sum rules derived from the forward doubly virtual Compton scattering (VVCS) $\gamma^* N\rightarrow\gamma^* N$, where both photons have the same spacelike virtuality $q^2=-Q^2<0$~\cite{Pascalutsa:2014zna,Lensky:2017dlc}. 
In particular, these sum rules provide relationships involving the slope of these GPs and properties of the proton target, such as the anomalous magnetic moment and the squared Pauli radius, along with the RCS spin polarizabilities $\gamma_{E1M2}$ and $\gamma_{E1E1}$ and two quantities from VVCS. Therefore, the quantities in  these relations are all observed in a different process, but they are not yet fully constrained by experimental measurements. 
Awaiting for a direct experimental verification, the sum rules have been used to infer predictions for the slope of the GPs,  using  the available information on the Pauli radius~\cite{Bernauer:2013tpr}, the recent extractions of  the RCS spin polarizabilities at MAMI~\cite{Paudyal:2019mee,Martel:2014pba} and the empirical information from MAID for the VVCS quantities~\cite{Lensky:2017dlc}. Such predictions turned out to be consistent, within the large uncertainties, with the estimates for the slope of the GPs  from both  DRs and BChPT. 
Finally, for the $P^{(L1,M2)1}$ and $P^{(M1,L2)1}$ GPs, we observe that the BChPT predictions are similar, respectively, to the next-order and leading-order calculation  within HBChPT. These GPs at $Q^2=0$ are proportional, respectively, to the RCS polarizabilities  $\gamma_{E1M2}$ and $\gamma_{M1E2}$. In Fig.~\ref{fig-theo-gp-2}, we show the corresponding values for the GPs at the real photon point (cf. Table~\ref{tab-six-dipole-gps}), using the experimental RCS results of Ref.~\cite{Paudyal:2019mee} given by the weighted average of the values extracted within fixed-$t$ subtracted DRs~\cite{Pasquini:2007hf,Drechsel:1999rf} and covariant BChPT~\cite{Lensky:2009uv} (cf. Table 1 of Ref.~\cite{Paudyal:2019mee}). 
The RCS value of $P^{(M1,L2)1}$ is consistent with the DR predictions and the next-order results of HBChPT, but not with the covariant BChPT value. The experimental value for  $P^{(L1,M2)1}$ at $Q^2=0$  is compatible with the positive value predicted from DRs, at variance with the estimates from both HBChPT and BChPT. However, at larger $Q^2$ the different theoretical predictions are very similar, showing a rather flat behavior in $Q^2$.

DR calculations of the structure functions $\pllptte$ and $\plt$ are discussed in sect.~\ref{sec-experimental-results-sfs}, cf. Figs.~\ref{fig-exp-sf-2} and \ref{fig-theo-sf-3}, together with the prediction of the models for $\ptt$  shown in Fig.~\ref{fig-theo-sf-2}. We mention here that the $\pi^0$-exchange in the $t$-channel, although being an important contribution to several of the vector GPs, does not contribute to the  $\ptt$ structure function, i.e. the specific combination of  $P^{(M1,M1)1}(Q^2)$ and $P^{(L1,M2)1}(Q^2)$ given in Eq.~(\ref{formula-sfs-combinations-of-gps-1}). This is because the $\pi^0$-pole contribution is vanishing in the combination of amplitudes $(F_5 + 4F_{11})$ (cf. Refs.~\cite{Lensky:2016nui,Drechsel:2002ar}).

\subsection{The ($\epgreact$) cross section and the GP effect}
\label{sec-the-epg-cross-section-and-the-gp-effect}

We will limit our study to the cross section calculated in two formalisms: LEX and DR, which provide the most useful interface with experiments. However, other descriptions of the $\epgreact$ cross section exist, such as the effective Lagangian model~\cite{Vanderhaeghen:1996iz}. Most of the other theoretical approaches focus on modeling the GPs (cf. Sect.~\ref{sec-theoretical-models-for-nucleon-gps}) and are less directly connected to the GP extraction from experiments. In this section we discuss only the non-polarized case. The polarized case will be mentioned in Sect.~\ref{sec-other-experimental-results}.

The GP effect can be defined as the part of the cross section that contains the polarizability contribution, normalized to the part that does not contain it (i.e., $\sigbhb$). Depending on the theoretical approach, one gets:

- using the LEX: GP effect =  $( \siglex - \sigbhb ) / \sigbhb = ( \Phi \cdot \qpr ) \cdot \Psi_0 / \sigbhb $ ,

- using the DR model: GP effect =  $( \sigdr - \sigbhb ) / \sigbhb $ .

\vskip 2mm

\begin{figure}[tb]
\begin{center}
\begin{minipage}[t]{12 cm}
\epsfig{file=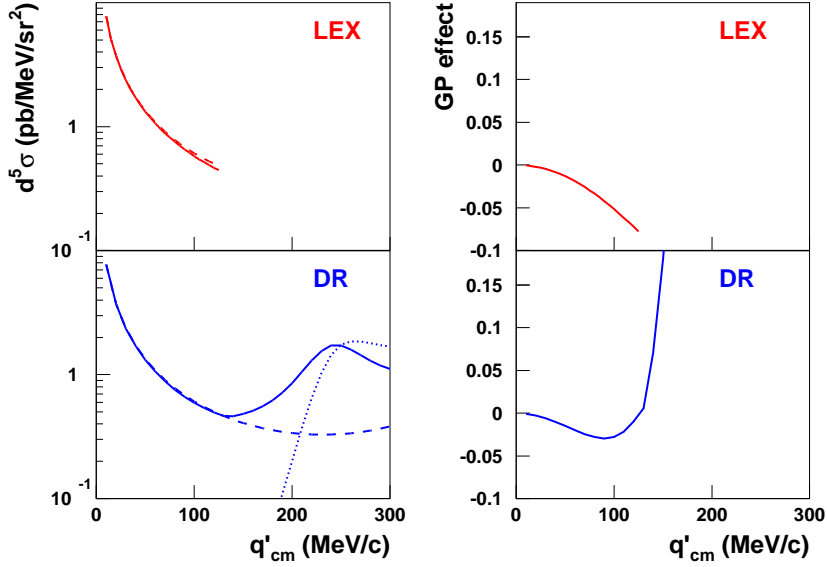,scale=0.6}
\end{minipage}
\begin{minipage}[t]{16.5 cm}
\caption{
(color online) $\qpr$-dependence of the cross section and the GP effect at fixed $Q^2$ = 0.2 GeV$^2$, $\epsilon$ = 0.9, $\thcm$ = 100$^{\circ}$ and $\phicm$ = 180$^{\circ}$. Top-left plot: the LEX cross section (solid) and BH+Born cross section (dashed). Top-right plot: the GP effect from the LEX, as defined in Sect.~\ref{sec-the-epg-cross-section-and-the-gp-effect}. The curves are drawn only up to the pion production threshold. Bottom-left plot: the DR cross section; full calculation (solid), BH+Born (dashed) and the non-Born contribution alone (dotted). Bottom-right plot: the GP effect from DRs, as defined in Sect.~\ref{sec-the-epg-cross-section-and-the-gp-effect}. In all cases, the polarizability effect is calculated with the following structure functions values: $\pllptte$ = 17.6 GeV$^{-2}$ and $\plt$ = -5.3 GeV$^{-2}$, close to the experimental ones at $Q^2$ = 0.2 GeV$^2$. They correspond to $\aeq = 3.83 \cdot 10^{-4}$ fm$^3$, $\bmq = 1.96 \cdot 10^{-4}$ fm$^3$ in the DR model.
\label{fig-cs-and-gpeffect-fig1}}
\end{minipage}
\end{center}
\end{figure}

\begin{figure}[tb]
\begin{center}
\begin{minipage}[t]{12 cm}
\epsfig{file=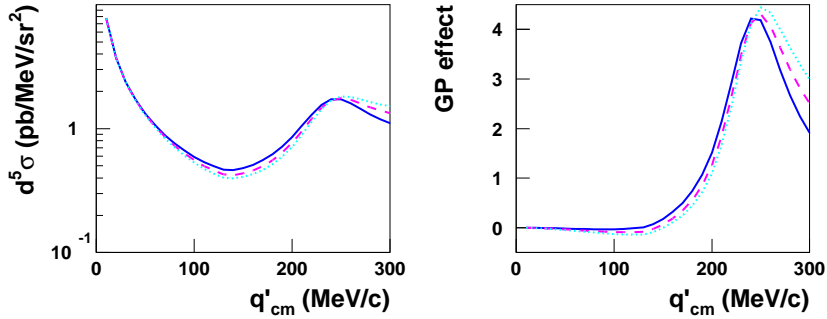,scale=0.6}
\end{minipage}
\begin{minipage}[t]{16.5 cm}
\caption{
(color online) Same kinematics as in Fig.~\ref{fig-cs-and-gpeffect-fig1}. Left plot: the DR cross section for three different choices of the free parameters $(\la, \lb)$: (0.61,0.61) GeV for the solid curve (which is the case of the previous figure); (0.80,0.80) GeV for the dashed curve; (1.0, 1.0) GeV for the dotted curve. Right plot: the GP effect from DR, corresponding to these three choices of $(\la, \lb)$.
\label{fig-cs-and-gpeffect-fig2}}
\end{minipage}
\end{center}
\end{figure}

\begin{figure}[tb]
\begin{center}
\begin{minipage}[t]{12 cm}
\epsfig{file=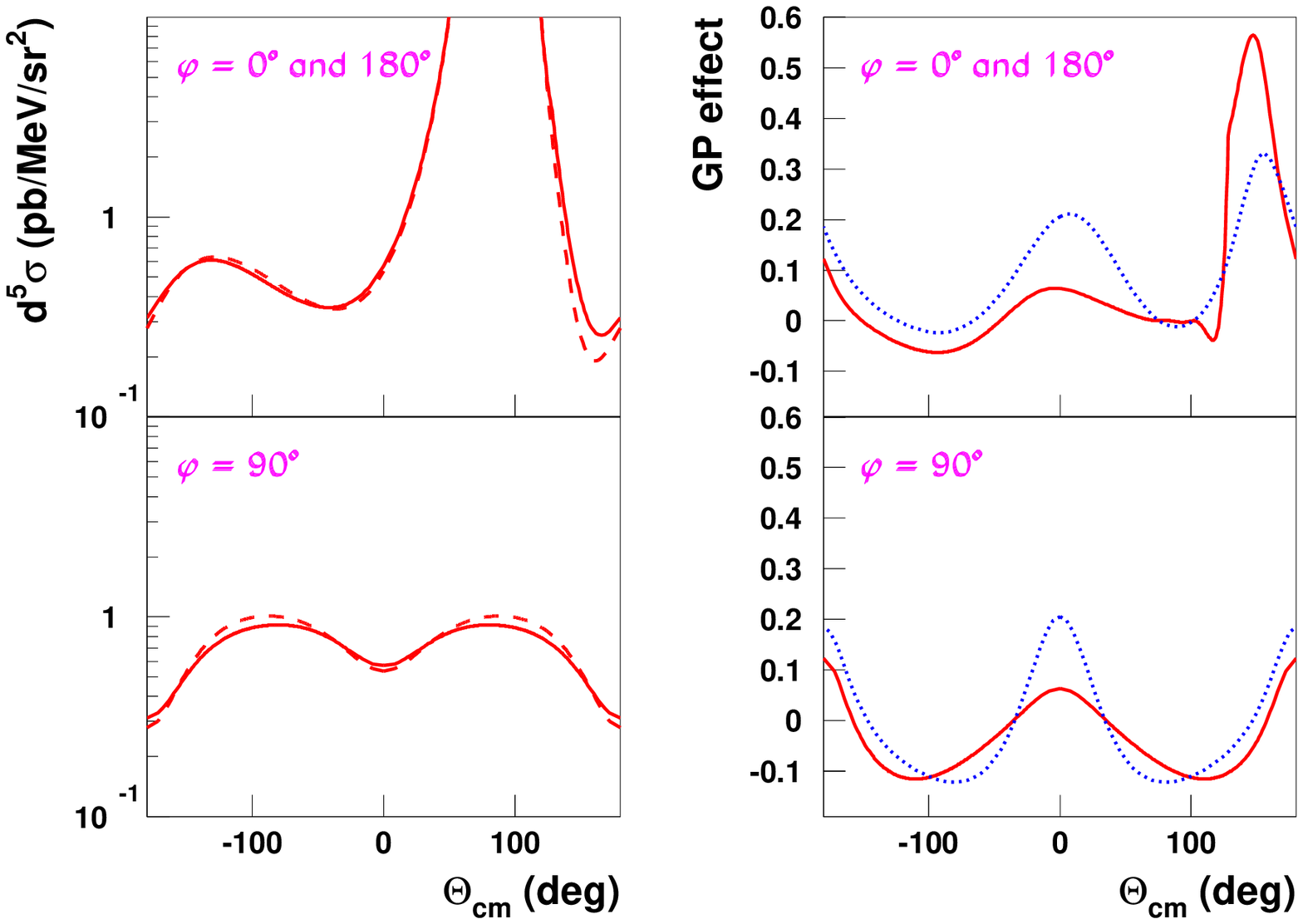,scale=0.6}
\end{minipage}
\begin{minipage}[t]{16.5 cm}
\caption{(color online) Angular dependence at fixed $Q^2$ = 0.2 GeV$^2$, $\epsilon$ = 0.9 and $\qpr$ = 110 MeV/c. Top-left plot: the LEX cross section (solid) and BH+Born cross section (dashed) for in-plane kinematics (DR cross section not shown). The sharp rise is due to the BH peaks. The $\thcm$ angle is set negative by convention in the $\phicm$-hemisphere opposite to the BH peaks.  Top-right plot: the GP effect from the LEX (solid) and from DR (dotted) in these in-plane kinematics (the small ditch in the solid curve near $\thcm = +110^{\circ}$ is an artefact due to limited numerical precision right in the BH peaks). Bottom-left and bottom-right: same as the top plots,  this time at  $\phicm$ = 90$^{\circ}$. Structure functions have the same values as in Fig.~\ref{fig-cs-and-gpeffect-fig1}.
\label{fig-cs-and-gpeffect-angular}}
\end{minipage}
\end{center}
\end{figure}

\begin{figure}[tb]
\begin{center}
\begin{minipage}[t]{15 cm}
\epsfig{file=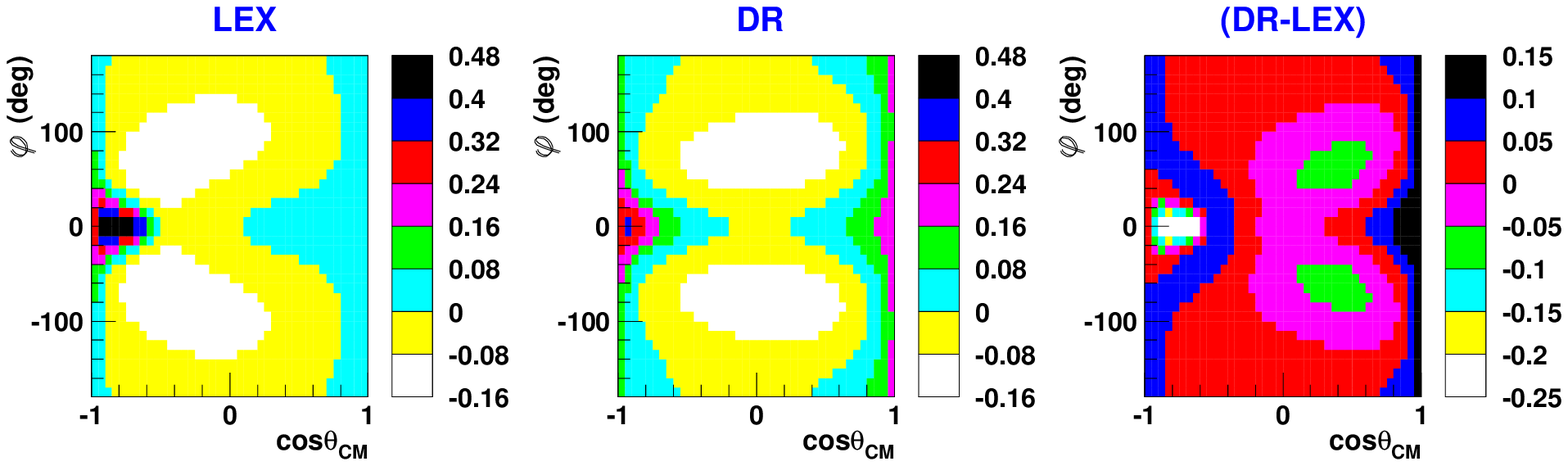,scale=0.8}
\end{minipage}
\begin{minipage}[t]{16.5 cm}
\caption{(color online) The GP effect in the 2D-plane ($\cthcm, \phicm$) at fixed $Q^2$ = 0.2 GeV$^2$, $\qpr$ = 110 MeV/c and  $\epsilon$ = 0.9. The GP effect from the LEX (left plot), from DR (middle plot) and their difference (right plot), also equal to $( \sigdr - \siglex ) / \sigbhb $. The structure functions have the same values as in Fig.~\ref{fig-cs-and-gpeffect-fig1}.
\label{fig-gpeffect-2Dcomplete}}
\end{minipage}
\end{center}
\end{figure}

\begin{figure}[tb]
\begin{center}
\begin{minipage}[t]{12 cm}
\epsfig{file=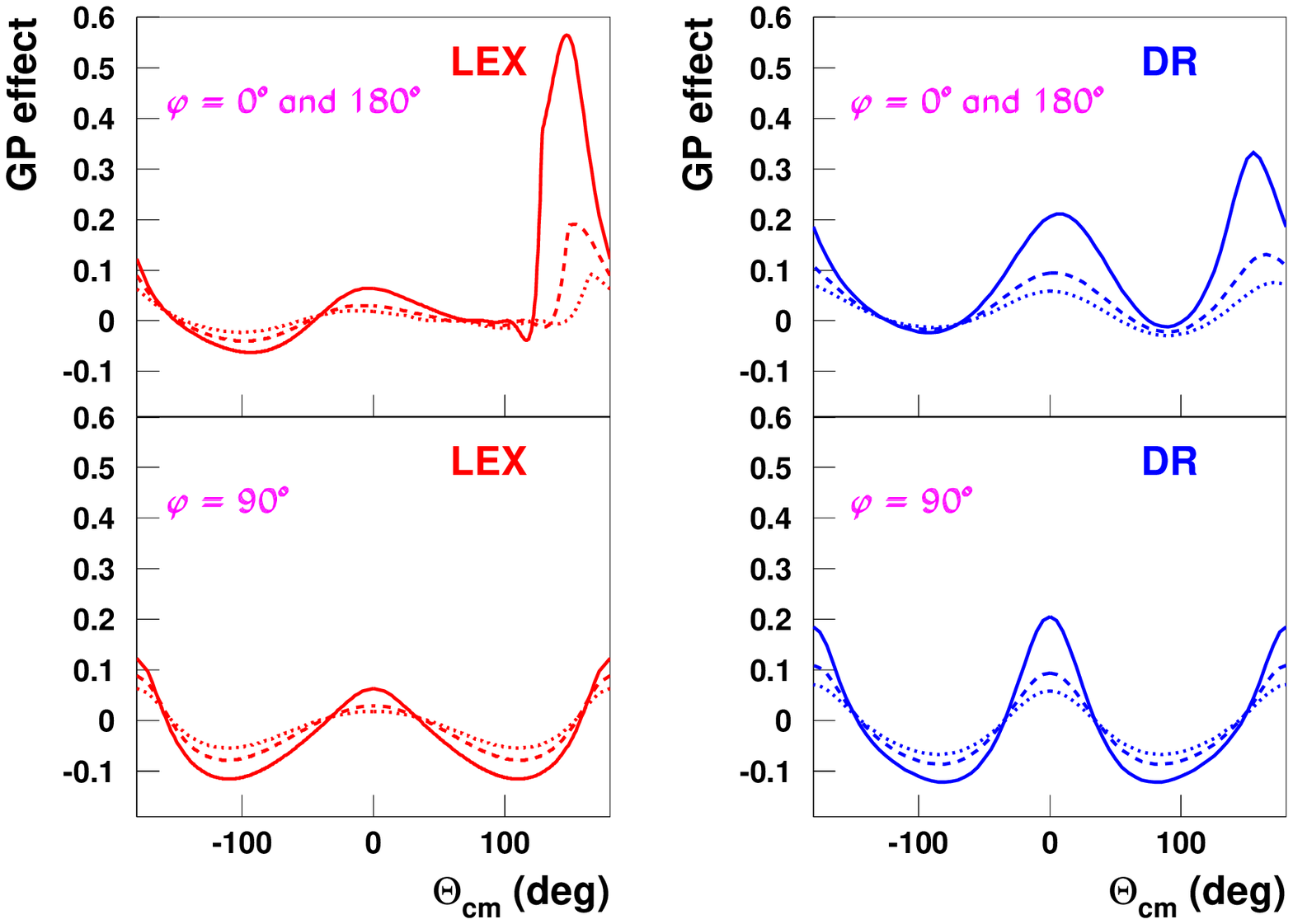,scale=0.6}
\end{minipage}
\begin{minipage}[t]{16.5 cm}
\caption{(color online) Dependence in $\epsilon$ at fixed $Q^2$ = 0.2 GeV$^2$ and $\qpr$ = 110 MeV/c. Left plots: the GP effect from the LEX, in-plane (top) and at  $\phicm$ = 90$^{\circ}$ (bottom) for three values of $\epsilon$: 0.5 (dotted), 0.7 (dashed) and 0.9 (solid). Right plots: same study for the GP effect from DR.  The structure functions have the same values as in Fig.~\ref{fig-cs-and-gpeffect-fig1}. 
\label{fig-cs-and-gpeffect-epsilon}}
\end{minipage}
\end{center}
\end{figure}

The main features of the cross section and GP effect are illustrated in Figs.~\ref{fig-cs-and-gpeffect-fig1} to \ref{fig-cs-and-gpeffect-epsilon}, as a function of the VCS kinematical variables, varying them one at a time.


Figure~\ref{fig-cs-and-gpeffect-fig1} shows a typical $\qpr$-behavior. The cross section is of bremsstrahlung type ($\sim  1/\qpr$), and hence infra-red divergent at the origin. It is well-known that this divergence is compensated by another infra-red divergent term: the virtual radiative correction to electron-proton elastic scattering. In any case the very low $\qpr$-region $(\le 10$ MeV/c) is of no interest for VCS; it contains no information about the GPs, since $d \sigma \to \sigbhb $ when $\qpr \to 0$.
The GP effect in the LEX approach, as defined above, is roughly quadratic in $\qpr$ (cf. top-right plot of Fig.~\ref{fig-cs-and-gpeffect-fig1}). The DR cross section has a more complex behavior. The $\Delta(1232)$ resonance, which is incorporated through the resonant $\pi N$ intermediate states, shows up as a broad bump (cf. bottom-left plot of Fig.~\ref{fig-cs-and-gpeffect-fig1}) due to the non-Born contribution
\footnote{We remind that the total c.m. energy  is given by: $\sqrt{s} = \qpr + \sqrt{ \qprpowtwo + \mnucleon ^2 }$. The interference of the non-Born and the BH+Born amplitudes slightly shifts the resonance peak (nominally at $W=1232$ MeV, or $\qpr = 259$ MeV/c) to the left.}. 
As a consequence, the GP effect from DR may differ noticeably from the LEX one, as soon as $\qpr \ge 50$ MeV/c (cf. bottom-right plot of Fig.~\ref{fig-cs-and-gpeffect-fig1}). Another important feature is that the sensitivity of the cross section to the GPs is enhanced in the region above the pion production threshold. This last property is better seen in Fig.~\ref{fig-cs-and-gpeffect-fig2}, where DR calculations are shown for different sets of parameters $(\la, \lb)$. The sensitivity to the GPs is manifest in the differences between the possible shapes of the resonance bump.

An example of the behavior as a function of the angles $\thcm$ and $\phicm$ is given in Fig.~\ref{fig-cs-and-gpeffect-angular}. The GP effect has a complex angular dependence, implying that experimentally one cannot integrate over too wide regions in $(\cthcm , \phicm )$, otherwise the fine variations of the GP effect are missed. In this figure one can already notice that there are few angular regions where the GP effects from LEX and from DR agree to  better than a few percent of the BH+Born cross section (more on this in Sect.~\ref{sec-further-comments-on-the-higher-order-term-of-the-lex}).

Figure~\ref{fig-gpeffect-2Dcomplete} provides a more complete view, showing the full complexity of the GP effect in the 2D $(\cthcm , \phicm )$ phase space. The difference of behavior between the LEX and DR calculations (left and middle plots) appears clearly; one will also note the rapid variation of the GP effect at very backward angles (near $\cthcm = -1$), a region to handle with care for GP extraction.
 Figures~\ref{fig-cs-and-gpeffect-fig1} to \ref{fig-gpeffect-2Dcomplete} demonstrate that the LEX and DR approaches provide two significantly different descriptions of the $\epgreact$ cross section.

 Figure~\ref{fig-cs-and-gpeffect-epsilon} shows examples of the $\epsilon$-dependence of the GP effect. The effect increases with $\epsilon$, more than linearly, and in all cases it is advantageous to work at high $\epsilon$ for a GP extraction (cf. Table~\ref{tab-exp-3-sf} for the $\epsilon$-values of the experiments).

\vskip 3mm 


The figures presented in this section can serve as guidelines for designing VCS experiments.

\begin{figure}[tb]
\begin{center}
\begin{minipage}[t]{12 cm}
\epsfig{file=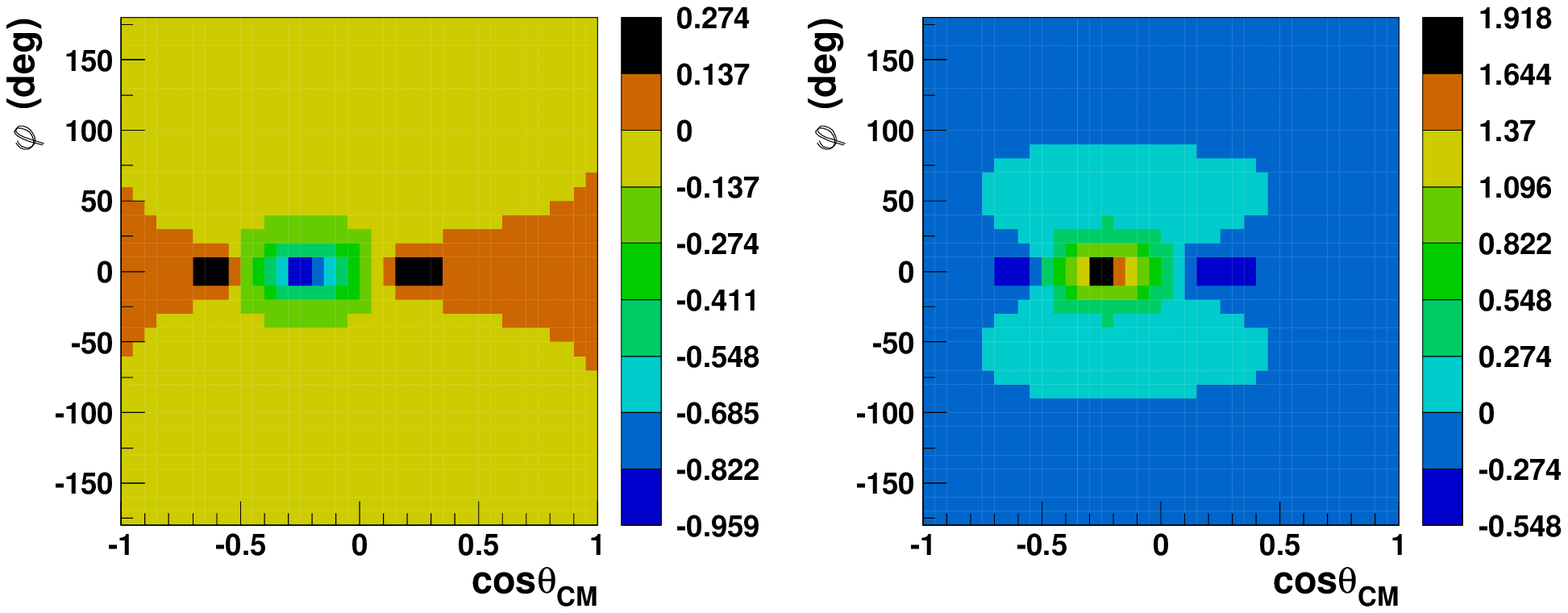,scale=0.6}
\end{minipage}
\begin{minipage}[t]{16.5 cm}
\caption{ (color online) The  $\vll$ (left) and $\vlt$ (right) coefficients of the LEX formula of Eq.~(\ref{lexformula-1}) (see the Appendix for their definition),  in the 2D-plane $(\cthcm, \phicm )$ at $Q^2 =0.2$ GeV$^2$, $\qpr=110$ MeV/c and $\epsilon$ = 0.9. 
\label{fig-vll-vlt}}
\end{minipage}
\end{center}
\end{figure}

For an experiment aimed at extracting GPs using the LEX below the pion threshold, we can summarize the main prescriptions as follows. The GP effect increases with $\qpr$  and  $\epsilon$, and is more easily measurable at high values:  $\qpr \sim$ 100 MeV/c and $\epsilon$ close to 1. 
The GP effect has a strong dependence on the  $\thcm$ and $\phicm$ angles, due to the angular variations of the $\vll$ and $\vlt$ coefficients in the $\Psi_0$ term, which are shown in Fig.~\ref{fig-vll-vlt}. The GP effect ranges from $\sim$ -15\% to +15\% in most of the $(\cthcm, \phicm )$ phase space, except at very backward $\thcm$ where the behavior is most complex. 
The choice of the optimal region in  $(\cthcm, \phicm )$  for a good LEX fit is not an easy task, and every experiment explored differently promising conditions. First, one should obviously avoid the region of the BH peaks, due to their lack of sensitivity to the GPs and the very rapid variation of the cross section. 
Second, one should have a sufficiently large lever arm in $\vll$ and $\vlt$
\footnote{This lever arm is determined by the measured range in the angles  $\thcm$ and $\phicm $.}, 
because they are the weighing coefficients of the fitted structure functions. Note that in some angular regions, these coefficients vanish: e.g., $\vll$ = 0 at $\thcm=0^{\circ}$ and 180$^{\circ}$, $\vlt$ = 0 on a continuous ``ring'' passing near the point  $( \thcm=90^{\circ}, \phicm=90^{\circ})$ (cf. Fig.~\ref{fig-vll-vlt}). These regions are more specifically sensitive to either $\pllptte$ or $\plt$. 
Third,  if one applies the LEX fit in its usual truncated form, one should try to make sure that the higher-order terms $\ohigher$ are small enough to be neglected. If this is not the case, the LEX fit should be modified such as to take into account the higher-order contributions beyond the GPs we want to extract.

\smallskip
\smallskip
\smallskip


The development of the DR formalism opened the path for a second family of VCS experiments that focus on higher c.m. energy, namely the $\Delta(1232)$ region. An advantage in this case is that the sensitivity to the GPs is enhanced in the region above the pion production threshold. Furthermore,  the analysis of the data below and above the pion threshold through the DR framework is similar. On the other hand, in this region one cannot utilize the LEX, the extraction of the GPs has to rely solely on the DR approach. The $N\rightarrow\Delta$ transition amplitudes are used as an input in the DR calculation. These amplitudes have been extensively studied in the past two decades; they have been determined within an experimental uncertainty that ranges between a few to $10\%$, depending on the amplitude, and they are also associated with a theoretical model uncertainty which is typically of a similar magnitude to the experimental one. A detailed treatment of their experimental and model uncertainties is involved in the data analysis. A corresponding uncertainty is introduced to the extracted GPs but the effect is small compared to the overall level of the uncertainty. 
Similarly to the measurements below the pion production threshold, one has to be careful to avoid the BH peaks where the sensitivity to the GPs gets greatly suppressed, while one has to also deal with the rapid variation of the cross section within the measured acceptance. A feature that has been employed in this type of experiments is the measurement of the cross section azimuthal asymmetries, which offer sensitivity to the GPs while they allow for the suppression of part of the systematic uncertainties. 
Ideally one should aim at measuring a sufficiently large lever arm in $\thcm$, a range of about $50^{\circ}$ that is restricted by the BH peak. Lastly, one should give special consideration to the selection of the beam energy. One may find it beneficial to employ higher beam energies, as this will increase the sensitivity to the GPs and the cross section rate. For example, while measuring at a specific momentum transfer in the regime of $Q^2$ $\approx~0.5$ GeV$^2$ one could increase the counting rate by 25\% if one doubles the beam energy from 2 GeV to 4 GeV. Nevertheless, by working through this exercise one has to be careful, since moving the spectrometer to smaller angles will introduce a higher level of accidental rates which can be a limiting factor on the beam current and on the experimental dead time. At the same time one has to consider the resolution effects as, e.g., the momentum of the electron arm is increased.

\subsection{VCS extension to the $N \rightarrow \Delta$ program}
\label{sec-N-to-Delta-program}

The study of the $N \rightarrow \Delta$ transition involves an ongoing experimental and theoretical effort of nearly three decades, and has contributed significantly to the understanding of the nucleon structure and dynamics~\cite{Alexandrou:2012da}. One of the highlights of this program involves the exploration of the two quadrupole amplitudes, i.e., the electric quadrupole $(E2)$ and the Coulomb quadrupole $(C2)$, which allow one to investigate the presence of non-spherical components in the nucleon wave function~\cite{Sparveris:2004jn,Blomberg:2015zma}. It is the complex quark-gluon and meson cloud dynamics of hadrons that gives rise to these components, which, in a classical limit and at large wavelengths, will correspond to a "deformation". 
The spectroscopic quadrupole moment provides the most reliable and interpretable measurement of the presence of these components. For the proton it vanishes identically because of its spin 1/2 nature; instead, the signature of such components is sought in the presence of resonant quadrupole amplitudes in the $\gamma^* N\rightarrow \Delta$ transition. The ratios of the electric and Coulomb amplitudes with respect to the magnetic dipole amplitude, respectively, the EMR and the CMR, are routinely used to present the relative magnitude of the amplitudes of interest. Non-vanishing resonant quadrupole amplitudes will signify the presence of non-spherical components in either the proton or in the $\Delta (1232)$, or more likely, in both; moreover, their $Q^2$-evolution is expected to provide insights on the mechanism that generates them.

The experimental program has focused primarily on the dominant $\pi^0$ and $\pi^+$ excitation channels, due to the favourable branching ratios of approximately 66$\%$ and 33$\%$, respectively. For these channels, although the procedure of extracting the quadrupole amplitude signal from the measured cross sections is rather straightforward, the isolation of $E2$ and $C2$ becomes challenging since there are non-resonant background contributions that are coherent with the resonant excitation of the $\Delta (1232)$  and of the same order of magnitude. 
One has to constrain these interfering background processes (Born terms, tails of higher resonances, etc) to purely isolate the resonant quadrupoles, but doing so with a large number of background amplitudes is nearly impossible. As a result, the amplitudes are extracted with model errors which are often poorly known and rarely quoted. The effect of the physical background amplitudes, the treatment, and the control of the corresponding model uncertainties to the transition form factors represent an open front in this experimental program. 
The photon channel can be instrumental towards this direction, offering a valuable alternative to explore the same physics signal. One difficulty lies into the small branching ratio of approximately 0.6$\%$, which explains why experiments have so far relied primarily on the measurement of the two single-pion channels. In the photon channel, extracting the signal of interest from the measured cross sections is not as straightforward. 
Whereas the pion-electroproduction cross section can be factorized into a virtual photon flux and a sum of partial cross sections that contain the signal of interest, this is not possible for the photon-electroproduction process; the detected photon can emerge not only from the de-excitation of the $\Delta (1232)$, but also from the incoming or scattered electron, i.e., from the Bethe-Heitler process. The VCS reaction $\gamma^*p\rightarrow \gamma p$ amplitude also contains a Born component. 
The non-Born amplitude contains the physics of interest, which includes the GPs as well as the resonant amplitudes of the N$\rightarrow \Delta$ transition. The DR framework has made possible to analyze this type of experimental measurements and steps have been taken towards that direction in recent years. These measurements carry significant scientific merit since the resonant amplitudes are isolated within different theoretical frameworks in the two (pion and photon) channels, the background contributions are of different nature for the two cases and therefore present different theoretical problems. Thus, important tests of the reaction framework and of the model uncertainties of the world data are offered by the comparison of the results from the two different channels.

\subsection{QED Radiative corrections to VCS}
\label{sec-qed-radiative-corrections-to-vcs}

Before the advent of VCS as a dedicated research field, the $\epgreact$ reaction was just seen as being part of the radiative corrections to the $(ep \to ep)$ elastic process, namely, the internal bremsstrahlung part, dominated by radiation from the electron lines (i.e.,  the BH graphs of Fig.~\ref{fig-theo-graphs-1}). For the measurement of GPs, the $\epgreact$ reaction becomes a specific process, being subject to its own radiative effects. It is therefore crucial to understand and handle the radiative corrections  to the  $\epgreact$ reaction, as their effect on the cross section is of the same order as the GP effect itself.

This topic has been studied extensively in Ref.~\cite{Vanderhaeghen:2000ws} and also in several Thesis works~\cite{MarchandPhD:1998,LhuillierPhD:1997,FriedrichPhD:2000}. The radiative corrections to $\epgreact$ have been developed in deep analogy with radiative corrections to elastic scattering $(ep \to ep)$, and are globally as important as for the elastic process. This section summarizes the main features for the VCS case.

The calculation includes all graphs contributing to order $(\aqed ^4)$ in the cross section. One distinguishes between virtual corrections, that imply the exchange of a supplementary virtual photon, and real radiative corrections that imply the emission of a supplementary real photon. 
All infra-red divergences cancel when combining the soft-photon emission processes, from (virtual+real) corrections. One is then left with:  a  virtual correction $\delta_{V}$, plus  a real internal correction $\delta_R$ (i.e., a real radiation coming from any line of the $\epgreact$ graph), plus a real external correction $\delta '$ (i.e., a real radiation coming from another nucleus  in the target). The $\delta_{V}$ term is independent of acceptance cuts, and cannot be calculated analytically. Some parts of it, namely the vertex corrections to the BH graphs, have to be evaluated numerically, a task that required  most developments and innovative tricks w.r.t. the elastic case. The result exhibited a remarkable continuity with the virtual correction in the elastic case. 
The $\delta_{V}$ term  varies slowly with $Q^2$ and is almost constant as a function of the  other variables. The $\delta_R$ term is divided into an acceptance-dependent part, $\delta_{R1}$, and an acceptance-independent part, $\delta_{R2}$. 
The $\delta '$ term is treated in a classical way as in~\cite{Tsai:1971qi}. 
The  $\delta_{R2}$ term is analytical, varying slowly with $Q^2$ and  being almost constant as a function of the  other variables, except near the BH peaks. The  $\delta_{R1}$ term has large variations in the whole $(\qpr , \cthcm , \phicm)$ phase space. The  $\delta_{R1}$ and  $\delta '$  corrections are usually implemented in the simulation of the experiment, at the event generation level, in order to create a realistic radiative tail and apply properly the experimental cuts~\cite{Janssens:2006vx}.

 One gets for the ``radiatively corrected`` cross section the formal expression~\cite{Vanderhaeghen:2000ws}:
%
%
\bea
\sigexp ^{corrected}  =  \sigexp ^{raw} \cdot F_{rad} \ , \ \  \ \mbox{with} \  \ \ 
 F_{rad}  =  [ 1 + \delta_{V} + \delta_{R1} + \delta_{R2} + \delta ' ] ^{-1} ,
\eea 
%
%
where $\sigexp ^{raw}$ is the raw measured cross section, and $\sigexp ^{corrected}$ is the one that can be compared to the theory. The radiative corrections can also be exponentiated. Smaller terms can also be added, such as the two-photon exchange contribution, $\delta_1$, and radiative corrections at the proton side, $\delta_2^{(0)}$~\cite{Vanderhaeghen:2000ws}.

Numerically, for instance at $Q^2$ = 0.3 GeV$^2$ and kinematics close to those of the  first VCS experiment at MAMI, one has~\cite{MarchandPhD:1998}: 
two large terms, $ \delta_{V} \simeq -0.16$ and $\delta_{R2} \simeq +0.22$, 
and a small term, $(\delta_1+\delta_2^{(0)}) \simeq -0.01$, 
so that $1/( 1 + \delta_{V} + \delta_{R2} +\delta_1+\delta_2^{(0)}) \simeq 0.95$. The terms  $\delta_{R1}$ and $\delta '$ are also large, but their value depends on experimental cuts. With a cut on the maximal soft-photon energy $\Delta E_s$ = 15 MeV, one gets: $\delta_{R1} \simeq -0.17$ and $\delta ' \simeq -0.07$.

The accuracy on the  $F_{rad}$ factor is estimated to be $\pm$ 1-2\%. As in many other fields, asymmetry measurements in VCS are much less affected by radiative corrections than absolute cross-section measurements.

\section{Experiments}
\label{sec-experiments}

Dedicated VCS experiments have been performed since 1995 at various electron accelerators: MAMI, MIT-Bates, and JLab. All experiments used the same technique: an electron beam, a liquid hydrogen target, and the detection of the two charged particles $e'$ and $p'$ in magnetic spectrometers.  The exclusive reaction is then identified using the missing mass squared of the undetected photon. Complementary information can be found in previous reviews~\cite{dHose:2000vgm,Fonvieille:2004rb,HydeWright:2004gh,dHose:2006bos,Downie:2011mm,Hagelstein:2015egb}; details on apparatus and physics analyses can be found in the Thesis works 
\cite{RochePhD:1998,LhuillierPhD:1997,FriedrichPhD:2000,MarchandPhD:1998,LaveissierePhD:2001,JaminionPhD:2000,DegrandePhD:2000,TodorPhD:2000,JutierPhD:2001,BourgeoisPhD:2005,BensafaPhD:2006,JanssensPhD:2007,DoriaPhD:2008,BericicPhD:2015,CorreaPhD:2016,BenaliPhD:2016,BlombergPhD:2016} 
and the experimental publications 
\cite{Roche:2000ng,dHose:2006bos,Laveissiere:2004nf,Fonvieille:2012cd,Bourgeois:2006js,Bourgeois:2011zz,Bensafa:2006wr,Sparveris:2008jx,Doria:2015dyx,Janssens:2008qe,Blomberg:2019caf,Bericic:2019faq}.

Sect.~\ref{sec-types-of-experiments} gives a panel of all VCS experiments performed so far and the type of obtained results. Sections~\ref{sec-experiments-dedicated-to-gps} to \ref{sec-further-comments-on-the-higher-order-term-of-the-lex}  are devoted to the measurement of the structure functions $\pllptte$ and $\plt$ and the scalar GPs of the proton from unpolarized data, a topic which has seen the longest and most continuous developments. The other VCS experiments and results will be summarized in Sect.~\ref{sec-other-experimental-results}.

\subsection{Types of experiments}
\label{sec-types-of-experiments}

Due to the limited  acceptance of spectrometers, each experiment has been performed at isolated values of $Q^2$ and $\epsilon$. Regions in $W$ both below and above the pion production threshold have been explored, and some experiments used polarization degrees of freedom. Tables~\ref{tab-exp-1} and  \ref{tab-exp-2} summarize the characteristics of each experiment.

\begin{table}[h]
\begin{center}
\begin{minipage}[t]{16.5 cm}
\caption{Type of VCS experiments (in chronological order).
}
\label{tab-exp-1}
\end{minipage}
\begin{tabular}{|c|c|c|c|c|c|}
\hline
Laboratory & polarization & $Q^2$ (GeV$^2$) & c.m. energy W  & $\epsilon$ & data taking \\
\hline
MAMI-I & unpolarized & 0.33   & $< \pi$ threshold & 0.62 &
 1995-1997   \\
JLab  & unpolarized & 0.92, 1.76 & up to  1.9  GeV & 0.95, 0.88 &
  1998  \\
MIT-Bates & unpolarized & 0.057   & $< \pi$ threshold & 0.90 &
  2000  \\
MAMI-II & $\vec e$ & 0.35   & $\Delta(1232)$ & 0.48 &
 2002-2004   \\
MAMI-III & unpolarized & 0.06   & $\Delta(1232)$ & 0.78  & 2003    \\
MAMI-IV & $\vec e$ and $\vec p \, '$ & 0.33   &  $< \pi$ threshold & 0.645 &
 2004-2006  \\
MAMI-V  & unpolarized & 0.20   &  $\Delta(1232)$ & 0.77 &
 2012 \\
MAMI-VI  & unpolarized & 0.1, 0.2, 0.45   &  $< \pi$ threshold & 0.91, 0.85, 0.63 &
 2011-2015  \\
\hline
\end{tabular}
\end{center} 
\end{table}

\begin{table}[h]
\begin{center}
\begin{minipage}[t]{16.5 cm}
\caption{VCS experiments: type of results. Actual results are given in Tables~\ref{tab-exp-3-sf} and \ref{tab-exp-4-gp}. }
\label{tab-exp-2}
\end{minipage}
\begin{tabular}{|c|c|c|c|}
\hline
Laboratory &  observables & results & ref. \\
\hline
MAMI-I &  cross sections & structure functions and GPs & \cite{Roche:2000ng,dHose:2006bos}  \\
JLab  &   cross sections & structure functions and GPs & \cite{Laveissiere:2004nf,Fonvieille:2012cd} \\
MIT-Bates &  cross sections &  structure functions and GPs & \cite{Bourgeois:2006js,Bourgeois:2011zz} \\
MAMI-II &  beam spin  asymmetry & asym. for $\vec e p \to e p \gamma$ and $\vec e p \to e p \pi^0$ & \cite{Bensafa:2006wr}  \\
MAMI-III &  cross sections & $M_{1+}^{3/2}$ in $\gamma^* N \to \Delta$ transition & \cite{Sparveris:2008jx}  \\
MAMI-IV & double spin asymmetry and  &  polarized struct. funct. $\pltperp$ and & \cite{Doria:2015dyx}  \\
    &  cross sections & structure functions and GPs & \cite{Janssens:2008qe}  \\
MAMI-V &  cross sections and asym. & electric GP and CMR & \cite{Blomberg:2019caf}  \\
MAMI-VI & cross sections & structure functions and GPs & \cite{Bericic:2019faq}  \\
\hline
\end{tabular}
\end{center} 
\end{table}

\subsection{Experiments dedicated to the scalar GPs and structure functions}
\label{sec-experiments-dedicated-to-gps}
 
Here we give a brief overview of the VCS experiments which have measured the structure functions $\pllptte$ and $\plt$ and the scalar GPs of the proton. 
%
%
The MAMI-I experiment~\cite{Roche:2000ng,dHose:2006bos} was the really pioneering one, in which all experimental aspects, from design to analysis, were established for the first time, including, e.g., radiative corrections (see Sect.~\ref{sec-qed-radiative-corrections-to-vcs}), dedicated Monte-Carlo simulations, LEX fit methods, etc. The covered range in $\qpr$ was complete, but data were limited to in-plane angles.
%
%
The JLab experiment~\cite{Laveissiere:2004nf,Fonvieille:2012cd} explored the highest photon virtualities so far, in the range 1-2 GeV$^2$, and found very small GPs, indicating a fast fall-off with $Q^2$. This experiment was also the first one to show that GP extractions both below and above the pion production threshold, using respectively the LEX and DR formalisms, gave consistent results.
%
%
The MIT-Bates experiment~\cite{Bourgeois:2006js,Bourgeois:2011zz} exploited out-of-plane kinematics  more specifically, and  made measurements at the smallest $Q^2$ so far (0.057 GeV$^2$), enabling the first estimation of the mean-square radius of the electric GP (see Sect.~\ref{sec-mean-square-radii}). This experiment was also the first one to evidence a bias in the LEX fit (see Sect.~\ref{sec-experimental-results-sfs}).
%
%
At this stage the $Q^2$-dependence of the observables started to  appear as non-trivial, especially for $\pllptte$ and $\aeq$,  showing an enhancement at  $Q^2$ =  0.33 GeV$^2$ w.r.t. the other  data points (see Figs.~\ref{fig-exp-sf-1} and \ref{fig-exp-ab-1}).
Measurements were repeated at this $Q^2$ during the MAMI-IV experiment~\cite{Janssens:2008qe} (ignoring the double polarization information), at angular kinematics very similar to MAMI-I, and confirmed the results previously found.


This situation, with scarce data points and a puzzling $Q^2$-behavior, motivated the need for new measurements. Two recent experiments performed in the intermediate $Q^2$-range brought new elements of answer.


The MAMI-V experiment~\cite{Blomberg:2019caf} determined the electric GP at $Q^2$ = 0.20 GeV$^2$ from cross section and asymmetry measurements in the $\Delta(1232)$ resonance region. In parallel the experiment offered an important first measurement of the $N \rightarrow \Delta$ quadrupole amplitude through the photon channel (cf. Sect.~\ref{sec-dr-fits-in-the-first-resonance-region}), thus providing a stringent control to the model uncertainties of the $N \rightarrow \Delta$ transition world data.

 
The MAMI-VI experiment~\cite{Bericic:2019faq} was performed at three values of  $Q^2$: 0.10, 0.20 and 0.45 GeV$^2$. The structure functions $\pllptte$ and $\plt$ and the scalar GPs were extracted with good precision from cross-section data below the pion production threshold. Out-of-plane kinematics were designed at each $Q^2$, in the line of the MIT-Bates experiment, but covering a larger angular phase space (cf. Fig.~\ref{fig-ohigherdr-allexp}). Polarizability fits were performed with and without a novel bin selection method that was aimed at suppressing the higher-order terms of the LEX (see Sect.~\ref{sec-lex-fits}). This was another inheritance from the MIT-Bates experiment.


The new  experiment E12-15-001 at JLab~\cite{Sparveris:2016} acquired data recently and is currently at the early stages of the data analysis (see Sect.~\ref{sec-e12-15-001-experiment}). The experiment aims to determine the two scalar GPs in the range $Q^2$ = 0.3 GeV$^2$ to $Q^2$ = 0.75 GeV$^2$ through cross section and asymmetry measurements in the nucleon resonance region. The experiment will rely on the extraction of the GPs through the DR framework. The higher energy employed in these measurements offers an enhanced sensitivity to the GPs, and the results will provide a direct cross check to the MAMI-I and MAMI-IV results where the enhancement of the electric GP with $Q^2$ was previously observed.

\subsection{Extraction methods for the scalar GPs and structure functions}
\label{sec-extraction-methods}

As for the extraction of polarizabilities in RCS, the extraction of GPs in VCS is not direct and requires a fit, made within a theoretical framework. Experiments use two different frameworks: a model-independent one based on the LEX, and a model-dependent one based on Dispersion Relations. These two formalisms have different domains of validity in $W$.


The LEX formalism is valid only below the pion production threshold.  Indeed, in Ref.~\cite{Guichon:1995pu} the VCS amplitude is considered as real, a property that holds only for $W < (\mnucleon + m_{\pi})$. As soon as hadronic intermediate states other than the nucleon can be created on-shell, starting by a nucleon plus a pion, the VCS amplitude acquires an imaginary part and the LEX formalism~\cite{Guichon:1995pu} is not valid anymore. The LEX fit (on unpolarized data) at fixed $\qcm$ and $\epsilon$, only yields the two structure functions $\pllptte$ and $\plt$, and individual GPs are not accessed
\footnote{The particular case of the LEX fit in doubly polarized VCS, where the six lowest-order GPs can in principle be disentangled, will be discussed in Sect.~\ref{sec-other-experimental-results}.}. 
The scalar GPs can be deduced only if one subtracts from these two structure functions their spin-dependent part, i.e., $\ptt$ and $\plt \, _{spin}$.  In absence of any available measurements of the spin GPs, this subtraction relies on a model calculation.

In essence, since $\sigbhb$ is the cross section without any polarizability effect, the structure functions  $\pllptte$ and $\plt$ are always obtained by fitting the deviation of $\sigexp$ from $\sigbhb$. The difference $(\sigexp - \sigbhb)$, or more precisely the quantity $\deltam = (\sigexp - \sigbhb)/ ( \Phi  \cdot \qpr)$, plays therefore a special role in the LEX fits
\footnote{This quantity is noted $({\cal{M}}_0^{exp} - {\cal{M}}_0^{BH+Born})$ in Ref.~\cite{Guichon:1998xv}.}.


We have seen in Sect.~\ref{sec-the-dr-model} that Dispersion Relations provide a very appropriate and efficient formalism to analyze VCS experiments both below and above the pion production threshold. The imaginary part of the VCS amplitude is a central ingredient of the model, entering dispersive integrals saturated by $\pi N$ intermediate states. As a key feature, the existence of free parameters in the model, related to the unconstrained part of $\aeq$ and $\bmq$ (cf. Eq.~(\ref{asymp})), allows us to perform an experimental fit in order to extract these scalar GPs. On the other side, the spin GPs are fully constrained in the model and cannot be fitted. 
The formalism is suited for all values of $W$ up to $\sim$ 1.3 GeV, i.e. slightly above the $\pi \pi N$ threshold, thus covering most of the $\Delta(1232)$ resonance region.

\subsubsection{LEX fits}
\label{sec-lex-fits}

The LEX fit in its standard form is based on the comparison of a set of measured cross sections, $\sigexp$, at fixed $\qcm$ and $\epsilon$, to the expression of $\siglex$ in Eq.~(\ref{lexformula-1}). Neglecting the $\ohigher$ term means that the quantity  $\deltam$ defined above is assumed to have no dependence on $\qpr$. The fit is thus most simple, consisting in a linear $\chi^2$-minimization of $\deltam$ with two free parameters, the structure functions $\pllptte$ and $\plt$:
$$ 
\chi^2_{\mathrm{LEX}} \ = \ \sum_{\mathrm{exp. points} \  i=1}^N \ 
{\bigg [} \ 
{ \displaystyle  \  \deltam (i)
  \ - \ 
( \vll (i) \cdot  ( \pllptte ) + \vlt (i) \cdot \plt )
\over 
\displaystyle        \Delta \deltam (i) _{\mathrm{stat}}
}
\ {\bigg ]} ^2 \ \ .
$$

This standard LEX fit has two virtues: model-independence and simplicity. However, we still have at present time a limited understanding of its validity. Other variants of the LEX fit exist, addressing in several ways the possible $\qpr$-evolution of the cross section.


In Refs.~\cite{dHose:2006bos,RochePhD:1998} two variants were investigated on the cross-section data of the MAMI-I experiment: i) a linear $\qpr$-dependence of $\deltam$; ii) a more complete form of the (BH+Born)-(non-Born) interference term, in which all six lowest-order GPs are free parameters. This study essentially concluded that the observed $\qpr$-dependence of $\deltam$ was always weak, but the obtained structure functions nevertheless showed some sensitivity to these different fitting hypotheses.


More recently, in the MAMI-VI experiment~\cite{Bericic:2019faq}, another variant of the LEX fit was considered, by selecting only the regions in phase space where the terms $\ohigher$ in Eq.~(\ref{lexformula-1}) are  small enough to be neglected. 
To this aim, one uses the DR model, in which the cross section includes all orders in $\qpr$.  By subtracting the $\siglex$ cross section of Eq.~(\ref{lexformula-1}) from the DR cross section $\sigdr$, one isolates just the higher-order terms of the LEX. The quantity  $\ohigherdr =  ( \sigdr - \siglex ) / \sigbhb $ is calculated at the kinematics of every cross-section point. Then, keeping only the points where $ \vert \ohigherdr \vert $ is smaller than a few percent, one performs the standard LEX fit as defined above. 
Although model-dependent, this estimator  $\ohigherdr$ can be useful to improve the reliability of the LEX fit. Such a case has already been observed for the results at $Q^2$ = 0.1 GeV$^2$~\cite{Bericic:2019faq}. Values reported in Tables~\ref{tab-exp-3-sf} and  \ref{tab-exp-4-gp} for the MAMI-VI experiment are obtained using this phase-space selection criterion.

\subsubsection{DR fits below the pion production threshold}
\label{sec-dr-fits-below-pion-threshold}

VCS data below $W = \mnucleon + m_{\pi}$ can always be analyzed in terms of the two formalisms, LEX and DR, and most experiments performed this double analysis. The DR fit is based on the comparison of the measured cross section to the one calculated by the DR model. In practice this fit is less straightforward than the LEX fit, because the structure functions or GPs do not appear in a simple analytic form in the model cross section. One usually has to scan the whole phase space of the free parameters of the model. Two-dimensional grids, either in $(\la, \lb )$ or in $( \aeq, \bmq )$, have been used to this aim. One builds a $\chi^2$ at each node of the grid: 
$$ 
\chi^2_{\mathrm{DR}} \ = \ \sum_{\mathrm{exp. points} \  i=1}^N \ 
{\bigg [} \ 
{ \displaystyle     \sigexp (i) - \sigdr (i, \mathrm{node})
\over 
\displaystyle        \Delta \sigexp (i) _{\mathrm{stat}}
}
\ {\bigg ]} ^2
$$
and finds the minimum numerically. The minimization provides the values of the scalar GPs and the structure functions $\pllptte$ and $\plt$ as well.

\subsubsection{Systematic errors}
\label{sec-systematic-errors}

In most VCS experiments, the results are dominated by systematic errors, which are larger than the statistical ones by a factor $\sim$ 2 to 4 (or sometimes more)
\footnote{There are only two exceptions: the MIT-Bates experiment and the JLab experiment, data set ``I-b''.}. 
Statistical errors are given by the fit itself, typically by the size of the contour at $(\chi^2_{min}+1)$. For the systematic errors, several sources of uncertainty are well identified:  1) the experimental luminosity and detector efficiencies (triggering, tracking, etc.); 2) radiative corrections to VCS; 3) the choice of proton form factors in the calculation of $\sigbhb$; 4) the solid angle calculation by Monte-Carlo; 5) the limited knowledge of spectrometer optics and experimental offsets; 6) the fitting assumption, or model uncertainty. 
Sources 1) to 3) are generally considered as acting as a global normalization uncertainty on the cross section, while the other sources may not be so global and may induce point-to-point distortions of the angular distributions. 
Overall, it is hard to reduce the total systematic error on the cross sections below the $\pm$ 3-4\% level.

%
%

Although data at low $\qpr$ do not bring much information about the GPs, they can help in reducing the total systematic error. Indeed, for $\qpr \le 50$ MeV/c, the GP effect is very small, of the order of 1\% (resp. 2\%) at $\qpr$ = 25 (resp. 45) MeV/c. In this $\qpr$-range the GP effect is therefore not fully negligible, but it is well under control, even if evaluated with approximate GP values. In these conditions the $\ohigher$ terms vanish and the measured cross section must match the theoretical $\siglex$. The test consists in fitting the global renormalization factor $F_{norm}$ that realizes this matching. This is done by comparing $\sigexp ' =  F_{norm} \times \sigexp$ with $\siglex$
\footnote{Note that this test cannot be performed at higher $\qpr$ because there the GP effect and a normalization effect on the cross section completely mix, or interfere.}. 
The factor $F_{norm}$ found at low $\qpr$ is then applicable to the remaining part of the data set, at higher $\qpr$,  considering that it corrects for global systematic errors which affect every cross-section value in the same way (i.e., sources 1 to 3 mentioned above). The effect of such systematic errors is thus greatly reduced by adopting this renormalization procedure. 
Every VCS experiment having low-$\qpr$ data utilized them to test the absolute normalization of the cross section, and to renormalize it if needed
\cite{Roche:2000ng,dHose:2006bos,Laveissiere:2004nf,Fonvieille:2012cd,Bourgeois:2006js,Bourgeois:2011zz,Bericic:2019faq}. By this method, the overall normalization uncertainty was reduced to, e.g.,  $\pm$ 1\%  in the MAMI-I experiment~\cite{Roche:2000ng,FriedrichPhD:2000} and in the MAMI-VI experiment~\cite{Bericic:2019faq}.

%
%

To apply this renormalization procedure (and more generally to extract GPs) one has to make a choice for the proton form factors $(G_E^p, G_M^p)$, which enter the BH+Born cross section. Different form-factor parametrizations can vary by several percent at intermediate $Q^2$, inducing variations of $\sigbhb$ which can be even larger. The fitted value of $F_{norm}$ therefore depends directly on the form-factor choice. This ``floating normalization'' may seem dangerous at first sight for the stability of the polarizability fit, but actually it is a way to eliminate almost completely the form-factor dependency of the physics results. The reason is the following one. The key feature is that, when changing the value of $G_E^p$ or $G_M^p$, $\sigbhb$ just scales globally, to a good approximation
\footnote{This scaling effect is indeed constant in the $(\qpr, \thcm, \phicm)$ phase space, at fixed $\qcm$ and $\epsilon$, to better than $\pm$ 1\% (see for instance~\cite{CorreaPhD:2016} or~\cite{RochePhD:1998}).}. 
Therefore the normalization factor  $F_{norm}$ of the low-$\qpr$ test will scale accordingly, and in the polarizability fit, $\sigexp$ and $\sigbhb$ will both be rescaled by the same factor. Due to the nature of $\deltam$
\footnote{Being a small number obtained by the difference of two large numbers ($\sigexp$ and $\sigbhb$), the quantity $\deltam$ remains essentially unchanged when both cross sections are rescaled by the same factor, in contrast to the case where $\sigbhb$ is rescaled and not $\sigexp$.}, 
this process stabilizes the fit. In other words,  with a proper normalization of  $\sigexp$ relative to $\sigbhb$, made possible using low-$\qpr$ data, the polarizability fit can concentrate on the important feature which is the  shape  of $(\sigexp - \sigbhb)$ versus $(\qpr, \thcm, \phicm)$, without being disturbed by scale effects.

It was shown in several thesis works~\cite{RochePhD:1998,CorreaPhD:2016,BericicPhD:2015,BenaliPhD:2016} that, if one follows this procedure, the results in terms of structure functions and GPs become essentially independent of the form-factor choice. On the contrary, it was not the case in the MAMI-IV experiment~\cite{Janssens:2008qe}; due to the lack of cross-section data at low-$\qpr$, an explicit and non-negligible error is quoted as coming from the proton form factors, mostly for $\plt$.

To conclude, this whole argumentation on normalization holds to a precision of about $\pm$ 1\% of the cross sections, but not better. Therefore VCS analyses can reach a systematic error due to absolute normalization uncertainty as low as $\pm$ 1\%, but it seems presently an irreducible limit.

\subsubsection{DR fits in the first resonance region}
\label{sec-dr-fits-in-the-first-resonance-region}

The DR model is the only tool to extract GPs from data in the region of the nucleon resonance. The model can be used in different ways to obtain information about GPs, and several types of experimental analyses have been performed in this $W$-region.

\smallskip
\smallskip
\smallskip


One can perform a DR fit as described in Sect.~\ref{sec-dr-fits-below-pion-threshold} to extract the scalar GPs, i.e., by scanning the whole phase space of the $(\la, \lb)$ parameters. Such an analysis was done in the JLab experiment~\cite{Fonvieille:2012cd,LaveissierePhD:2001} with the data set I-b at $Q^2$ = 0.92 GeV$^2$ and $W$ mostly above the pion production threshold, in the range [1-1.28] GeV. Cross sections were measured at backward polar c.m. angle ($\cthcm = -0.975$ to $-0.650$) and full azimuthal coverage. 
This first DR fit in the nucleon resonance region proved to be competitive: its results were in very good agreement with the ones obtained by the same experiment at $W <  (\mnucleon + m_{\pi})$, and they had significantly smaller error bars, mostly for the systematics (cf. the JLab-Ib results compared to the JLab-Ia results in Tables~\ref{tab-exp-3-sf} and \ref{tab-exp-4-gp}).

\smallskip
\smallskip
\smallskip


The first JLab VCS experiment thus demonstrated great success in extracting the scalar GPs from measurements in the nucleon resonance region, opening up the path for more measurements of this type. 
The aim of these experiments was further extended to also include the study of the $N \rightarrow \Delta$ transition form factors (cf. Sect.~\ref{sec-N-to-Delta-program}). The first such experiment was MAMI-III~\cite{Sparveris:2008jx}. The experiment was initially designed to measure the $H(e,e^{'}p)\pi^0$ reaction in the nucleon resonance region and thus the selection of the kinematics, as well as the experiment beam time, was not optimized for the simultaneous measurement of the photon channel. The experiment offered limited sensitivity to the scalar GPs, but was nevertheless successful into making a first exploration of the transition form factors through the photon channel.

\begin{figure}[tb]
\begin{center}
\begin{minipage}[t]{15 cm}
\epsfig{file=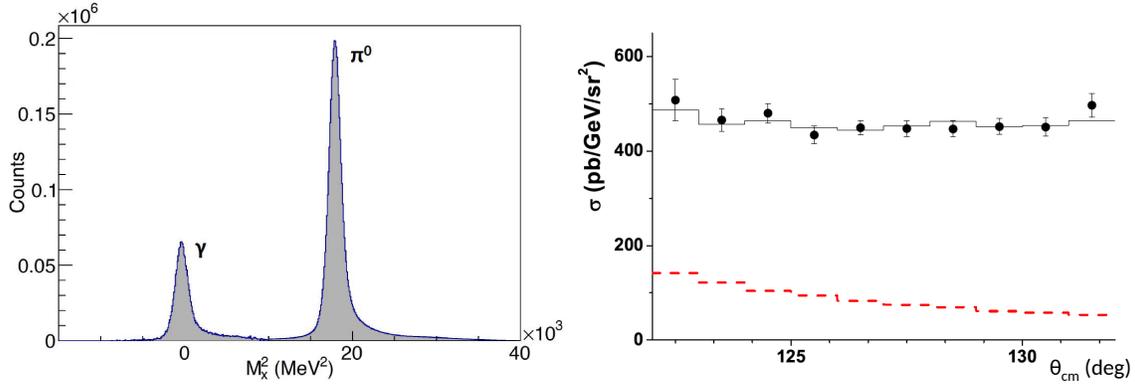,scale=0.6}
\end{minipage}
\begin{minipage}[t]{16.5 cm}
\caption{
(color online) Plots from the MAMI-V experiment. Left panel: The missing mass spectrum offers a clear separation of the photon and of the pion peaks. Right panel: The measured cross sections as a function of $\thcm$. The solid curve corresponds to the full DR calculation~\cite{Pasquini:2000pk} and the dashed curve to the BH+Born term. Figure taken from Ref.~\cite{Blomberg:2019caf}.
\label{f1n}}
\end{minipage}
\end{center}
\end{figure}

\begin{figure}[tb]
\begin{center}
\begin{minipage}[t]{15 cm}
\epsfig{file=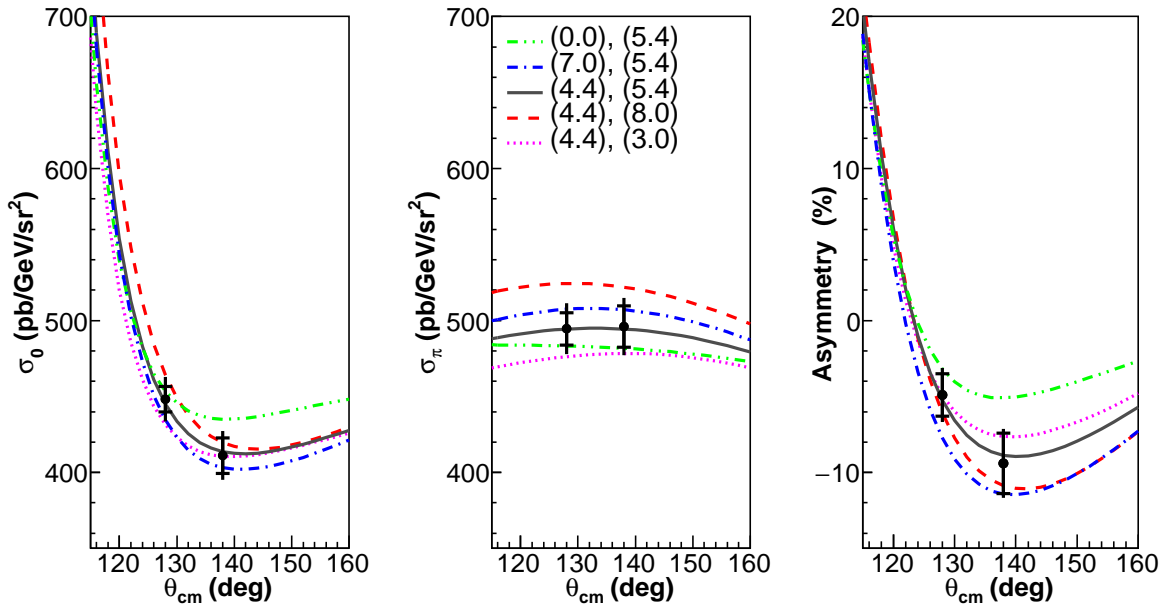,scale=0.80}
\end{minipage}
\begin{minipage}[t]{16.5 cm}
\caption{
(color online) Cross sections and asymmetries measured at $Q^2=0.20$ GeV$^2$ from the MAMI-V experiment. The DR calculation~\cite{Pasquini:2000pk} is also shown with different variations of $\aeq$ and of CMR to exhibit the sensitivity of the measurements to these amplitudes. Chosen values are indicated as numbers in parenthesis in the middle plot. The first number corresponds to the CMR (in percent) and the second one to the value of $\aeq$ (in $10^{-4}$ fm$^3$). Figure taken from Ref.~\cite{Blomberg:2019caf}.
\label{f2n}}
\end{minipage}
\end{center}
\end{figure}

The next step forward involved MAMI-V~\cite{Blomberg:2019caf}, a dedicated experiment that focused on the parallel extraction of the scalar GPs and of the $N \rightarrow \Delta$ transition form factors. A new feature that was introduced in this experiment is the measurement of the cross section azimuthal asymmetries of the type 
$ ( d\sigma_{\phicm = 180^{\circ}} - d\sigma_{\phicm = 0^{\circ}} )  / 
  ( d\sigma_{\phicm = 180^{\circ}} + d\sigma_{\phicm = 0^{\circ}} )  $,  
which offers sensitivity to the physics signal, while, at the same time, allows for the suppression of part of the systematic uncertainties. In these measurements the BH+Born contribution accounts for $\approx 20\%$ of the total cross section (see Fig.~\ref{f1n}), while the primary sources of systematic uncertainties involve the uncertainties in the momenta and the angles of the two spectrometers, the luminosity, the knowledge of the acceptance, and the radiative corrections. The combined uncertainty coming from the solid angle, the luminosity, and the radiative corrections is of the order of $\approx \pm 2.5 \%$ to $\pm 3 \%$  for these measurements, while the part that depends on the uncertainties in the momenta and the angles of the two spectrometers varies on a per setting basis. The statistical uncertainty on the cross section is smaller, ranging between $1.5 \%$ and $2 \%$. 
In the DR fits one has to consider different parametrizations for the proton form factors in the analysis, since these quantities enter the calculation of the BH+Born cross section, as well as the uncertainty in the knowledge of the resonant amplitudes. These two types of uncertainties are of about the same level. In these experiments one measures parasitically the pion electroproduction cross section within the spectrometer acceptance, which is at least an order of magnitude larger compared to the VCS one. As this cross section is well known, it offers a valuable normalization measurement for these experiments. At the same time, one has to be careful with the small tail of the pion events that could contaminate the missing mass peak of the photon channel. The correction from such contributions is rather small and introduces an uncertainty in the cross section at the order of $0.1\%$. 
The measurement of the azimuthal asymmetries helps to reduce the effect of the systematic uncertainties in the fitting of the scalar GPs, and for MAMI-V the $\aeq$ systematic uncertainty was found nearly double compared to the statistical one. The sensitivity of these measurements to the electric GP is exhibited in Fig.~\ref{f2n}. 
The sensitivity to the Coulomb quadrupole amplitude is also explored in the figure, and a detailed discussion about it will be presented in Sect.~\ref{sec-N-to-delta-multipoles}.

\subsection{Results for the structure functions $\pllptte$ and $\plt$}
\label{sec-experimental-results-sfs}


The structure functions are the observables directly fitted from experiments in the LEX approach. As seen in Sect.~\ref{sec-the-low-energy-theorem}, $\pllptte$ as well as $\plt$ are combinations of scalar and spin GPs. Table~\ref{tab-exp-3-sf} collects the values of $\pllptte$ and $\plt$ extracted by the various experiments. Most results come in ``doublets'', under the form of a LEX fit and a DR fit, performed on the same cross-section data, and presented in two successive lines of  Table~\ref{tab-exp-3-sf}. Strictly speaking, the  results of LEX and DR fits are comparable only at the level of these structure functions, not at the level of the GPs, since the LEX fit does not access the latter directly.

From Table~\ref{tab-exp-3-sf} it is clear that each laboratory explored a $Q^2$-range of its own, going from very low (MIT-Bates) to intermediate (MAMI) and high $Q^2$ (JLab). It is also clear that the (total) error of the measurements decreases less rapidly with $Q^2$ than the observables themselves. At high $Q^2$ ($\sim$ 1 GeV$^2$ and above) it becomes increasingly difficult to measure structure functions or GPs which are significantly non-zero (within their error bar). This is evidenced by the JLab results~\cite{Laveissiere:2004nf} at $Q^2$ = 0.92 and 1.76 GeV$^2$.

\begin{table}[h]
\begin{center}
\begin{minipage}[t]{16.5 cm}
\caption{Experimental results for the structure functions $\pllptte$ and $\plt$  using LEX and DR fits, ordered by increasing values of $Q^2$. The first error is statistical and the second one is systematic, except in the special cases in footonote. The first four lines are values at  $Q^2$ = 0  deduced from various fits of the RCS polarizabilities $\ale$ and $\bem$, converted to structure functions using the expressions of the Appendix.
}
\label{tab-exp-3-sf}
\end{minipage}
\begin{tabular}{|l|c|c|ccc|ccc|c|}
\hline
Experiment &  $Q^2$  & $\epsilon$ & \multicolumn{3}{c|}{ $\pllptte$ } & \multicolumn{3}{c|}{ $\plt$  } & type of  \\
nomenclature &  (GeV$^2$) & \  &  \multicolumn{3}{c|}{ (GeV$^{-2}$) } & \multicolumn{3}{c|}{ (GeV$^{-2}$) } &  analysis \\ 
\hline
RCS 1$^{\rm a}$    &     0    &     & 81.0   & $\pm$ 2.0     &  {\scriptsize $\mp$ 2.7$\pm$ 2.0}  & -5.35   & $\pm$  1.33   & {\scriptsize  $\pm$1.33$\pm$1.33}  & \cite{OlmosdeLeon:2001zn} \\
RCS 2$^{\rm a}$    &     0    &     & 71.29   & $\pm$ 2.34   & {\scriptsize $\pm$1.34$\pm$2.01}  & -10.54   & $\pm$  1.17   & {\scriptsize $\mp$0.67$\pm$1.00}   & \cite{McGovern:2012ew} \\
RCS 3$^{\rm b}$    &     0    &     & 80.52   & $^{+3.22} _{-3.61}$   &   & -5.92   &   $^{+1.80} _{-1.74}$   &  & \cite{Pasquini:2019nnx} \\
RCS 4$^{\rm b}$    &     0    &     & 74.97   & $\pm$ 2.68   &   & -8.37   & $\pm$  1.34   &  & \cite{Tanabashi:2018oca} \\
%
\hline
MIT-Bates &  0.057 & 0.90 &     54.5  & $\pm$ 4.8 &  $\pm$ 3.4 &   -20.4 & $\pm$  2.9  & $\pm$ 1.4  & LEX \cite{Bourgeois:2006js} \\
MIT-Bates &  0.057 & 0.90 &     46.7  & $\pm$ 4.9 &  $\pm$ 3.4 &   -8.9 & $\pm$  4.2  & $\pm$ 1.4  & DR \cite{Bourgeois:2006js} \\
\hline
MAMI-VI & 0.10  & 0.91 &     33.15  & $\pm$  1.53  & $\pm$  4.53 & -8.54  & $\pm$   0.60  & $\pm$   1.62     & LEX \cite{Bericic:2019faq}  \\
MAMI-VI & 0.10  & 0.91 &     35.95  & $\pm$  1.80  & $\pm$  5.21 & -9.03  & $\pm$   0.98  & $\pm$   1.82     & DR \cite{Bericic:2019faq}  \\
MAMI-VI & 0.20  & 0.85 &     14.57 & $\pm$   0.55  & $\pm$  3.47 & -5.37  & $\pm$   0.33  & $\pm$   1.25  & LEX \cite{Bericic:2019faq}  \\
MAMI-VI & 0.20  & 0.85 &     14.94  & $\pm$   0.60  & $\pm$   4.06  & -5.31  & $\pm$   0.44  & $\pm$   1.40    & DR \cite{Bericic:2019faq}  \\
MAMI-V$^{\rm c,d}$  & 0.20  & 0.77 &  24.2  & $\pm$  2.3  & $\pm$  5.0  &               &              &                & DR \cite{Blomberg:2019caf} \\
MAMI-I$^{\rm a}$ &  0.33 & 0.62  &    23.7  & $\pm$  2.2  & {\scriptsize $\pm$0.6$\pm$4.3} & -5.0  & $\pm$  0.8  &  {\scriptsize $\pm$1.1$\pm$1.4} & LEX \cite{Roche:2000ng}  \\
MAMI-I$^{\rm e}$ &  0.33 & 0.62  &    23.2  & $\pm$  3.0  &  & -3.2  & $\pm$  2.0  &  & DR \cite{dHose:2006bos}  \\
MAMI-IV$^{\rm f}$ &  0.33 & 0.645  &    27.4   & $\pm$ 1.9   & $\pm$ 3.0  & -6.8  & $\pm$  0.7  & $\pm$  2.2  & LEX \cite{Janssens:2008qe} \\
MAMI-VI & 0.45  & 0.63 &     4.21  & $\pm$   0.65  & $\pm$   2.24    &   -1.00  & $\pm$   0.37  & $\pm$   0.50  & LEX \cite{Bericic:2019faq}  \\
MAMI-VI & 0.45  & 0.63 &     4.10  & $\pm$   0.62  & $\pm$  2.48  &  -1.36  & $\pm$   0.29  & $\pm$   0.40   & DR \cite{Bericic:2019faq}  \\
\hline
JLab-Ia     & 0.92 & 0.95 &     1.77  & $\pm$   0.24  & $\pm$   0.70 &  -0.56   & $\pm$  0.12  & $\pm$   0.17   & LEX  \cite{Laveissiere:2004nf} \\
JLab-Ia     & 0.92 & 0.95 &      1.70  & $\pm$  0.21  & $\pm$  0.89  &  -0.36  & $\pm$  0.10  & $\pm$  0.27     & DR  \cite{Laveissiere:2004nf} \\
JLab-Ib$^{\rm c}$    & 0.92 & 0.95 &      1.50  & $\pm$  0.18  & $\pm$  0.19 &  -0.71  & $\pm$  0.07  & $\pm$  0.05      & DR  \cite{Laveissiere:2004nf} \\
JLab-II     & 1.76 & 0.88 &      0.54  & $\pm$  0.09  & $\pm$  0.20  &  -0.04  & $\pm$  0.05  & $\pm$  0.06     & LEX  \cite{Laveissiere:2004nf} \\
JLab-II     & 1.76 & 0.88 &       0.40  & $\pm$  0.05  & $\pm$  0.16  &  -0.09  & $\pm$  0.02  & $\pm$  0.03  & DR  \cite{Laveissiere:2004nf} \\
\hline
\end{tabular}
\begin{minipage}[t]{16.5 cm}
\vskip 0.5cm
\noindent
%
%
$^{\rm a}$ Two separate systematic errors.  \\
$^{\rm b}$ Total error only.  \\
$^{\rm c}$ DR fit performed in the resonance region. \\
$^{\rm d}$ These values are unpublished. We have deduced them from the electric GP of Table~\ref{tab-exp-4-gp} using the DR formalism.  \\
$^{\rm e}$ Statistical error only. \\
$^{\rm f}$ Values for one selected choice of proton form factors (parametrization of Ref.~\cite{Hohler:1976ax}). \\
\end{minipage}
\end{center} 
\end{table}

\begin{figure}[tb]
\begin{center}
\begin{minipage}[t]{12 cm}
\epsfig{file=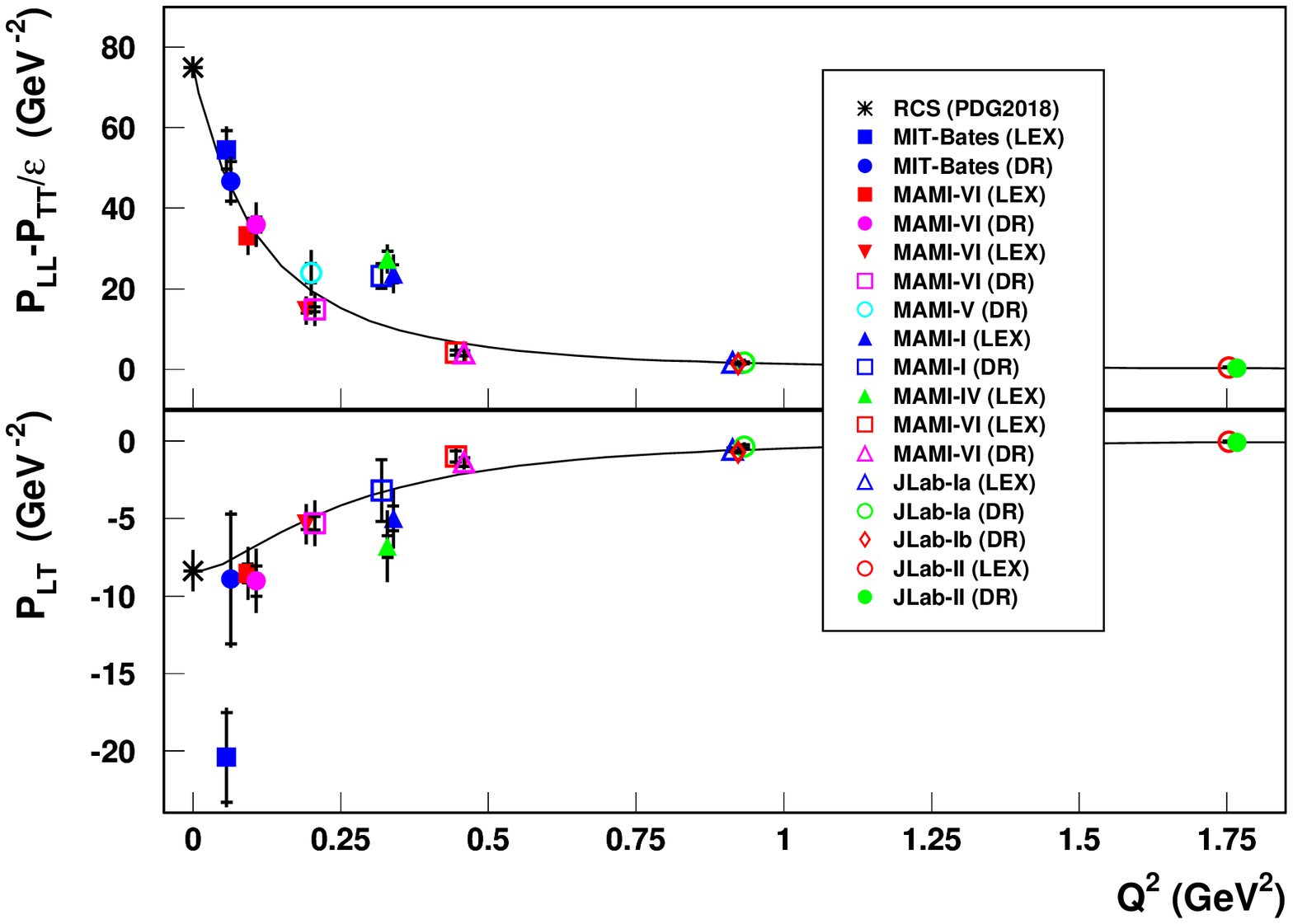,scale=0.7}
\end{minipage}
\begin{minipage}[t]{16.5 cm}
\caption{
(color online) World data for the structure functions $\pllptte$ and $\plt$ as a function of $Q^2$. The nomenclature follows the one of Table~\ref{tab-exp-3-sf}, including the RCS point~\cite{Tanabashi:2018oca}. Inner (resp. outer) error bars are statistical (resp. total). Some points are slightly shifted for visibility. The solid curve is the DR model calculation for ($\la = \lb = 0.7$ GeV) and $\epsilon = 0.65$~\cite{Pasquini:2000pk}. 
\label{fig-exp-sf-1}}
\end{minipage}
\end{center}
\end{figure}


Figure~\ref{fig-exp-sf-1} depicts the whole set of values of $\pllptte$ and $\plt$ reported in Table~\ref{tab-exp-3-sf}. Note that the structure functions $\pllptte$ and $\plt$ at $Q^2$ = 0 are simply proportional to the RCS polarizabilities $\ale$ and $\bem$ (cf. the Appendix).

Thanks to the most recent experiments, the $Q^2$-region below 0.5 GeV$^2$ in Fig.~\ref{fig-exp-sf-1} starts now to be densely filled, with precise data, and a rather  consistent $Q^2$-picture emerges. The vast majority of points agree well with a smooth behavior, in remarkably good agreement with a typical DR model calculation (solid curve), in which a single dipole shape is assumed for the unconstrained part of the scalar GPs (cf. Eq.~(\ref{asymp}))
\footnote{Note that the experimental points correspond to various values of $\epsilon$, while the DR curve is calculated at a fixed $\epsilon$ = 0.65. However the comparison between theory and experiment is not affected, since in the DR model the $\ptt$ structure function is very small (see Fig.~\ref{fig-theo-sf-2}). A DR curve calculated at $\epsilon$ = 0.9 would be almost exactly superimposed to the one of Fig.~\ref{fig-exp-sf-1}.}.  
We remind that the DR model does not by itself predict the electric and magnetic GPs, and that the DR curve in Fig.~\ref{fig-exp-sf-1} is obtained using $(\la, \lb)$ parameters fitted from experimental data. The dipole ansatz was introduced in the model only as a practical way to parametrize a shape in $Q^2$; any other shape could be used instead. Experimental fits are made at each $Q^2$ separately, and independently of any assumption on the global $Q^2$-dependence. Yet, in view of the present data, this dipole ansatz seems to be not too far from reality.


There are two exceptions to the overall smooth $Q^2$-behavior in Fig.~\ref{fig-exp-sf-1}. 
The first one is the MIT-Bates LEX point for $\plt$, lying at a large negative value. It is well understood in terms of a bias in the LEX fit~\cite{Bourgeois:2006js,Bourgeois:2011zz}, due to the competition of the lowest-order and higher-order GP terms of the low-energy expansion
\footnote{In the in-plane kinematics of the MIT-Bates experiment, at $\thcm=90^{\circ}$ and $\phicm=180^{\circ}$, the lowest-order GP term of the LEX was exceptionally small, due to a near-perfect cancellation between $\vll \cdot ( \pllptte )$ and $(\vlt \cdot \plt )$. The $\ohigher$ terms were then non-negligible, and ignoring them in the LEX fit was too approximate. This first evidence of a problem in a LEX fit was instructive for posterior designs, e.g., of the MAMI-VI experiment. }. 
The DR fit in this experiment is obviously better behaved than the LEX fit, and  gives a more sensible value of $\plt$. 
The second exception consists in the three MAMI points for $\pllptte$ at $Q^2=0.33$ GeV$^2$, all lying over the general trend. No experimental or analysis bias has been identified in the two experiments (MAMI-I and IV), and this enhancement remains presently unexplained.

\begin{figure}[tb]
\begin{center}
\begin{minipage}[t]{15 cm}
\epsfig{file=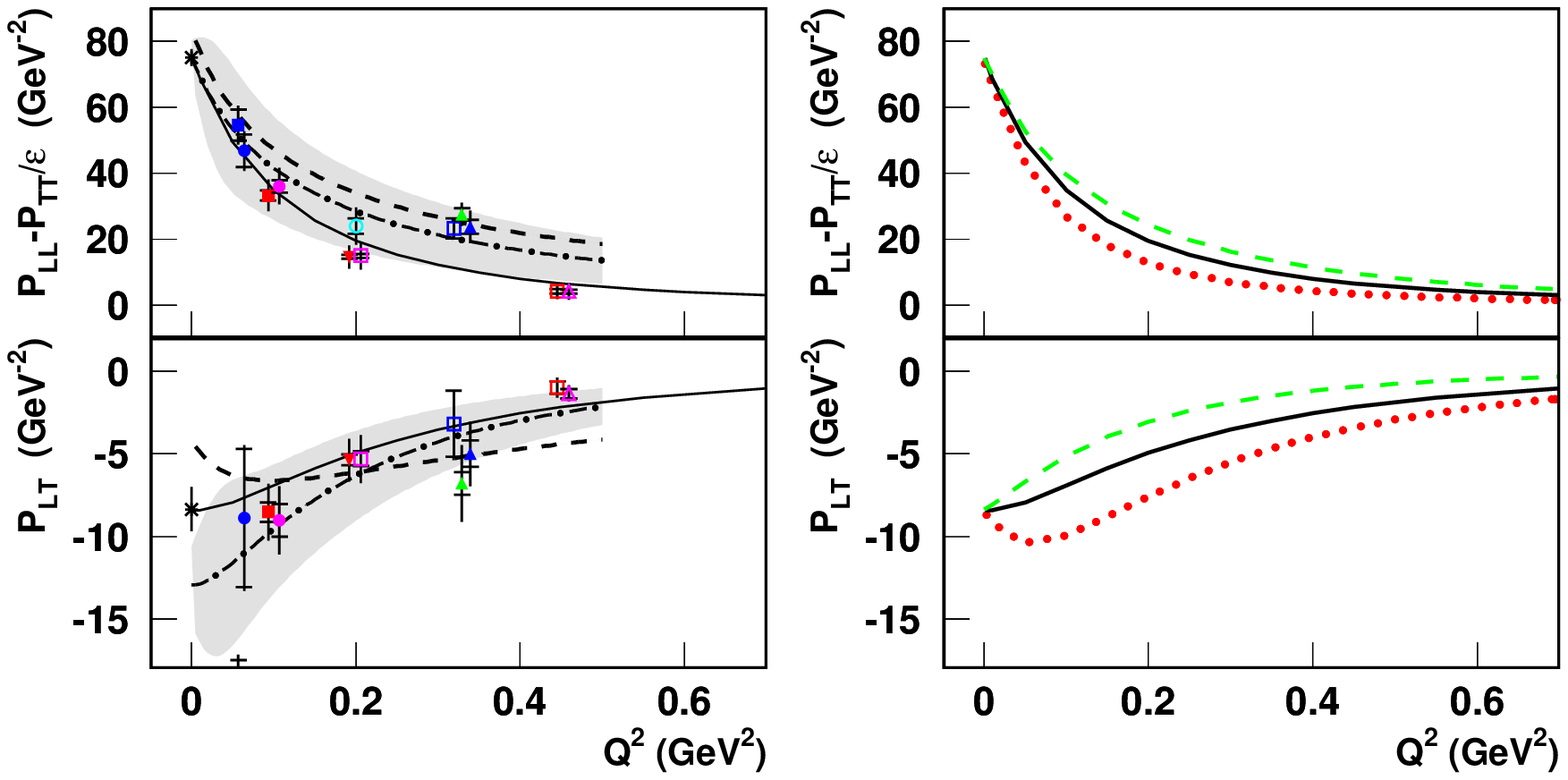,scale=0.8}
\end{minipage}
\begin{minipage}[t]{16.5 cm}
\caption{
(color online) Left plots: the structure functions $\pllptte$ and $\plt$ in the intermediate $Q^2$-range and three different model predictions. The experimental data follow the nomenclature of Fig.~\ref{fig-exp-sf-1}. The solid curve is the DR model calculation for ($\la = \lb = 0.7$ GeV). The dashed curve is the HBChPT ${\cal O}(p^3)$ calculation of Ref.~\cite{Hemmert:1999pz}. The dashed-dotted curve with its error band (shaded area) is from covariant BChPT at ${\cal O}(p^3)+{\cal O}(p^4/\bar\Delta)$~\cite{Lensky:2016nui}. Right plots: the DR model calculation~\cite{Pasquini:2000pk} for three different choices of the dipole mass parameter: $(\la = \lb)$ = 0.5 GeV (dotted curve), 0.7 GeV (solid curve), 0.9 GeV (dashed curve). Model calculations use $\epsilon =0.65$.
\label{fig-exp-sf-2}}
\end{minipage}
\end{center}
\end{figure}


The left part of Fig.~\ref{fig-exp-sf-2} shows three model calculations of $\pllptte$ and $\plt$ in the lower part of the $Q^2$-range, where ChPT is applicable. Historically, the HBChPT ${\cal O}(p^3)$ calculation of Ref.~\cite{Hemmert:1999pz} appeared as very successful in describing the VCS observables, due to its good agreement with the measurements from the MAMI-I and MIT-Bates experiments. In particular, the large values of  $\ptt$ and $\plt \, _{spin}$ predicted by this model, i.e., the spin-dependent part of the two extracted structure functions, were a key ingredient to reproduce the MAMI-I results
\footnote{To view the spin-dependent part of $\pllptte$ and $\plt$ as predicted by HBChPT ${\cal O}(p^3)$, we address the reader to Ref.~\cite{dHose:2006bos}, Fig.9. This contribution is the difference between the solid curve (scalar+spin) and the dashed curve (scalar only) in the right plots of the figure.}. 
The HBChPT calculation was then pushed one order further, but only for the spin GPs~\cite{Kao:2004us,Kao:2002cn}. The result showed a severe lack of convergence: see for instance the HBChPT ${\cal O}(p^3)$ and ${\cal O}(p^4)$ calculations of $\ptt$ in Fig.~\ref{fig-theo-sf-2}. In addition, such a ${\cal O}(p^4)$ calculation does not exist yet for the scalar GPs. Therefore it appears hard to draw any firm conclusion from the comparison of HBChPT with VCS experiments, and the agreement mentioned above with the MAMI-I results is possibly accidental.

\begin{figure}[tb]
\begin{center}
\begin{minipage}[t]{15 cm}
\epsfig{file=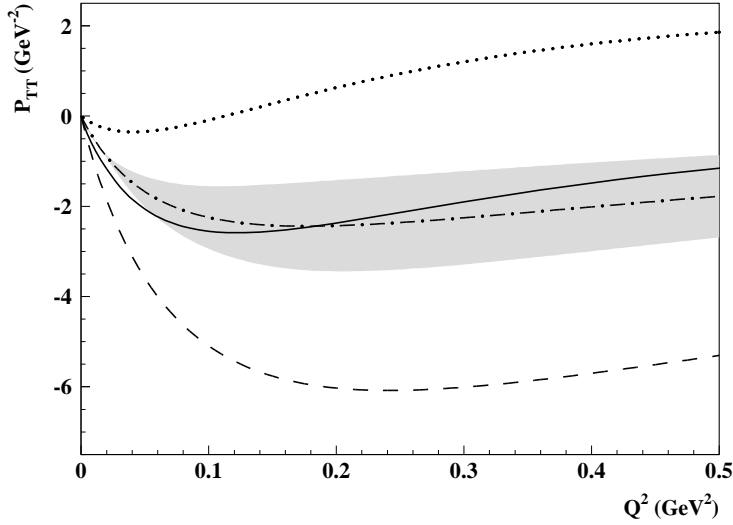,scale=0.4}
\end{minipage}
\begin{minipage}[t]{16.5 cm}
\caption{
Model predictions for the $\ptt$ structure function: HBChPT at ${\cal O}(p^3)$ \cite{Hemmert:1999pz} (dashed curve), HBChPT at ${\cal O}(p^4)$ \cite{Kao:2002cn} (dotted curve), Dispersion Relations~\cite{Pasquini:2000pk} (solid curve), and covariant BChPT  at ${\cal O}(p^3)+{\cal O}(p^4/\bar\Delta)$ \cite{Lensky:2016nui} (dashed-dotted curve, with its error band as a shaded area). 
\label{fig-theo-sf-2}}
\end{minipage}
\end{center}
\end{figure}

A new calculation of VCS observables is provided by the recently developed covariant BChPT~\cite{Lensky:2016nui}. The structure function $\pllptte$ calculated by this model is in better agreement with the data than HBChPT, although it still lies above the most recent experimental points (Fig.~\ref{fig-exp-sf-2}). The structure function $\plt$ from  covariant BChPT reproduces well the VCS data, although being at tension with the RCS data. The theoretical uncertainty  of this model (shaded area) is quite large, but hopefully it can be reduced in the future.

The right part of Fig.~\ref{fig-exp-sf-2} shows the sensitivity in the DR model when one changes the free parameters $(\la,\lb)$. We stress that, in the VCS analyses, these parameters have always been fitted independently for each experimental data set.  The fitted values have all been found in a remarkably narrow range, [0.5,0.8] GeV for $\la$ and $\lb$ (with the exception of  the data at $Q^2=0.33$ GeV$^2$). Throughout this article we have considered $(\la = \lb$ = 0.7 GeV) as an average reference, compatible with most of the data points.

\begin{figure}[tb]
\begin{center}
\begin{minipage}[t]{15 cm}
\epsfig{file=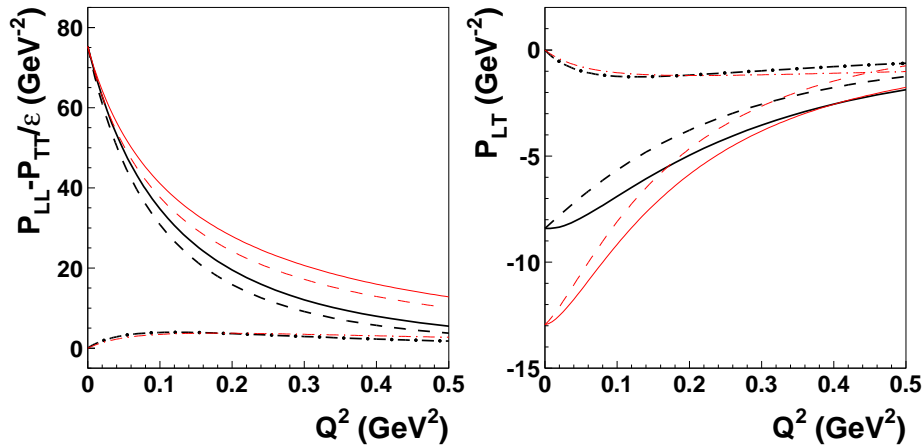,scale=0.32}
\end{minipage}
\begin{minipage}[t]{16.5 cm}
\caption{
(color online) The structure functions that are accessed by the experiments, in terms of their scalar and spin-dependent parts, for two models: DR  (thick black curves)~\cite{Pasquini:2000pk,Drechsel:2002ar} and covariant BChPT (thin red curves)  \cite{Lensky:2016nui}. Dashed curves are for the scalar part, dashed-dotted curves for the spin-dependent part, and solid curves for the sum. The DR calculation uses $(\la=\lb=0.7$ GeV). In the left plot one has $\epsilon=0.65$, so that the dashed-dotted curves correspond to $(- \ptt / 0.65)$.
\label{fig-theo-sf-3}}
\end{minipage}
\end{center}
\end{figure}

Finally, we show in Fig.~\ref{fig-theo-sf-3}  the scalar and spin-dependent parts of the measured structure functions, as calculated by two models: DRs and covariant BChPT
\footnote{We remind that in the DR model, spin GPs and their combinations are fully predicted.}. 
In these two calculations, the spin-dependent part (dashed-dotted curves) is very similar, and of small magnitude (in contrast to HBChPT ${\cal O}(p^3)$, not shown here). The dominance of the scalar part means that $\pllptte$ and $\plt$ give an almost direct picture of the electric and magnetic GPs, at least at low and intermediate $Q^2$. At higher $Q^2$ it is less and less true; in the DR model for instance, the spin part tends to contribute as much as the scalar part to the measured structure functions when $Q^2$ reaches 1 GeV$^2$ and above
\footnote{This can be seen, e.g., in Ref.~\cite{Fonvieille:2012cd}, Fig.20, zoomed inserts.}.

\subsection{Results for the scalar GPs}
\label{sec-experimental-results-gps}


Table~\ref{tab-exp-4-gp} collects the values of the proton scalar GPs extracted by the various experiments. These values are a direct output of the fit only in the case of DR analyses; in the case of LEX analyses, only $\pllptte$ and $\plt$ are fitted, and their spin-dependent part must be subtracted. To this aim we use the DR model, for several reasons: its good reliability, and its validity over the full $Q^2$-range, a feature that no other model provides. The obtained picture of the scalar GPs is thus model-dependent (but consistently), and would be different if one used another model for this subtraction.

We have reported four sets of polarizability values at $Q^2=0$ in Table~\ref{tab-exp-4-gp} in order to reflect the present state of knowledge of $\ale$ and $\bem$ in RCS. The error bar quoted for these polarizabilities by the PDG~\cite{Tanabashi:2018oca} is quite small and  may not reflect the actual spread between the various analyses~\cite{Schumacher:2005an,McGovern:2012ew,Lensky:2015awa,Pasquini:2019nnx}. An intense effort is ongoing experimentally and theoretically to pin down the RCS polarizabilities; see, e.g., Refs.~\cite{Hagelstein:2015egb,Pasquini:2018wbl,Pasquini:2019nnx,Krupina:2017pgr}.

\begin{table}[h]
\begin{center}
\begin{minipage}[t]{16.5 cm}
\caption{Experimental results for the electric and magnetic GPs of the proton using LEX and DR fits, ordered by increasing values of $Q^2$. The first error is statistical and the second one is systematic, except in the special cases in footonote. DR values are directly fitted; LEX values are obtained indirectly (see text). The first four lines are the RCS values of $\ale$ and $\bem$ deduced from various fits. }
\label{tab-exp-4-gp}
\end{minipage}
\begin{tabular}{|l|c|ccc|ccc|c|}
\hline
Experiment &  $Q^2$  & \multicolumn{3}{c|}{ $\aeq$ } & \multicolumn{3}{c|}{ $\bmq$  } & type of  \\
nomenclature &  (GeV$^2$)   &  \multicolumn{3}{c|}{ ($10^{-4}$ fm$^3$) } & \multicolumn{3}{c|}{ ($10^{-4}$ fm$^3$) } &  analysis \\ 
\hline
RCS 1$^{\rm a}$    &     0    &     12.1   & $\pm$ 0.3     &  {\scriptsize $\mp$0.4$\pm$0.3}  &             1.6   & $\pm$  0.4   & {\scriptsize $\pm$0.4$\pm$0.4}  & \cite{OlmosdeLeon:2001zn} \\
RCS 2$^{\rm a}$    &     0    &     10.65   & $\pm$ 0.35   & {\scriptsize $\pm$0.2$\pm$0.3}  &          3.15   & $\mp$  0.35   & {\scriptsize $\pm$0.2$\mp$0.3}  & \cite{McGovern:2012ew} \\
RCS 3$^{\rm b}$    &     0    &     12.03   & $^{+0.48} _{-0.54}$   &   &          1.77   &   $^{+0.52} _{-0.54}$   &  & \cite{Pasquini:2019nnx} \\
RCS 4$^{\rm b}$    &     0    &     11.2   & $\pm$ 0.4   &   &          2.5   & $\pm$  0.4   &  & \cite{Tanabashi:2018oca} \\
%
\hline
MIT-Bates$^{\rm d}$ &  0.057  &     9.22  & $\pm$ 0.85 &  $\pm$ 0.60 &         6.76 &  $\pm$  1.02  & $\pm$ 0.49     & LEX \cite{Bourgeois:2006js} \\
MIT-Bates &  0.057  &     7.85  & $\pm$ 0.87 &  $\pm$ 0.60 &         2.69 &  $\pm$  1.48  & $\pm$ 0.49     & DR \cite{Bourgeois:2006js} \\
\hline
MAMI-VI & 0.10   &     6.06  & $\pm$  0.30  & $\pm$  0.90 &          2.82  & $\pm$  0.23  & $\pm$ 0.63     & LEX \cite{Bericic:2019faq}  \\
MAMI-VI & 0.10   &     6.60  & $\pm$  0.36  & $\pm$  1.03 &          3.02  & $\pm$  0.38  & $\pm$ 0.72     & DR \cite{Bericic:2019faq}  \\
MAMI-VI & 0.20   &     3.02 & $\pm$   0.14  & $\pm$  0.87 &          2.01  & $\pm$  0.16  & $\pm$ 0.61     & LEX \cite{Bericic:2019faq}  \\
MAMI-VI & 0.20  &      3.11  & $\pm$  0.15  & $\pm$  1.02 &          1.98  & $\pm$  0.22  & $\pm$ 0.68     & DR \cite{Bericic:2019faq}  \\
MAMI-V$^{\rm c}$  & 0.20  & 5.3  & $\pm$  0.6  & $\pm$  1.3 &               &              &                & DR \cite{Blomberg:2019caf} \\
MAMI-I$^{\rm d,a}$ &  0.33  &      6.90  & $\pm$  0.72  & {\scriptsize $\pm$0.19$\pm$1.41} &  2.52  & $\mp$  0.51  & {\scriptsize $\mp$0.69$\mp$0.88} & LEX  \cite{Roche:2000ng} \\
MAMI-I$^{\rm d,e}$ &  0.33  &  6.73   &  $\pm$  0.82  &              &  1.39  &  $\mp$  1.26  &               & DR \cite{dHose:2006bos}  \\
MAMI-IV$^{\rm d,f}$ &  0.33  & 8.14   &  $\pm$  0.63  &  $\pm$  0.99  & 3.64  &  $\mp$  0.43  &  $\mp$  1.37  & LEX \cite{Janssens:2008qe}    \\
MAMI-VI & 0.45  &     0.92  & $\pm$   0.26  & $\pm$   0.92    &      0.19  & $\pm$   0.28  & $\pm$   0.38  & LEX \cite{Bericic:2019faq}  \\
MAMI-VI & 0.45  &     0.87  & $\pm$   0.25  & $\pm$  1.01  &         0.47  & $\pm$   0.22  & $\pm$   0.30   & DR \cite{Bericic:2019faq}  \\
\hline
JLab-Ia     & 0.92  &     1.09  & $\pm$   0.21  & $\pm$   0.60 &        0.42   & $\pm$  0.18  & $\pm$   0.26   & LEX  \cite{Laveissiere:2004nf} \\
JLab-Ia     & 0.92  &      1.02  & $\pm$  0.18  & $\pm$  0.77  &        0.13  & $\pm$  0.15  & $\pm$  0.42     & DR  \cite{Laveissiere:2004nf} \\
JLab-Ib$^{\rm c}$    & 0.92  &  0.85  & $\pm$  0.15  & $\pm$  0.16 &     0.66  & $\pm$  0.11  & $\pm$  0.07      & DR  \cite{Laveissiere:2004nf} \\
JLab-II     & 1.76  &      0.82  & $\pm$  0.20  & $\pm$  0.44  &      -0.06  & $\pm$  0.17  & $\pm$  0.20     & LEX  \cite{Laveissiere:2004nf} \\
JLab-II     & 1.76  &       0.52  & $\pm$  0.12  & $\pm$  0.35  &      0.10  & $\pm$  0.07  & $\pm$  0.12  & DR  \cite{Laveissiere:2004nf} \\
\hline
\end{tabular}
\begin{minipage}[t]{16.5 cm}
\vskip 0.5cm
\noindent
%
%
$^{\rm a}$ Two separate systematic errors. \\
$^{\rm b}$ Total error only.  \\
$^{\rm c}$ DR fit performed in the resonance region. \\
$^{\rm d}$ These values are unpublished. We have deduced them from the structure functions of Table~\ref{tab-exp-3-sf} using the DR formalism. \\
$^{\rm e}$ Statistical error only. \\
$^{\rm f}$ Values for one selected choice of proton form factors (parametrization of Ref.~\cite{Hohler:1976ax}). \\
\end{minipage}
\end{center} 
\end{table}

\begin{figure}[tb]
\begin{center}
\begin{minipage}[t]{12 cm}
\epsfig{file=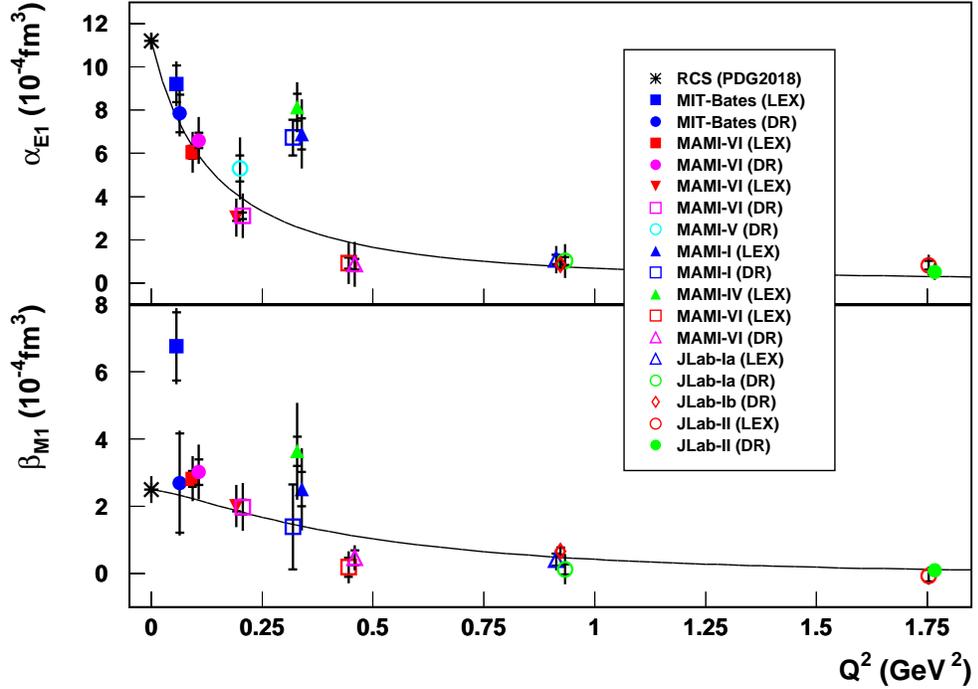,scale=0.7}
\end{minipage}
\begin{minipage}[t]{16.5 cm}
\caption{
(color online) World data for the electric and magnetic GPs of the proton as a function of $Q^2$. The nomenclature follows the one of Table~\ref{tab-exp-4-gp}, including the RCS point~\cite{Tanabashi:2018oca}. Inner (resp. outer) error bars are statistical (resp. total). Some points are slightly shifted for visibility. The solid curve is the DR model calculation for ($\la = \lb = 0.7$ GeV).
\label{fig-exp-ab-1}}
\end{minipage}
\end{center}
\end{figure}


Figure~\ref{fig-exp-ab-1} displays the data of Table~\ref{tab-exp-4-gp} for $\aeq$ and $\bmq$. Apart from a change of sign when going from $\plt$ to the magnetic GP,  Figs.~\ref{fig-exp-sf-1} and \ref{fig-exp-ab-1} show great similarities in shape. This is due to the smallness of the spin-dependent part in $\pllptte$ and $\plt$, especially in the DR model (cf. Fig.~\ref{fig-theo-sf-3}).

Over the explored $Q^2$-range, there is again a good overall agreement of the experimental data of Fig.~\ref{fig-exp-ab-1} with a smooth fall-off, as described by the DR model calculation (solid curve) already shown in Fig.~\ref{fig-exp-sf-1}
\footnote{This DR curve goes by construction through the RCS point of the PDG2018~\cite{Tanabashi:2018oca}.}. 
One observes again the exception of the MAMI-I and MAMI-IV data points at $Q^2=0.33$ GeV$^2$, mostly for the electric GP. This enhancement of the data for the electric GP is more pronounced than the corresponding one for $\pllptte$; it simply originates from the presence of the proton electric form factor in $\pll$
\footnote{We remind that $\pll = {\mathrm constant} \cdot \aeq \cdot G_E^p(Q^2)$. If $\aeq$ would fall off like a dipole, then $\pll$ would fall off like a double dipole.}. 
For the  MAMI-I experiment, the DR fit~\cite{dHose:2006bos}, which is a priori the most reliable fit, gives a value of $\bmq$ in smooth agreement with the general trend, but still gives a high value of $\aeq$, that is also confirmed by the MAMI-IV experiment. In the absence of explanation on the experimental side, this localized enhancement of the electric GP has effectively to be considered as the signal  of a physics mechanism not yet understood. In Ref.~\cite{Gorchtein:2009qq}, a new parametrization of the unknown asymptotic contribution to the DR calculation of $\alpha_{E1}(Q^2)$   has indeed been proposed to take into account this local ``bump''.

More measurements would be necessary in the region of  $Q^2=0.33$ GeV$^2$ to investigate further this puzzling behavior. The new JLab VCS experiment E12-15-001~\cite{Sparveris:2016} is presently exploring the intermediate $Q^2$-range of  0.3-0.7 GeV$^2$ (see Sect.~\ref{sec-e12-15-001-experiment}) and  will bring crucial elements of answer. 
If confirmed, a non-smooth $Q^2$-dependence of $\pllptte$ will call for really unusual explanations. As a side remark, one may wonder if such an anomaly (if real) can have its origin either in $\pll$ or in $\ptt$. From a formal point of view, both origins are possible, since the two structure functions are not measured separately (see Sect.~\ref{sec-access-spin-gps} about perspectives for $\ptt$ measurements). However, from a physical point of view, $\ptt$ is an unlikely candidate. Indeed, such a local variation in $Q^2$ would correspond to a long-range structure of the nucleon (due to Fourier transform properties). The simplest contribution, i.e.  the single-$\pi^0$ exchange in the $t$-channel,  is excluded since it does not contribute to $\pll$ nor to $\ptt$  (cf. Sect.~\ref{sec-the-dr-model}). The next contribution would be a two-pion $(J = I = 0)$ state, i.e. an effective $\sigma$-meson, which would point to $\pll$ due to its scalar nature.


The smallness of $\bmq$  w.r.t. to $\aeq$ makes it difficult to determine the magnetic GP with a good precision (in relative value); this is illustrated, e.g., by the spread of the three data points a $Q^2=0.33$ GeV$^2$ in Fig.~\ref{fig-exp-ab-1}. Nevertheless a consistent behavior tends to emerge from the world data, namely thanks to the precise data points from the MAMI-VI experiment. The low-$Q^2$ measurements suggest the existence of an extremum of $\bmq$, weakly pronounced, in the region near $Q^2=0.1$ GeV$^2$; but the exact shape depends crucially on the actual RCS value, which is under debate{~\cite{Hagelstein:2015egb,Pasquini:2018wbl,Pasquini:2019nnx,Krupina:2017pgr}. The DR model as shown in Fig.~\ref{fig-exp-ab-1} accounts rather well for the measurements of the magnetic GP over the full $Q^2$-range, with just a single dipole ansatz parametrizing the unconstrained part of $\bmq$.

\begin{figure}[tb]
\begin{center}
\begin{minipage}[t]{15 cm}
\epsfig{file=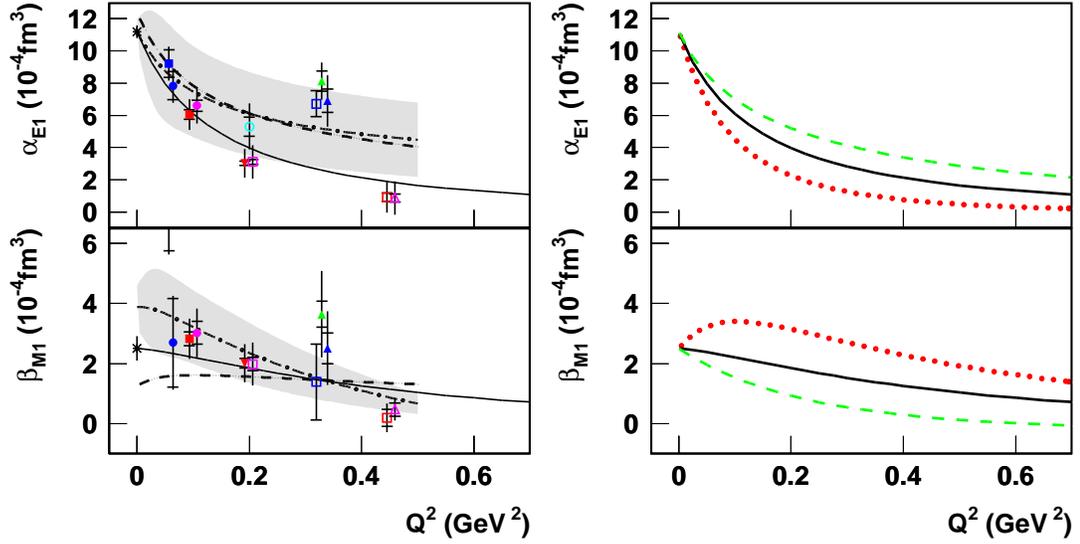,scale=0.8}
\end{minipage}
\begin{minipage}[t]{16.5 cm}
\caption{
(color online) Left plots: the electric and magnetic GPs  in the intermediate $Q^2$-range and three different model predictions. The experimental data follow the nomenclature of Fig.~\ref{fig-exp-ab-1}. The solid curve is the DR model calculation for ($\la = \lb = 0.7$ GeV). The dashed curve is the HBChPT ${\cal O}(p^3)$ calculation of Ref.~\cite{Hemmert:1999pz}. The dashed-dotted curve with its error band (shaded area) is from covariant BChPT~\cite{Lensky:2016nui}. Right plots: the DR model calculation for three different choices of the dipole mass parameter: $(\la = \lb)$ = 0.5 GeV  (dotted curve), 0.7 GeV (solid curve), 0.9 GeV (dashed curve). 
\label{fig-exp-ab-2}}
\end{minipage}
\end{center}
\end{figure}


The left plots of Fig.~\ref{fig-exp-ab-2} show three model calculations of $\aeq$ and $\bmq$, in complete analogy to~Fig.~\ref{fig-exp-sf-2}. As already mentioned in Sect.~\ref{sec-the-dr-model}, the HBChPT~\cite{Hemmert:1999pz} and covariant BChPT~\cite{Lensky:2016nui} calculations are in close agreement for the electric GP. However, due to their different calculation of $\ptt$ (cf. Fig.~\ref{fig-theo-sf-2}), the two models differ visibly for  $\pllptte$ (cf. Fig.~\ref{fig-exp-sf-2}). 
Secondly, in the right plots of Fig.~\ref{fig-exp-ab-2} we note that the presence of an extremum of the magnetic GP at low $Q^2$ depends on the value of $\lb$ in the DR model. For example, $\lb= $ 0.5 GeV generates an extremum, but the DR curve at this $\lb$ overshoots most of the data points at higher $Q^2$ ($\ge$ 0.4 GeV$^2$). 
This suggests that, at least for $\bmq$, the single-dipole ansatz could be replaced by another $Q^2$-parametrization, which would better describe the data. The present DR description of the scalar GPs is nevertheless quite satisfactory. The delicate balance between the large diamagnetic and paramagnetic parts of $\bmq$ (cf. Fig.~\ref{fig-theo-ab-1}) can already be well-tuned in the model. In addition, the description of $\aeq$, with an almost completely dominant [asymptotic + beyond $\pi N$] component behaving as a pure dipole, is able to reproduce most of the data.

\subsection{Mean-square polarizability radii}
\label{sec-mean-square-radii}

Mean-square radii are a basic measure of the extension of spatial distributions. Similarly to form factors, the mean-square radius of $\aeq$ and $\bmq$ is obtained from the slope of the electric and magnetic GPs at $Q^2=0$. One has for instance:
\bea
\langle r _{\ale}^2 \rangle \ = \  { -6  \over  \ale (0) } \ \cdot \ 
{ d \over dQ^2 } \aeq {\bigg \vert} _{Q^2=0} \ . 
\label{eq-radii}
\eea
These radii can be determined using a DR fit to the very-low $Q^2$ data in VCS. Such a work was presented in Ref.~\cite{Bourgeois:2011zz}, based on the two experimental data points: RCS and the MIT-Bates measurement at $Q^2=0.057$ GeV$^2$. With the addition of the MAMI-VI measurements at $Q^2=0.1$ GeV$^2$ and the new RCS values of Ref.~\cite{Tanabashi:2018oca}, it is appropriate to give here an update of these mean-square polarizability radii. We use the same method as described in Ref.~\cite{Bourgeois:2011zz}, and the  new results are reported in Table~\ref{tab-radii-2}, for the full DR calculation and for the separate $\pi N$ and asymptotic contributions. 

The error bars for the total results and for the asymptotic contributions take into account the uncertainties from the RCS experimental value (in the denominator of Eq.~(\ref{eq-radii})) and from the fit of the $\la$ and $\lb$ parameters to the  VCS data points. On the other side, the $\pi N$ contribution is fully determined by the dispersion integrals and the corresponding error bar reflects only the uncertainties on the RCS polarizabilities. The total results for  $\ale$ are consistent, within the error bars,  with the first determination of~\cite{Bourgeois:2011zz}. However, we find a more pronounced contribution from the  $\pi N$ channel w.r.t. the asymptotic term. 
Furthermore, the mean-square electric polarizability radius is much larger than the mean-square charge  radius (which is about $0.77$ fm$^2$), showing the effect of the deformation of the meson cloud of the proton under the influence  of an external electric field. The results for the mean-square magnetic radius are much better constrained than in the determination of~\cite{Bourgeois:2011zz},  thanks to the recent low-$Q^2$ measurements at MAMI. The  total result comes from a delicate cancellation between a large negative asymptotic (diamagnetic) contribution (equal to $-3.9$ fm$^2$) and a large positive $\pi N$ (paramagnetic) contribution (equal to 2.7 fm$^2$), with the dominance of the diamagnetic term associated with the long-distance effects of the pion cloud.

\begin{table}[h]
\begin{center}
\begin{minipage}[t]{16.5 cm}
\caption{
The mean-square radii of  the electric and magnetic polarizabilities of the proton, with their total error bar. The determination is made for the full DR calculation as well as for the separate components, $\pi N$ and  asymptotic.
}
\label{tab-radii-2}
\end{minipage}
\begin{tabular}{|r|rl|rl|}
\hline
\  &  \multicolumn{2}{|c|}{$\rsqaz$ \ (fm$^2$)}  & \multicolumn{2}{|c|}{$\rsqbz$ \ (fm$^2$)} \\
\hline 
\ & \ & \ & \ & \\
Full Dispersion  & $1.70$  & $\ ^{+0.33}_{-0.24}$   & $-1.24$  & $\ ^{+1.38}_{-1.86}$   \\
\ & \ & \ & \ & \\
Asymptotic       & $0.60$  & $\ ^{+0.32}_{-0.26}$   & $-3.91$  &  $\ ^{+1.47}_{-2.00}$   \\
\ & \ & \ & \ & \\
$\pi N$          & $1.10$  & $\ ^{+0.04}_{-0.04}$   & $ 2.67$  & $\ ^{+0.51}_{-0.37}$   \\
\ & \ & \ & \ & \\
\hline
\end{tabular}
\end{center} 
\end{table}

\subsection{Further comments on the $\ohigher$ term of the LEX}
\label{sec-further-comments-on-the-higher-order-term-of-the-lex}

We have seen in Sect.~\ref{sec-lex-fits} that the DR model provides a way to estimate the higher-order terms of the LEX expansion. Actually the low-energy theorem of Ref.~\cite{Guichon:1995pu} is not an expansion in $\qpr$ but an expansion in ($\qpr$/$\qcm$)
\footnote{From P. Guichon, private communication. It can also be seen from the detailed LEX expression in, e.g., Ref.~\cite{Guichon:1998xv}.}. 
Therefore when $\qcm$ decreases, or equivalently when $Q^2$ decreases, one may wonder how the validity of the LEX truncation evolves, and if the higher-order terms of the LEX become more important. The DR estimator introduced in Sect.~\ref{sec-lex-fits} allows us to study this question. We have built this quantity, i.e., the  higher-order terms $\ohigher$ as given by the DR model, divided by  $\sigbhb$, for various experimental conditions. Fig.~\ref{fig-ohigherdr-allexp} displays the result for seven different experiments, at the highest measured values of $\qpr$ below the pion production threshold (which are always around 100 MeV/c), in the 2D-plane $(\cthcm, \phicm )$. The DR estimator was evaluated using the fitted values of the structure functions in each case. The top plots show indeed the anticipated general trend: when $Q^2$ decreases, from right to left, the quantity $\ohigherdr$  tends to reach high values in wider regions in $(\cthcm, \phicm$)
\footnote{The angular variations of $\ohigherdr$ in the figure seem to be quite different from  right (high $Q^2$) to left (low $Q^2$),  but some similarities in pattern are hidden by the choice of a unique color map scale.}. 
The bottom plots show the angular phase-space where this estimator remains small ($<3$\%) compared to a typical first-order GP effect of 10-15\%. This region, displayed as a filled area, is the one where the LEX truncation to first order is a priori most reliable. These plots illustrate how the choice of angular kinematics can potentially impact the LEX fit. Namely, the filled area shrinks when $Q^2$ decreases, pointing to difficulties to do a proper LEX fit at very low  $Q^2$.

By looking at the points where cross sections have been measured (open black circles in the bottom plots) one sees that the various experiments made quite different choices, depending on the adopted strategy and the possibilities offered by the apparatus. Of course the DR estimator presented here should not be taken too strictly. However, as said in Sect.~\ref{sec-lex-fits}, appyling a selection criterion based on the $\ohigherdr$ quantity can be seen as a valuable tentative to improve the reliability of the LEX fit, at the price of introducing  a slight (DR-)model dependence.

\begin{figure}[tb]
\begin{center}
\begin{minipage}[t]{16.5 cm}
\epsfig{file=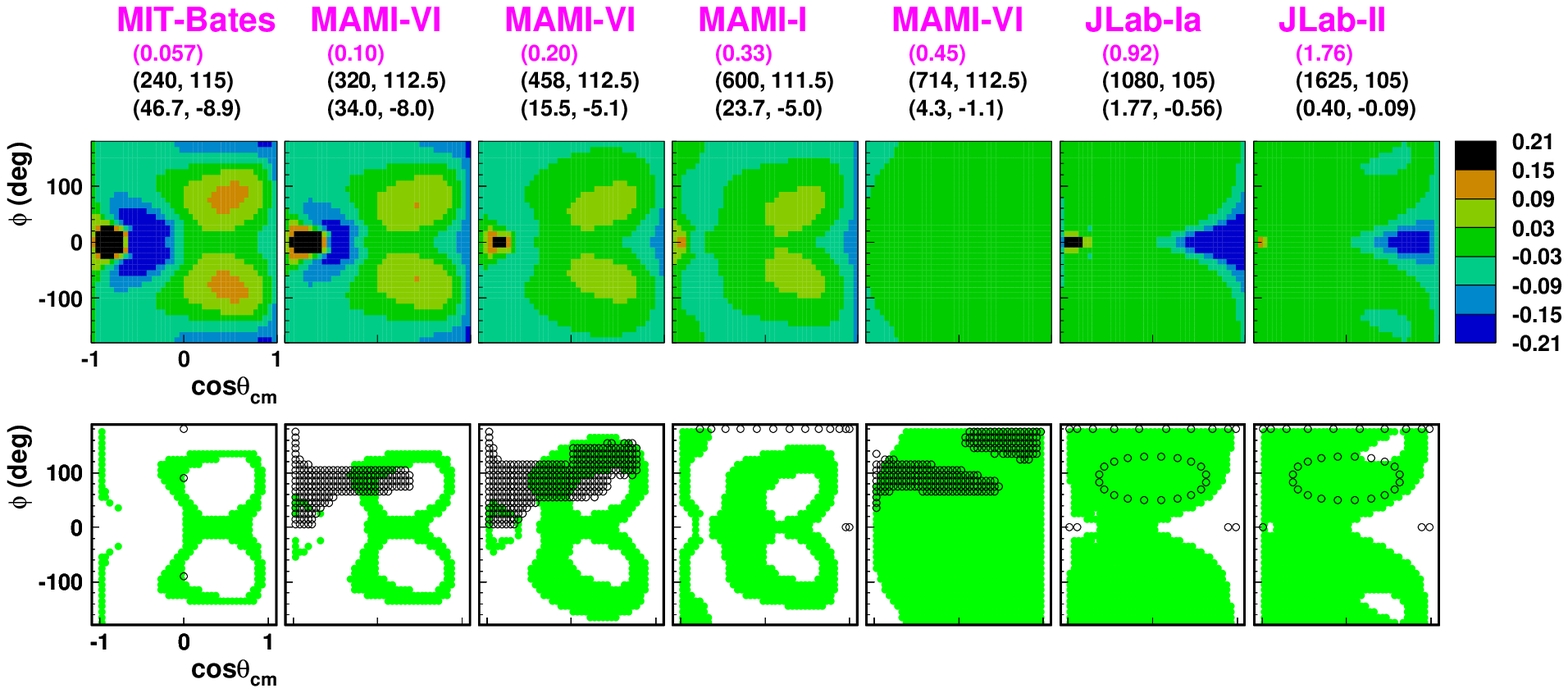,scale=0.85}
\end{minipage}
\begin{minipage}[t]{16.5 cm}
\caption{ 
(color online) The higher-order estimator $\ohigherdr = ( \sigdr - \siglex ) / \sigbhb $   (see text) for various VCS experiments, ordered by increasing $Q^2$ from left to right, and represented in the 2D plane ($\cthcm, \phi$). The calculation is made at the kinematics of the experiments, indicated by the three lines of numbers in parenthesis: $Q^2$ in GeV$^2$ (first line),  $\qcm$ and $\qpr$ in MeV/c (second line), and input values for the structure functions $\pllptte$ and $\plt$, in GeV$^{-2}$ (third line). The top plots are filled with the values of $\ohigherdr$ (truncated to the range [-21,+21]\%). In the bottom plots, the filled area (green online) is the region where $\vert \ohigherdr \vert $ is smaller than 3\%. The open circles (black online) in the bottom plots show where cross sections have been measured in each experiment.
\label{fig-ohigherdr-allexp}}
\end{minipage}
\end{center}
\end{figure}

\subsection{Other experimental results}
\label{sec-other-experimental-results}

The spectrum of low-energy VCS observables is wider than just the two structure functions $\pllptte$ and $\plt$ and the scalar GPs of the proton. Some VCS experiments have explored other observables, albeit in a less extensive way. Their results are summarized in this section.

\subsubsection{Beam single-spin asymmetry}
\label{sec-single-spin-asymmetry}


The beam single-spin asymmetry (beam SSA) in  VCS was first introduced in Ref.~\cite{Kroll:1995pv}, with a focus on the hard scattering regime. The main physics interest was to access non-trivial phases of QCD and to test the diquark model predictions.

The observable is the asymmetry $(d\sigma^+ - d\sigma^- ) / (d\sigma^+ + d\sigma^- )$, where  $d\sigma^+$ and $d\sigma^-$ are the photon electroproduction cross-sections with a longitudinally polarized electron beam of helicity $ + \frac{1}{2}$ and $ - \frac{1}{2}$. The numerator is equal to  $ {\cal I}m (T^{\mathrm{VCS}}) \cdot  {\cal R}e (T^{\mathrm{VCS}} + T^{\mathrm{BH}})$ and indicates that the beam SSA is proportional to the imaginary part of the VCS amplitude. Therefore  one must go above the pion production threshold to access this asymmetry. The first term,  ${\cal I}m (T^{\mathrm{VCS}}) \cdot  {\cal R}e (T^{\mathrm{VCS}})$, is purely due to VCS and measures the relative phase between longitudinal and transverse virtual Compton helicity amplitudes. The second term, ${\cal I}m (T^{\mathrm{VCS}}) \cdot  {\cal R}e (T^{\mathrm{BH}})$, is an interference term that measures the relative phases between the VCS and the BH amplitudes. In kinematics where BH dominates, this interference plays the role of an amplificator of the VCS contribution and enhances the asymmetry. Both terms of the numerator vanish at $\phicm=0^\circ$ and $180 ^{\circ}$, so one must go out-of-plane to access this asymmetry.


The  MAMI-II experiment~\cite{Bensafa:2006wr,BensafaPhD:2006} measured the beam SSA in the first resonance region, at  $W=1.2$ GeV, $Q^2=0.35$ GeV$^2$, $\phicm=220^{\circ}$ and $\thcm < 35^{\circ}$. Here the physics goal was different, and focusing on testing the input of the DR model, i.e., the calculation of ${\cal I}m (T^{\mathrm{VCS}})$ entering the DR integrals. 
The beam SSA resulting from the measurement was of small magnitude (below 10\%) and rather limited precision. These data showed an overall good agreement with the DR calculation, which used the $(\gamma^{(*)} N \to \pi N)$ multipoles of the MAID2003 analysis. The main finding was that, in the MAMI-II kinematics, the DR calculation had little sensitivity to the GPs, but had a good sensitivity to the two small longitudinal multipoles $S_{1+}$ and $S_{0+}$ in the $(p \pi^0)$ channel, i.e., involved in the  $(p \pi^0)$ intermediate state of the VCS process. 
Another finding was that the Beam SSA in the pion electroproduction reaction $( \eppionreact )$, that was measured simultaneously to $( \epgreact )$, provided supplementary constraints for possible adjustments of these small multipoles. Indeed in these measurements the two channels are coupled, since ${\cal I}m (T^{\mathrm{VCS}})$ is connected to the  $\pi N$ multipoles by unitarity.

\subsubsection{Double-spin asymmetry}
\label{sec-double-spin-asymmetry}


The case of doubly polarized VCS has been first studied theoretically in Ref.~\cite{Vanderhaeghen:1997bx}, and formulated in its final shape in Ref.~\cite{Guichon:1998xv} after downcutting the number of independent GPs from ten to six~\cite{Drechsel:1997xv,Drechsel:1998zm}. The observables, doubly polarized cross sections or asymmetries, need polarization on both the leptonic and hadronic sides. For the $(\vec e p \to e \vec p \gamma)$  process corresponding to a longitudinally polarized beam and the measurement of the final proton polarization, the double-spin asymmetry has the expression~\cite{JanssensPhD:2007}:
%
%
%
\bea
\begin{array}{lllll}
\picm & =  & 
{ \displaystyle \sigpol \, _{h=+1/2, s'_i \uparrow} \, -  \,  \sigpol \, _{h=+1/2, s'_i \downarrow}
\over
\displaystyle \sigpol \, _{h=+1/2, s'_i \uparrow} \,  +  \, \sigpol \, _{h=+1/2, s'_i \downarrow}
}  \ = \ 
{ \displaystyle \Delta \sigpol _i
\over 
\displaystyle 2 \sigpol _{unpol.}
} \ , 
\label{dsaformula-1}
\end{array}
\eea
%
%
%
where $s'_i$ is the projection of the final proton spin along the direction  $i=x,y$ or $z$ in the c.m. (cf. Fig.~\ref{fig-kinematics-1}), $h$ is the beam helicity and $\sigpol$ is a doubly polarized cross section. 
In contrast to the beam single-spin asymmetry, the double-spin asymmetry does not vanish below the pion production threshold. In this range of $W$, and in analogy with the unpolarized case, a low-energy theorem has been established for the polarized cross section difference $\Delta \sigpol _i$: 
%
%
%
\bea
 \Delta \sigpol _i =  \Delta \sigpol _{i, \mathrm{BH+Born}} + (\Phi \cdot \qpr ) \cdot \Delta {\cal M}  _{0i, \mathrm{NB}} + \ohigher \ .
\label{dsaformula-2}
\eea
%
%
%
In Eq.~(\ref{dsaformula-2}), $ \Delta \sigpol _{i, \mathrm{BH+Born}}$ contains no GPs and is entirely calculable. The first-order polarizability term $\Delta {\cal M}  _{0i, \mathrm{NB}}$ contains new combinations of the six lowest-order GPs, under the form of the structure functions $\pltz, \pltzprim$ and  $\pltperpprim$ (see the complete formulas in the Appendix). Together with the three structure functions of the unpolarized case $( \pll, \ptt$ and $\plt )$, they form a set of six independent structure functions, that is equivalent to the set of the six independent GPs. Therefore, by measuring the three proton polarization components $\picm  (i=x,y,z) $, this formalism opens up the possibility to disentangle all the lowest-order GPs, a perspective that looks of course very attractive.

For convenience, three more structure functions are introduced: $\pltperp , \pttperp$ and $\pttperpprim$, because they appear in the expression of $\pxcm$ and $\pycm$. They are simply linear combinations of the other structure functions (see the Appendix). Model calculations~\cite{Vanderhaeghen:1997bx} predict large double-spin asymmetries in typical MAMI kinematics, where the dominant contribution comes from the (BH+Born) process and is modulated by a few-percent effect coming from the GPs.


Double polarization observables  were explored in the one-and-only MAMI-IV experiment, at kinematics essentially similar to the ones of MAMI-I. The beam was longitudinally polarized and a focal-plane polarimeter (FPP) was used to measure the recoil proton transverse polarization components $( P_x^{fp} , P_y^{fp} )$. High statistics were needed, because one had to cut away the majority of protons, which scattered at too low angle ($< 9^{\circ})$ in the carbon analyzer of the FPP. 
A first analysis step consisted in fitting the c.m. polarizations $\picm \, (i=x,y,z)$ to the azimuthal distribution of events in the FPP. This analysis showed that  only $\pxcm$ and $\pycm$ could be adjusted, and that  $\pzcm$ had to be fixed to its theoretical (BH+Born) value.  The $\pycm$ component was very small and almost all the new information was carried only by $\pxcm$. 
A second step consisted in fitting individual GPs to the same FPP  distribution as above. The fit utilized an unbinned likelihood method, in which the quantities   $\Delta {\cal M}  _{0i, \mathrm{NB}} \, (i=x,y)$ are replaced by their analytical content in terms of  GPs. Unfortunately this fit was inconclusive. 
As a third step, a more conclusive fit was achieved when the quantities   $\Delta {\cal M}  _{0i, \mathrm{NB}} \, (i=x,y)$ were replaced by their analytical content in terms of the structure functions. By fixing $ \pttperp$, $\pttperpprim$ and $\pltperpprim$ to model predictions, the fit yielded the structure function $\pltperp$. The extracted value: $\pltperp = ( -15.4 \pm 3.3_{stat} \, ^{+1.5} _{-2.4} \, _{syst} )$ GeV$^{-2}$ is larger in magnitude than most model predictions. The polarizability effect in the data can be vizualized in Fig.~\ref{fig-pxcm} by the central solid curve, as compared to the BH+Born curve.

The challenges of this experiment and the complexity of the analysis were clearly one step higher than in an unpolarized experiment. The measured double-spin asymmetry turned out to be less sensitive than expected to the GPs, and left the disentangling of the six lowest-order GPs as a far-reaching goal. Improved measurements of that kind would require more data over a wider range of the kinematical variables.

\begin{figure}[tb]
\begin{center}
\begin{minipage}[t]{13.5 cm}
\epsfig{file=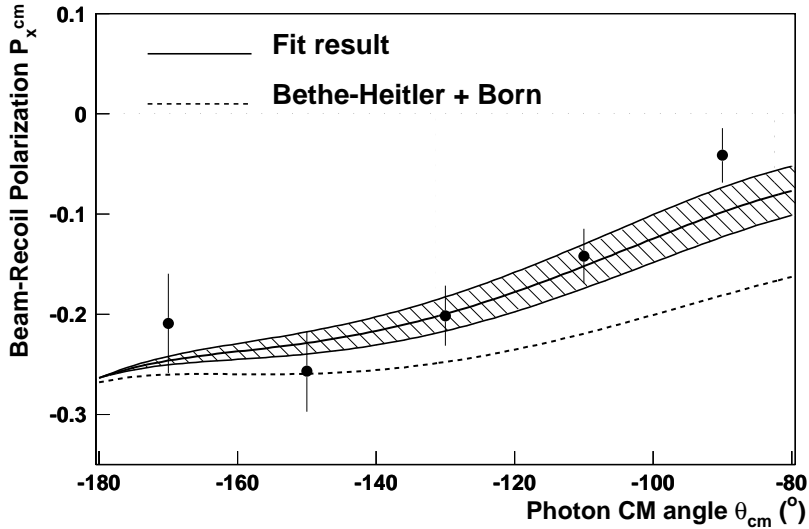,scale=0.6}
\end{minipage}
\begin{minipage}[t]{16.5 cm}
\caption{ 
The $\pxcm$ component of the recoil proton polarization as measured in the MAMI-IV experiment. The five points with their statistical error bar are the result of the first-step fit. The solid curve is the calculation of $\pxcm$ using the measured value of $\pltperp$, the shaded band representing the statistical uncertainty. The dashed curve shows the BH+Born calculation of $\pxcm$, i.e., without any GP effect. Figure taken from Ref.~\cite{Doria:2015dyx}.
\label{fig-pxcm}}
\end{minipage}
\end{center}
\end{figure}

\subsubsection{$N \to \Delta$  multipoles}
\label{sec-N-to-delta-multipoles}

\begin{figure}[tb]
\begin{center}
\begin{minipage}[t]{15 cm}
\epsfig{file=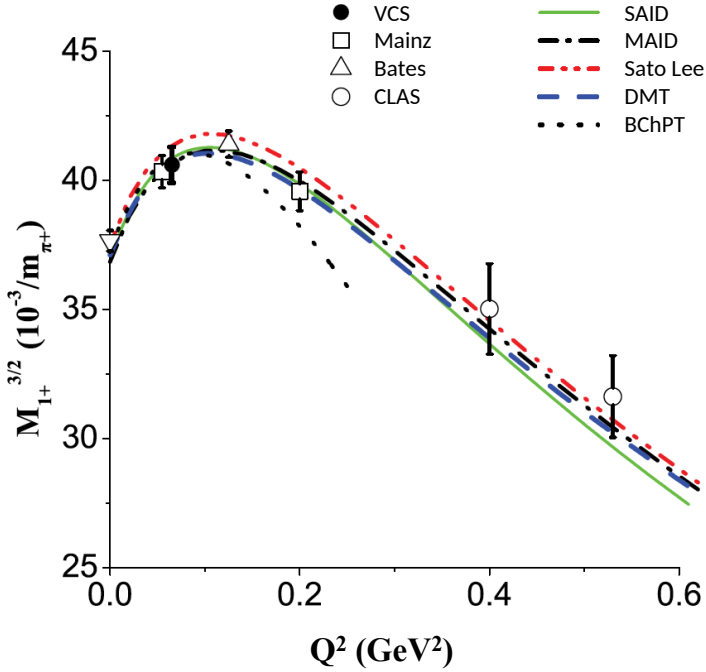,scale=0.7}
\end{minipage}
\begin{minipage}[t]{16.5 cm}
\caption{
(color online) The VCS result for the $M^{3/2}_{1+}$ (solid circle) from the MAMI-III experiment as compared to the world data from the measurements of the pion electroproduction channel (open symbols). The two results at $Q^2 = 0.06$ GeV$^2$ have been shifted by 0.005 GeV$^2$ to be distinguishable. The theoretical predictions of MAID~\cite{Drechsel:1998hk}, DMT~\cite{Kamalov:1999hs,Kamalov:2001qg},  SAID~\cite{Arndt:2002xv}, Sato and Lee~\cite{Sato:2000jf}, and   BChPT of Pascalutsa and Vanderhaeghen~\cite{Pascalutsa:2005ts,Pascalutsa:2005vq} are also shown. Figure taken from Ref.~\cite{Sparveris:2008jx}.
\label{f4n}}
\end{minipage}
\end{center}
\end{figure}

\begin{figure}[tb]
\begin{center}
\begin{minipage}[t]{15 cm}
\epsfig{file=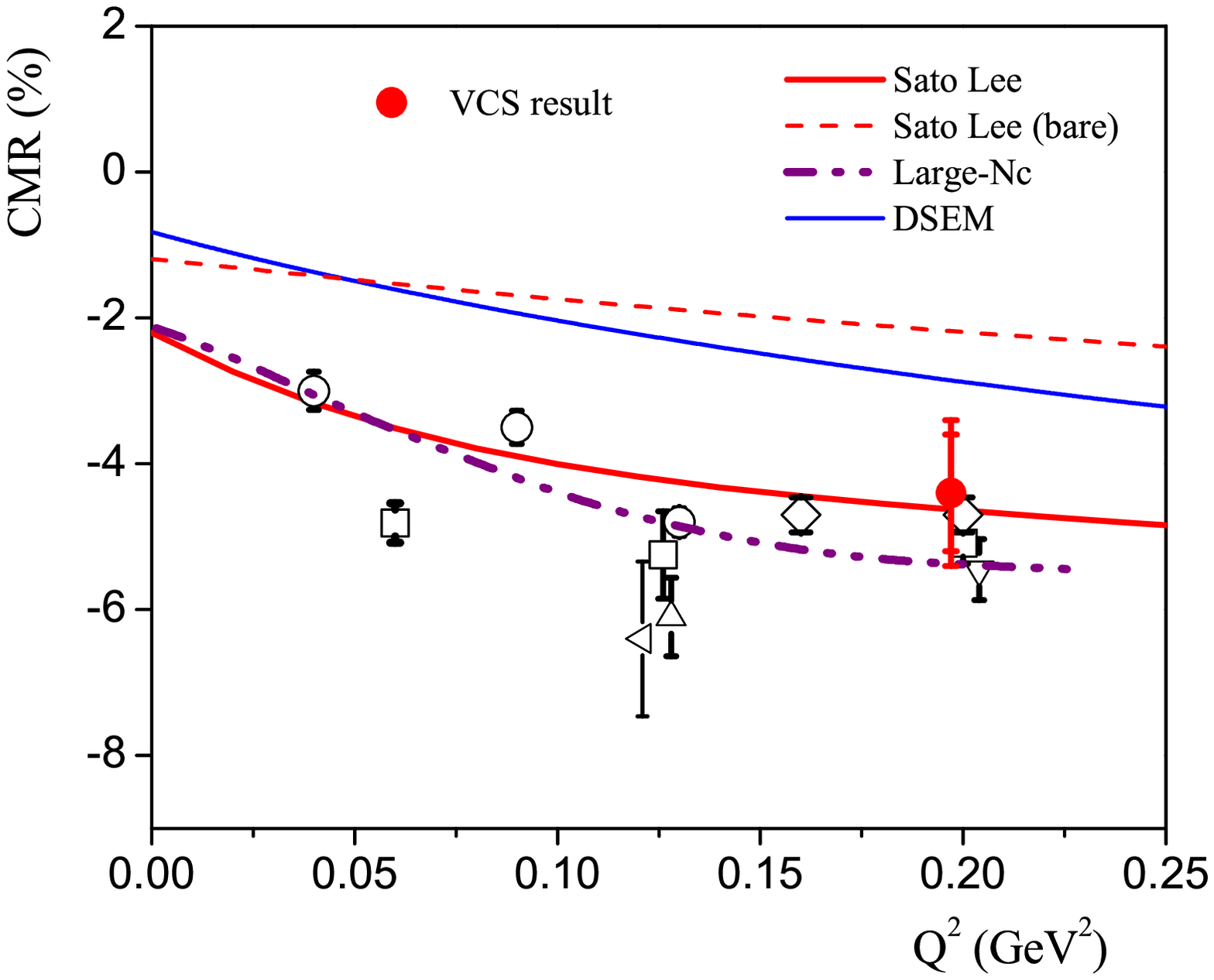,scale=0.7}
\end{minipage}
\begin{minipage}[t]{16.5 cm}
\caption{
(color online) The VCS result for the CMR (solid circle) from the MAMI-V experiment as compared to the pion channel measurements (open symbols). Figure taken from Ref.~\cite{Blomberg:2019caf}. The theoretical predictions of  Sato-Lee\cite{Sato:2000jf}, the large-Nc calculation~\cite{Pascalutsa:2007wz}, and the Dyson-Schwinger equation model (DSEM)~\cite{Segovia:2014aza} are also shown.
\label{f3n}}
\end{minipage}
\end{center}
\end{figure}

MAMI-III~\cite{Sparveris:2008jx} was the first experiment to achieve an exploration of the $N \rightarrow \Delta$ transition amplitudes through the photon channel. A first measurement of the dominant magnetic dipole amplitude in the transition was performed, and the result was found in excellent agreement to the corresponding result from the pion channel (see Fig.~\ref{f4n}). That was an important step for the $N \rightarrow \Delta$ program and the experience gained from these measurements provided guidance for planning the next measurement (MAMI-V) that would focus on the central part of this program, the quadrupole amplitudes in the transition. 
The MAMI-V experiment~\cite{Blomberg:2019caf} achieved the first extraction of the $N \rightarrow \Delta$ Coulomb quadrupole amplitude through the VCS channel. The measurement of the azimuthal asymmetries were proven very beneficial as the systematic uncertainties were constrained to a level comparable to the statistical ones.  The sensitivity of these measurements to the Coulomb quadrupole amplitude is exhibited in Fig.~\ref{f2n}. The Coulomb quadrupole was measured at $Q^2=0.20$ GeV$^2$, and the result CMR$_{(\mathrm{VCS})} = (-4.4 \pm 0.8_{stat} \pm 0.6_{sys})\%$ validated the pion channel world data, where the corresponding measurement is CMR $ = (-5.09 \pm 0.28_{stat+sys} \pm 0.30_{model})\%$ (see Fig.~\ref{f3n}). The results demonstrated that a good control of the model uncertainties was achieved, and gave further credence to the theoretical interpretation that the $\Delta (1232)$ resonance consists of a bare quark-gluon core and of a large pion-cloud contribution.

\section{Spatial density interpretation of the generalized polarizabilities}
\label{sec-spatial-density}

As described in Sect.~\ref{sec-from-rcs-to-vcs}, the $Q^2$ dependence of the GPs allows one to probe the spatial deformations of  the charge and magnetization densities, when the nucleon is subject to an external static electromagnetic field~\cite{Lvov:2001zdg,Gorchtein:2009qq}. The  formal connection between the GPs and the spatial densities of induced polarizations has been derived in Ref.~\cite{Gorchtein:2009qq}. 
In order to define proper spatial densities, i.e., with a true probabilistic interpretation without relativistic corrections,  one should consider the VCS process in a symmetric light-front frame, where the direction of the average nucleon momentum $\mathbf{P}=(\mathbf{p}+\mathbf{p}')/2$ is taken as the $\hat z$ axis and the momentum transfer to the nucleon $\Delta^\mu=(p'^\mu-p^\mu)$ is purely transverse. In this frame, the transverse components of the virtual photon momentum $\mathbf{q}_\perp$, with $Q^2=\vert \mathbf{ q}_\perp\vert^{2}$, are the conjugate variables to the transverse position $\mathbf{b}_{\perp}$, which  measures the transverse distance from the (transverse) center of momentum~\cite{Burkardt:2002hr,Soper:1972xc}.

In the following, we will  consider the polarization vector $\bm{\varepsilon}'_\perp$ of the outgoing photon corresponding with an applied electric field $\mathbf E\sim iq'_0\bm\varepsilon'_\perp$  that polarizes the charge distribution of the nucleon. Depending on the spin polarization of the nucleon, we have two different  induced polarization vectors.  They can be expressed in terms of GPs, and are functions only of the transverse photon momentum  $\mathbf q_\perp$. Therefore,  by a Fourier transform from $\mathbf q_\perp$ to $\mathbf b_\perp$, they provide a map of the deformation of the charge density in the transverse position space. The explicit relation between the GPs and the induced polarizations can be found in Ref.~\cite{Gorchtein:2009qq}. 
Figure~\ref{fig-spatial-density} shows the induced deformation in transverse-position space for an unpolarized  proton (left panel) and for a proton with the spin in the $\hat x$ direction aligned with the applied electric field (right panel), as calculated using the DR results of the GPs with the mass scale parameters $\la=0.73$ GeV and $\lb=0.63$ GeV~\cite{Pasquini:2018wbl}. 
In the  case of unpolarized proton, the polarization density displays a dipole pattern in the same direction of the applied field, mainly due to the contribution from the scalar GPs. The spatial extension  at the nucleon periphery strongly depends on the mass scales and the assumptions on the functional form of the asymptotic contributions. In the  case of a transversely polarized proton in the $\hat x$ direction, we observe a dipole deformation confined near the centre, and, on top of that, a quadrupole pattern with pronounced strength around 0.5 fm due to the electric GP.

\begin{figure}[tb]
\centerline{
\includegraphics[width = .42 \textwidth,angle=-90]{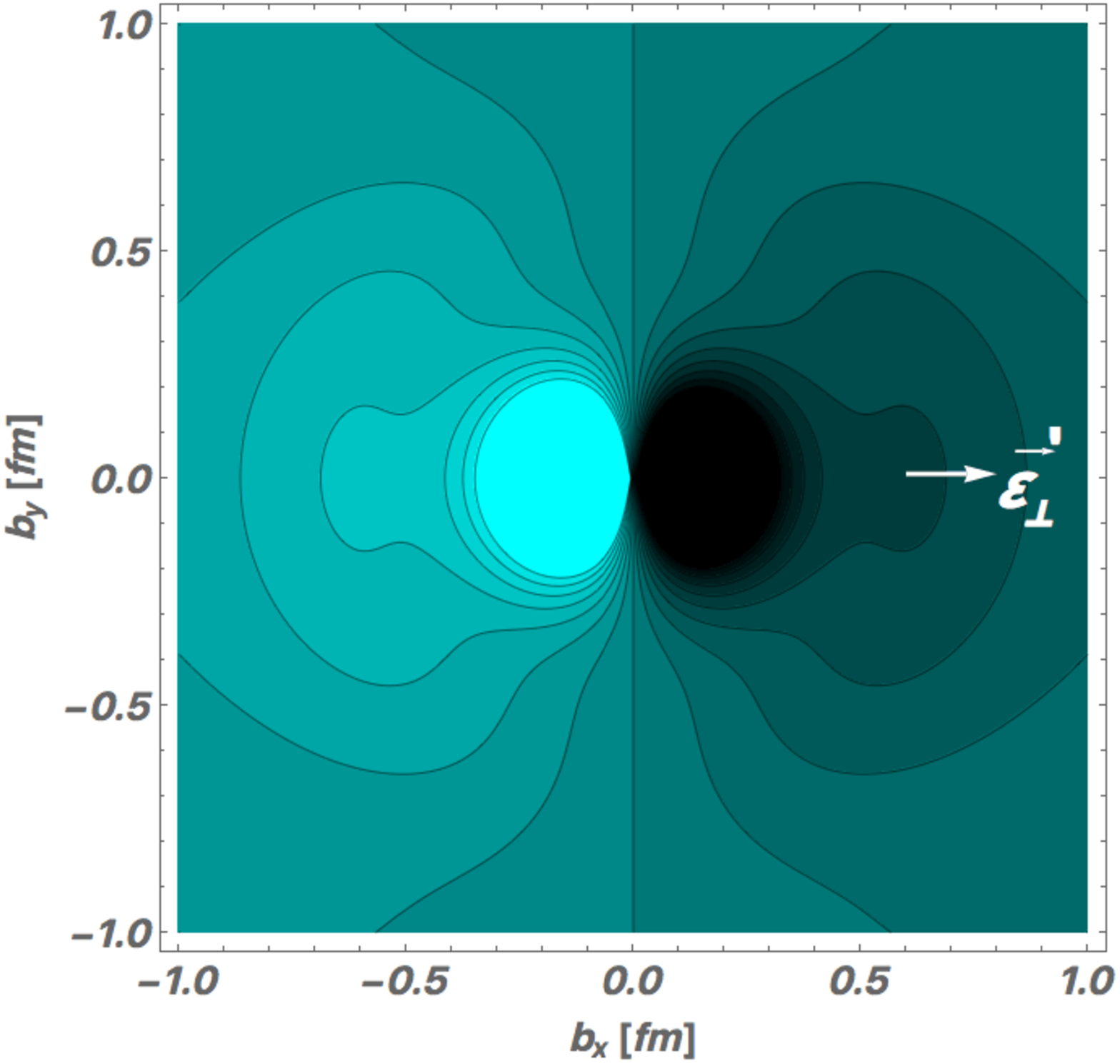}
\includegraphics[width = .42 \textwidth,angle=-90]{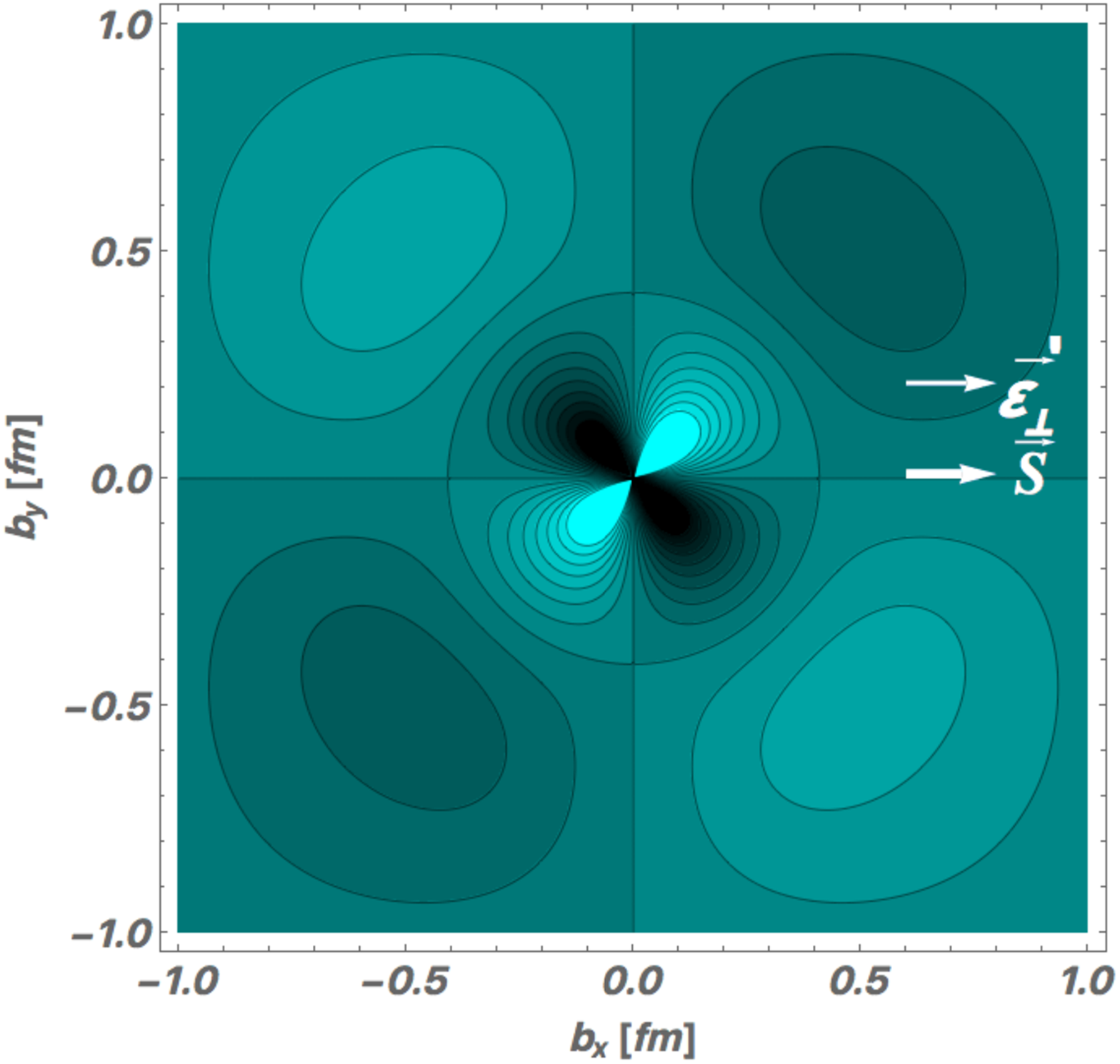}
}
\vspace{0.25 truecm}
\caption{ (color online) Induced polarization density  in an unpolarized  proton (left), and with spin $\boldmath{S}$ along the $\hat x$-axis (right), when submitted to an electromagnetic field with photon polarization along the $\hat x$-axis, as indicated. Light (dark) regions correspond to the largest (smallest) values. Figure taken from Ref.~\cite{Pasquini:2018wbl}.
}
\label{fig-spatial-density}
\end{figure}

\section{Conclusions and Outlook}
\label{sec-outlook-and-perspectives}

We should mention that, by focusing on the reaction $\vcsreactnucleon$, this review has covered only a part of the  ``polarizabilities'   world''. For sake of completeness, it may be useful to recall that polarizabilities in their most general form depend on the virtuality of the initial ($Q^2_i$) and final ($Q^2_f$) photons involved. ($Q^2_i = Q^2_f = 0 $) is the RCS case, while $(Q^2_i$ spacelike, $Q^2_f = 0$) is the VCS case of this review. Furthermore, ($Q^2_i$ spacelike or $Q^2_i= 0 $, $Q^2_f$ timelike) is the timelike Compton scattering case, and ($Q^2_i = Q^2_f$ = spacelike) is the VVCS case, studied experimentally via inclusive $(e,e')$ scattering
~\footnote{We wish to point out that  the inclusion of VCS data  in the generalized Baldin sum rule of VVCS, as was done in Fig.~4 of Ref.\cite{Sibirtsev:2013cga}, is inappropriate.}.

Discussing this entire field  is of course beyond the scope of the present review, and we restricted ourselves to discuss the VCS process at low energy. VCS offers a rich theoretical and experimental playground, that allows unique studies of the nucleon structure. Experiments conducted so far, in which data from MAMI have taken a prominent part, lead to a consistent picture of the electric and magnetic GPs of the proton, in the $Q^2$-range $\sim$ 0-2 GeV$^2$.  The data also raise questions, which will be addressed by the recently performed JLab experiment, in the intermediate $Q^2$ range. We have presented at length how the many facets of the DR model can be used in experimental VCS analyses. At the highest $Q^2$ it remains probably the only  approach to measure GPs with good precision. We hope that more dedicated VCS experiments will come to life in order to fill the gaps in our knowledge and understanding of the  nucleon GPs, including the spin GPs which will be a new challenge to the skills of experimentalists and theorists.

In this section we present a panel of ongoing and future developments in the field of VCS, covering both theoretical and experimental aspects.

\subsection{Theoretical front}
\label{sec-outlook-theory}

The joint experimental and theoretical efforts in the last years have allowed us to identify  a set of response functions that can be extracted from the Compton scattering process at different energy scales and in different kinematical conditions, and  have a clear interpretation in terms of structure properties of the nucleon. 
Low-energy Compton scattering provides information on global as well as spatially resolved electromagnetic properties of the nucleon in terms of  static and generalized polarizabilities, and, with increasing energy, allows us to study the effects of the nucleon excitation spectrum through dynamical (energy dependent) polarizabilities~\cite{Griesshammer:2001uw,Hildebrandt:2003fm,Pasquini:2017ehj}. 
Furthermore, the variation of the initial photon virtuality $Q^2$ allows one to probe a wide range of distance scales,  interpolating between hadronic degrees of freedom at low virtuality and partonic degrees of freedom at large virtuality. The unified description of the nucleon response functions in the whole $Q^2$ range is one of the main challenges of theoretical models.

Progress in this direction has recently been made in Refs.~\cite{Eichmann:2012mp,Eichmann:2011ec}, using  a Dyson-Schwinger/Faddeev approach. Unfortunately, some of the diagrams contributing to the process in this approach are numerically too hard to calculate. So far a practical solution has been proposed to give preliminary results only for the scalar GPs, although within certain approximations which violate gauge-invariance~\cite{Eichmann:2016tbi}.

A new formal approach to the description of the Compton scattering process has been recently addressed in Refs.~\cite{Belitsky:2012ch,Eichmann:2018ytt}. It can be viewed as a formalism that uses the same set of hadronic variables, the so-called Compton form factors, at large and low virtuality of the initial photon, and it has been suggested in~\cite{Belitsky:2012ch} to provide a unified framework for experimental studies of generalized parton distributions as well as generalized polarizabilities. 
Following this line, one could also explore the possibility to develop an unified dispersion relation formalism  for the Compton form factors in  different kinematical limits, connecting the existing DR approaches that deal separately with  either generalized parton distributions (see, e.g.,~\cite{Diehl:2007jb,Anikin:2007yh,Pasquini:2014vua}) or  polarizabilities~\cite{Pasquini:2018wbl,Drechsel:2002ar}.

Further progress in the DR approach for VCS at low energy could be to developing a subtracted DRs formalism along the lines of the subtracted DR framework used in RCS~\cite{Drechsel:1999rf,Pasquini:2007hf}. By choosing the subtraction point at the polarizability point $\nu=0,$ and $t=-Q^2$, one can write down the VCS amplitudes as the sum  of  $s$- and $t$-channel subtracted dispersion integrals, and  subtraction constants given in terms of the six leading-order GPs. The subtracted $s$-channel integrals can be evaluated through photo- and electro-production amplitudes, as described in this work, while the $t$-channel integrals can be saturated by $\pi \pi$ intermediate states in the $t$ channel $\gamma^*\gamma\rightarrow \pi \pi\rightarrow N\bar N$. The input for the subprocess $\gamma^*\gamma\rightarrow \pi \pi$ can  be taken, for example, from the recent dispersion analysis within a coupled-channel approach of Ref.~\cite{Danilkin:2018qfn}, while  the $\pi \pi\rightarrow N\bar N$ subprocess can be described as in the RCS case~\cite{Drechsel:1999rf}. The main advantage of  subtracted DRs for VCS is  that all the six GPs can be taken as free fit parameters to be adjusted to data. Furthermore, the model dependence introduced by the high-energy contributions to the dispersion integrals is considerably reduced.

\subsection{Experimental front}

\subsubsection{Experimental access to spin GPs}
\label{sec-access-spin-gps}

Although the polarizability phenomenon in the spin-dependent sector is not subtended by a simple and intuitive picture as in the scalar case, it is an essential piece of knowledge of nucleon structure, that calls for measurements. Unfortunately, investigating the nucleon spin GPs remains a virgin field so far; a few exploratory paths are outlined below.

A first perspective is offered by the $\ptt$ structure function, which is a combination of two spin GPs (cf. Eq.~(\ref{formula-sfs-combinations-of-gps-1})). The advantage is that $\ptt$ appears in the LEX, and can be in principle disentangled  from $\pll$ if one performs (unpolarized) measurements at several values of $\epsilon$. The difficulty of such an $\epsilon$-separation lies in the smallness of $\ptt$, according to the DR and covariant BChPT calculations of Fig.~\ref{fig-theo-sf-2}.

 Another strategy consists in combining unpolarized and doubly polarized observables in one experiment, at a single value of  $\epsilon$. With an unpolarized analysis yielding the structure function $\pllptte$ and a doubly polarized analysis yielding the structure function $\pltperp$, which is another combination of $\pll$ and $\ptt$ (cf. the Appendix), one can in principle separate  $\pll$ and $\ptt$. This method is discussed in Refs.~\cite{JanssensPhD:2007,DoriaPhD:2008} and was tried in the MAMI-IV experiment, but without significant results. The two correlation lines between $\pll$ and $\ptt$ obtained in this experiment, from the unpolarized and polarized analyses, turned out to be almost identical, due to the choice of kinematics.  Considerations on more optimal kinematics for such a separation can be found in Ref.~\cite{JanssensPhD:2007}.

Lastly, possible developments of the DR model, as exposed in Sect.~\ref{sec-outlook-theory}, offer potentially a new way to access spin GPs, by letting all six lowest-order GPs be free parameters. Experiments performed in the Delta resonance region could benefit from their higher sensitivity to GPs. Similarly to RCS, observables using polarization degrees of freedom would probably need to be investigated in order to find optimal measurements in the spin-GP sector.

\subsubsection{The E12-15-001 experiment at JLab}
\label{sec-e12-15-001-experiment}

The E12-15-001 experiment~\cite{Sparveris:2016} at JLab completed its first phase of data taking recently (July 2019). The experiment utilized the SHMS and HMS spectrometers~\cite{shms,hms} in Hall~C to detect, respectively, electrons and protons in coincidence, while the reconstructed missing mass has been used for the identification of the photon. An electron beam of energy $E=4.55$ GeV and a 10~cm liquid hydrogen target were employed for the measurements. 
The experiment aims to explore the GPs within the range of $Q^2=0.3$ GeV$^2$ to $Q^2=0.75$ GeV$^2$ in order to investigate the non trivial evolution of $\aeq$ with the momentum transfer, and to provide a precise measurement of $\bmq$. The experiment phase space covers the nucleon resonance region (see Fig.~\ref{f6n}) and thus the DR analysis framework will be utilized for the extraction of the GPs from the measured cross sections and azimuthal asymmetries. 
For the low-$Q^2$ settings the electron spectrometer was placed in a small angle of $\approx 8^{\circ}$ and the relatively high singles rates in conjunction with the large acceptance of the SHMS spectrometer have been the limiting factor of the beam current to about $30~\mu A$ for these settings. For the higher momentum transfer settings this limitation is relaxed as one can easily run at double the beam current.

The cross sections will be measured with a statistical uncertainty of about $\pm~1.5\%$, while the systematic uncertainties will be the dominating factor being roughly double compared to the statistical ones. The uncertainties of the beam energy and of the spectrometer angles will introduce a systematic uncertainty to the cross section ranging from $\pm~1\%$ to $\pm~2.5\%$, depending on the setting. 
Other sources of systematic uncertainties involve the target density, detector efficiency, acceptance, and target cell background, each one of which is expected to contribute to about $\pm~0.5\%$. Systematic uncertainties related to the target length, beam charge, dead~time corrections, and contamination of pions under the photon peak will also contribute, but to a smaller extent. The uncertainty due to the radiative corrections will be $\pm~1.5\%$, while various parametrizations for the form factors will be utilized in the analysis. For the asymmetries, the systematic uncertainties are still larger compared to the statistical ones, but not as dominant as in the case of the cross sections, and they are expected to be at the order of $\approx~1\%$, in absolute asymmetry magnitude. 
The extraction of the GPs will be performed by a DR fit to the measured cross sections and azimuthal asymmetries. The primary source of uncertainty for both the electric and the magnetic GP will be the systematic ones, while the statistical uncertainty for both GPs is expected to be $\approx~70\%$ of the systematic one.

In Fig.~\ref{f5n} the projected cross sections and asymmetries
are presented for $Q^2=0.65$ GeV$^2$. The solid (red) and dashed
(blue) curves correspond to a variation of the electric GP from 
$ \aeq = 4.8 \times 10^{-4}$~fm$^3$    ($ \bmq = 1.1 \times 10^{-4}$~fm$^3$) to
$ \aeq =1.5 \times 10^{-4}$~fm$^3$ ($ \bmq =1.1 \times 10^{-4}$~fm$^3$). 
A variation of  $\bmq$ from $=0.4 \times 10^{-4}$~fm$^3$ to 
$=1.6 \times 10^{-4}$~fm$^3$ is presented by the two, dotted and
dashed-dotted, green curves in the cross section figures. The same
variation in $ \bmq$ is represented through the light blue band in
the asymmetry figure. One can
observe that above $\thcm \approx 160^{\circ}$ the $\bmq$
variation is affecting both cross sections in a systematically similar
way, and this is reflected as a cancellation of
the effect in the azimuthal asymmetry (suppression of the light blue band in
the corresponding $\thcm$ range). The projected measurements 
for $\aeq$ are presented in Fig.~\ref{f7n}. The results will map 
 the momentum transfer signature of the two scalar GPs with high precision, 
and will offer a valuable 
cross check to the MAMI measurements at $Q^2=0.33$ GeV$^2$.

\begin{figure}[tb]
\begin{center}
\begin{minipage}[t]{15 cm}
\epsfig{file=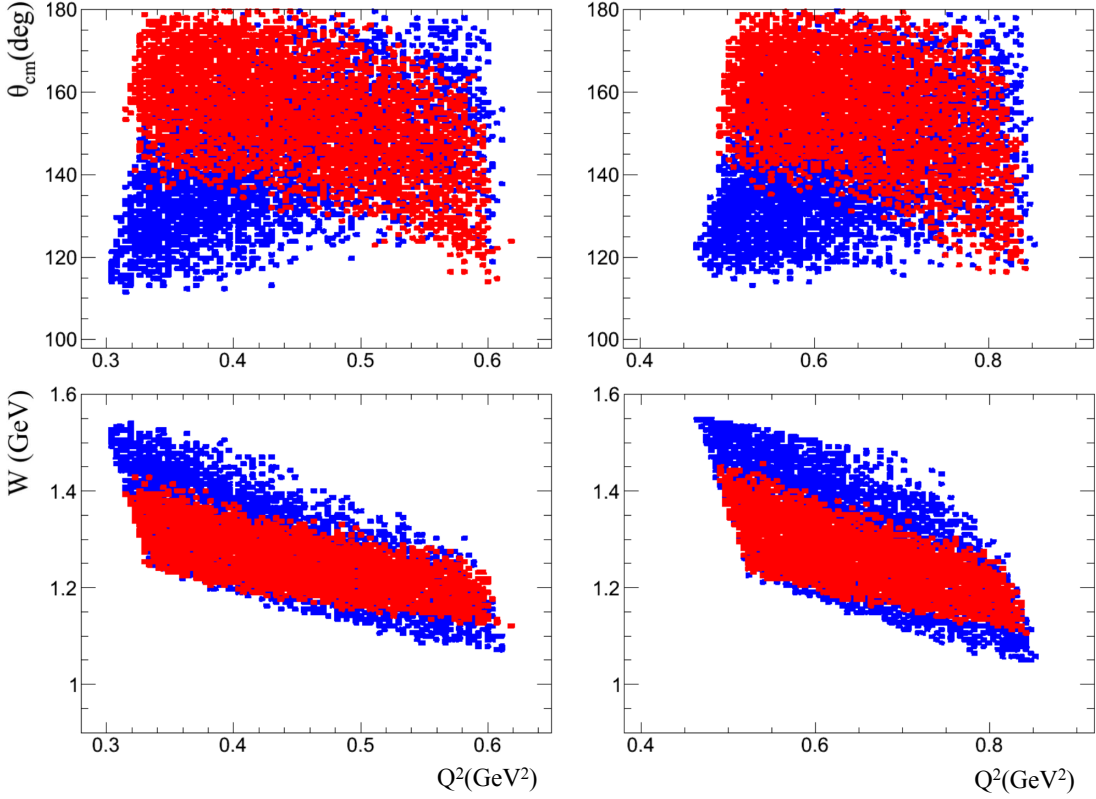,scale=0.6}
\end{minipage}
\begin{minipage}[t]{16.5 cm}
\caption{
(color online) Correlation of the phase space variables for a pair of measurements, at $\phicm=0^\circ$ (data in red) and $\phicm=180^\circ$ (data in blue). Left (top and bottom) and right (top and bottom) correspond to different settings of the JLab E12-15-001 experiment.
\label{f6n}}
\end{minipage}
\end{center}
\end{figure}

\begin{figure}[tb]
\begin{center}
\begin{minipage}[t]{16.5 cm}
\epsfig{file=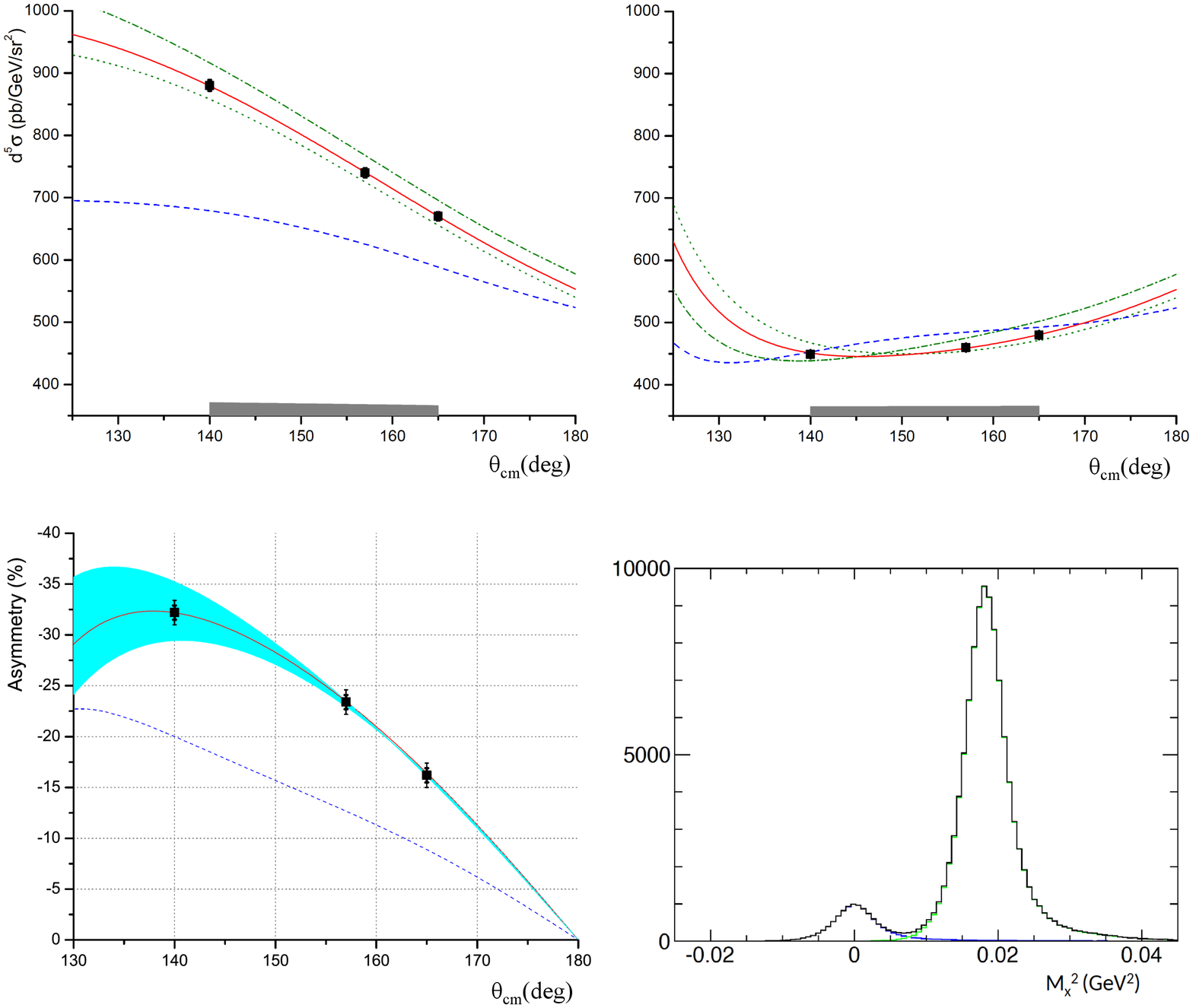,scale=0.55}
\end{minipage}
\begin{minipage}[t]{16.5 cm}
\caption{
(color online) Projected cross sections at $\phicm=0^{\circ}$ and  $180^{\circ}$ (top panels, right and left respectively) and azimuthal asymmetries (bottom left panel) at $Q^2=0.65$ GeV$^2$. The solid red (dashed blue) curve corresponds to
  $ \aeq = 4.8~10^{-4}$~fm$^3$, $ \bmq =1.1~10^{-4}$~fm$^3$ 
 ($ \aeq = 1.5~10^{-4}$~fm$^3$, $ \bmq =1.1~10^{-4}$~fm$^3$). 
A variation of $\bmq$ from $=0.4~10^{-4}$~fm$^3$ to $=1.6~10^{-4}$~fm$^3$ is presented through the two green curves (dotted, dashed-dotted) in the cross-section plots and as a light blue band in the asymmetry plot. The reconstructed missing mass spectrum is presented at the bottom right panel.
\label{f5n}}
\end{minipage}
\end{center}
\end{figure}

\begin{figure}[tb]
\begin{center}
\begin{minipage}[t]{15 cm}
\epsfig{file=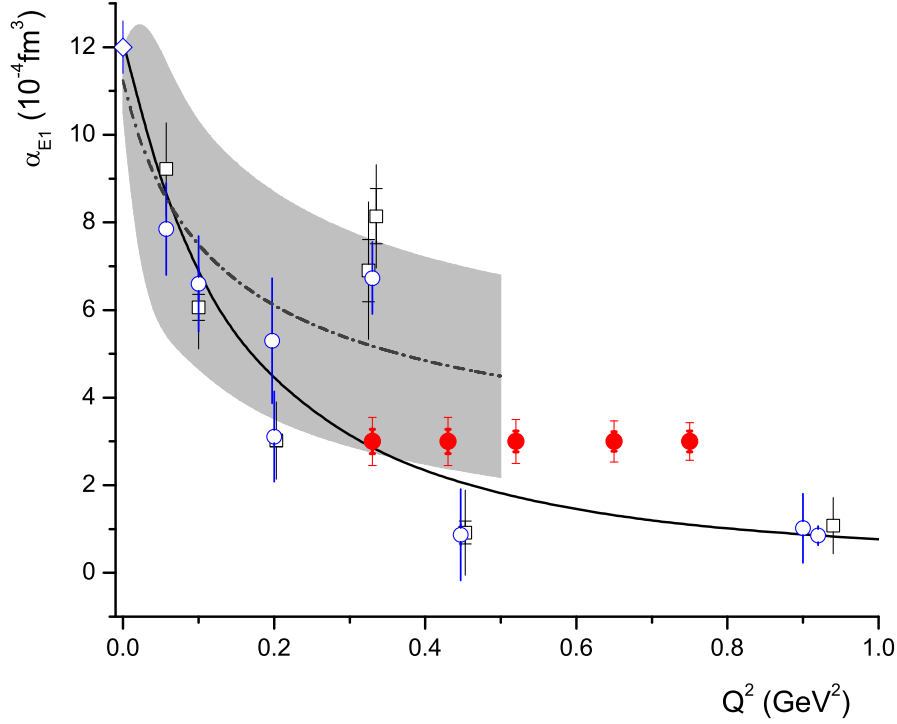,scale=0.8}
\end{minipage}
\begin{minipage}[t]{16.5 cm}
\caption{
(color online) The projected measurements for $\aeq$ from the E12-15-001 JLab experiment (filled cirles). The world data are shown as open symbols (open-circle: DR results, open-box: LEX results). The dashed-dotted curve corresponds to the result within covariant BChPT~\cite{Lensky:2016nui} and the solid curve shows the DR calculation with $\la = \lb = 0.7$ GeV.
\label{f7n}}
\end{minipage}
\end{center}
\end{figure}

\smallskip
\smallskip
\smallskip

{\bf \large Acknowledgments}

We wish to thank all people active in the VCS front: our theoretical colleagues  for continued interest and stimulating discussions, as well as our experimental colleagues from the  MAMI-A1, JLab and MIT-Bates Collaborations, for having made the VCS experiments possible and successful. Special thanks go to Vadim Lensky for providing the numerical results of the covariant BChPT calculation, Marc Vanderhaeghen for enlightening discussions and Luca Doria for some inputs to the manuscript. 
This work has been supported by the US Department of Energy award DE-SC0016577 and by the French CNRS-IN2P3.

\appendix
\section{Appendix}

This Appendix collects the expressions of a few elements entering the low-energy theorem for VCS, in the unpolarized and polarized cases. All formulas are taken from Ref.~\cite{Guichon:1998xv}.

\smallskip
\smallskip
\smallskip

\par\noindent $\bullet$ Cross section and phase-space factor: \newline
The $(\epgreact )$ cross section is defined  as: 
%
%
\bea
\begin{array}{lll}
 \displaystyle   {
d^5 \sigma 
\over 
 d k'_{\mathrm{lab}} d \cos \theta'_{e \, \mathrm{lab}} d \varphi'_{e \, \mathrm{lab}} d \cthcm d \phicm
} 
&  = &  ( \Phi \cdot \qpr ) \cdot {\cal M}  \ , \\
\end{array}
\eea
%
%
where the phase-space factor is defined as:  
%
%
\bea
\begin{array}{lll}
 ( \Phi \cdot \qpr ) & = &  
 \displaystyle { (2 \pi)^{-5} \over 64 \mnucleon }
\cdot
{k'_{\mathrm{lab}} \over k_{\mathrm{lab}} } 
\cdot
{s - \mnucleon ^2 \over s }  . \\
\end{array}
\eea
%
%

In this expression, one can factor out an explicit factor $\qpr$,  since  $(s - \mnucleon^2)$  is proportional to $\qpr$. The  complementary part of the phase-space factor,  \ $\Phi \ = \ { (2 \pi)^{-5} \over 64 \mnucleon } \cdot {k'_{lab} \over k_{lab} }  \cdot {2 \over \sqrt{s} } $ , remains finite when $\qpr$ tends to zero.

\smallskip
\smallskip
\smallskip

\par\noindent $\bullet$ $\vll$ and $\vlt$ coefficients: \newline
There are several notations in the literature for the coefficients which are in front of the structure functions in the LEX formula. Here we have used the following notations:
%
\bea
\begin{array}{lll}
 \vll &  = &   2 K_2 \cdot v_1 \cdot \epsilon  \\
 \vlt &  = &   2 K_2 \cdot (v_2 - 
{\displaystyle \qzerotild \over \displaystyle \qcm} 
\cdot v_3 ) \cdot \sqrt{ 2 \epsilon (1+\epsilon)}  \\
\end{array}
\eea
%
where $K_2, v_1 , v_2$ and $v_3$ are defined in Eqs.(98)-(100) of Ref.~\cite{Guichon:1998xv}, and  $\qzerotild$ is the virtual photon c.m. energy in the limit $\qpr \to 0$, given by:  $\qzerotild = \mnucleon - \sqrt{ \mnucleon^2 + \qcm ^2}$.

\smallskip
\smallskip
\smallskip

\par\noindent $\bullet$ Structure functions at $Q^2=0$: \newline
The expression of the measured structure functions $\pllptte$ and $\plt$ at $Q^2=0$ in terms of the RCS polarizabilities is obtained by applying Eq.~(\ref{formula-sfs-combinations-of-gps-1}) for a real incident photon, together with Eq.~(\ref{formula-aeq-bem}) and using the ``tilde'' variables:
%
%
\bea
\begin{array}{lll}
\pll (0) &  = & { 4 \mnucleon \over \aqed } \cdot \ale (0) \\
\ptt (0) & = & 0 \\
\plt (0) & = & { -2  \mnucleon \over \aqed } \cdot \bem  (0)  \\
\end{array}
\eea

\smallskip
\smallskip
\smallskip

\par\noindent $\bullet$ VCS with double polarization: \newline
Three new structure functions appear in the first-order GP term of the doubly polarized cross section. They are combinations of spin GPs only:
%
%
\bea
\begin{array}{lll}
\pltz        (Q^2) &  = & 
{ 3 \qtild \qcm \over  2 \qzerotild } 
 G_M^p (Q^2) \cdot P^{(L1,L1)1} (Q^2) \, - \, 
{ 3 \mnucleon \qcm \over \qtild   } 
 G_E^p (Q^2) \cdot P^{(M1,M1)1} (Q^2) \ , \\
\pltzprim    (Q^2) &  = & 
- { 3 \qtild \over 2 }  
 G_M^p (Q^2) \cdot P^{(L1,L1)1} (Q^2) \, + \, 
{ 3 \mnucleon \qcm^2 \over \qtild \qzerotild  } 
 G_E^p (Q^2) \cdot P^{(M1,M1)1} (Q^2) \ , \\
\pltperpprim (Q^2) &  = & 
{ 3 \qtild \qcm \over  2 \qzerotild } 
 G_M^p (Q^2) \cdot 
{\bigg (} \, 
P^{(L1,L1)1} (Q^2) \, - \, \sqrt{3 \over 2} \qzerotild  
 \cdot P^{(M1,L2)1} (Q^2) 
\, {\bigg )} \ . \\
\end{array}
\eea

The structure function  $\pltperpprim$ is the only one containing  the ``sixth GP'' $P^{(M1,L2)1}$, and it contributes only in out-of-plane kinematics. Three supplementary combinations are introduced:
%
%
\bea
\begin{array}{lll}
\pltperp        (Q^2) &  = & 
{ R  G_E^p(Q^2) \over  2  G_M^p(Q^2) } \cdot \ptt  (Q^2)  \, - \, 
{  G_M^p(Q^2) \over 2 R  G_E^p(Q^2)  }  \cdot \pll (Q^2)  \ , \\
\pttperp   (Q^2) &  = & 
{ G_M^p(Q^2) \over R  G_E^p(Q^2) } 
\cdot ( \, \pltz  (Q^2) - \plt (Q^2) \, )  \\
  \ &  = &  
- {\qcm \over 2}   G_M^p(Q^2) \cdot
{\bigg (} \, 
3  P^{(M1,M1)1} (Q^2)  + \sqrt{3 \over 2}  P^{(M1,M1)0} (Q^2) 
\, {\bigg )} \ , \\
\pttperpprim  (Q^2) &  = & 
{ G_M^p(Q^2) \over R  G_E^p(Q^2) }
\cdot ( \, \pltzprim  (Q^2) + {\qzerotild \over \qcm } \cdot \plt  (Q^2) \, ) \\
\ &   = &  
{\qcm \over 2}   G_M^p(Q^2) \cdot
{\bigg (} \, 
3 {\qcm \over \qzerotild } \cdot  P^{(M1,M1)1} (Q^2)  + \sqrt{3 \over 2}  {\qzerotild \over \qcm } \cdot P^{(M1,M1)0} (Q^2) 
\, {\bigg )} \ , \\
\end{array}
\eea
%
%
where $R= 2 \mnucleon / \qtild$. A first measurement of $\pltperp$ was provided by the MAMI-IV experiment.

\bibliographystyle{unsrt}
\bibliography{vcsbiblio}
\end{document}